\documentclass[twocolumn,preprintnumbers,showpacs,pre,floatfix,superscriptaddress,amsmath,amssymb]{revtex4}
\usepackage{epsfig}
\usepackage{subfigure}
\usepackage{amsmath}
\usepackage{color}
\usepackage{amssymb}
\usepackage{setspace}
\usepackage{graphicx}
\usepackage{dcolumn}
\usepackage{bm}
\usepackage{times}
\usepackage{enumerate}
\input epsf

\graphicspath{{figures/}}

\begin{document}

\author{Per Sebastian Skardal}
\email{skardals@gmail.com} 
\affiliation{Departament d'Enginyeria Inform\`{a}tica i Matem\`{a}tiques, Universitat Rovira i Virgili, 43007 Tarragona, Spain}
\affiliation{Department of Applied Mathematics, University of Colorado at Boulder, Colorado 80309, USA}

\author{Alain Karma}
\affiliation{Physics Department and Center for Interdisciplinary Research on Complex Systems, Northeastern University, Boston, Massachusetts 02115, USA}

\author{Juan G. Restrepo} 
\affiliation{Department of Applied Mathematics, University of Colorado at Boulder, Colorado 80309, USA}

\title{Spatiotemporal Dynamics of Calcium-Driven Cardiac Alternans}


\begin{abstract}
We investigate the dynamics of spatially discordant alternans (SDA) driven by an instability of intracellular calcium cycling using both amplitude equations  [P. S. Skardal, A. Karma, and J. G. Restrepo, Phys. Rev. Lett. {\bf 108}, 108103 (2012)] and ionic model simulations. We focus on the common case where the bi-directional coupling of intracellular calcium concentration and membrane voltage dynamics produces calcium and voltage alternans that are temporally in phase. We find that, close to the alternans bifurcation, SDA is manifested as a smooth wavy modulation of the amplitudes of both repolarization and calcium transient (CaT) alternans, similarly to the well-studied case of voltage-driven alternans. In contrast, further away from the bifurcation, the amplitude of CaT alternans jumps discontinuously at the nodes separating out-of-phase regions, while the amplitude of repolarization alternans remains smooth.  We identify universal dynamical features of SDA pattern formation and evolution in the presence of those jumps. We show that node motion of discontinuous SDA patterns is strongly hysteretic even in homogeneous tissue due to the novel phenomenon of ``unidirectional pinning": node movement can only be induced towards, but not away from, the pacing site in response to a change of pacing rate or physiological parameter. In addition, we show that the wavelength of discontinuous SDA patterns scales linearly with the  conduction velocity restitution length scale, in contrast to the wavelength of smooth patterns that scales sub-linearly with this length scale. Those results are also shown to be robust against cell-to-cell fluctuations owing to the property that unidirectional node motion collapses multiple jumps accumulating in nodal regions into a single jump. Amplitude equation predictions are in good overall agreement with ionic model simulations. Finally, we briefly discuss physiological implications of our findings. In particular, we suggest that due to the tendency of conduction blocks to form near nodes, the presence of unidirectional pinning makes calcium-driven alternans potentially more arrhythmogenic than voltage-driven alternans.
\end{abstract}


\maketitle

\section{Introduction}\label{sec1}

Each year sudden cardiac arrest claims over 300,000 lives in the United States, representing roughly half of all heart disease deaths, and making it the leading cause of natural death~\cite{Weiss2006CircRes,Christini2012,Karma2013AR}. Following several studies that linked beat-to-beat changes of electrocardiographic features to increased risk for ventricular fibrillation and sudden cardiac arrest~\cite{Adam1984JCE,Ritzenberg1984Nature,Smith1988Circ}, the phenomenon of ``cardiac alternans'' has been widely investigated~\cite{Pastore1999Circ,Pastore2000CircRes,Qu2000Circ,Rosenbaum2001JCE,Fox2002CircRes,Watanbe2001JCE,Sato2006CircRes,Sato2007BiophysJ,Echebarria2007EPJ,Hayashi2007BiophysJ,Mironov2008Circ,Ziv2009JPhys,Weiss2006CircRes,Karma2007PhysicsToday,Aistrup2009CircRes,Weiss2011CircRes}. At the cellular level, alternans originates from a period doubling instability of the coupled dynamics of the transmembrane voltage (V$_m$) and the intra-cellular calcium concentration ([Ca$^{2+}$]$_i$). This instability is typically manifested as a long-short-long-short sequence of action potential duration (APD)  accompanied by an in-phase (out-of-phase) large-small-large-small (small-large-small-large) sequence of peak calcium concentration (Ca).

At a tissue scale, cardiac alternans can be either spatially concordant, with the whole tissue alternating in-phase, or spatially discordant with different regions alternating out-of-phase. In two dimensions, those out-of-phase regions of period-two dynamics are separated by nodal lines of period-one dynamics, which reduce to points or nodes in one-dimension. In their pioneering study that evidenced spatially discordant alternans (SDA)~\cite{Pastore1999Circ}, Pastore {\it et al.} further demonstrated that SDA provides an arrhythmogenic substrate that facilitates the initiation of reentrant waves, thereby establishing a causal link between alternans at the cellular scale and sudden cardiac arrest. Subsequent research has focused on elucidating basic mechanisms of formation of SDA and conduction blocks promoted by SDA~\cite{Pastore2000CircRes,Qu2000Circ,Rosenbaum2001JCE,Fox2002CircRes,Watanbe2001JCE,Sato2006CircRes,Sato2007BiophysJ,Echebarria2007EPJ,Hayashi2007BiophysJ,Mironov2008Circ,Ziv2009JPhys}.  

\subsection{Voltage-driven alternans}

To date, our basic theoretical understanding of SDA is well developed primarily for the case where alternans is ``voltage-driven''~\cite{Echebarria2002PRL,Echebarria2007PRE,Dai2008SIAM,Dai2010ESAIM,Karma2013AR}, i.e., originate from an instability of the V$_m$ dynamics. For a one-dimensional cable of length $L$, the $V_m$ dynamics is governed by the well-known cable equation
\begin{align}
\partial_t V_m = D_V\partial_x^2 V_m - I_{ion}/C_m,\label{eq:cable}
\end{align}
where $D_V$ is the diffusion coefficient, $I_{ion}$ describes the total flux of ion currents, $C_m$ is the cell membrane capacitance, and by convention we assume the cable is periodically paced at the end $x=0$. While the cable equation provides in principle a faithful description of the $V_m$ dynamics, it does not allow an analytical treatment of the alternans bifurcation. A fruitful theoretical framework for characterizing this bifurcation has been the use of iterative maps first applied to the cell dynamics~\cite{Nolasco1968JAP,Guevara1984IEEE}  and formulated in terms of the APD restitution properties. This relation describes the evolution of APD for an isolated cell and is given by
\begin{align}\label{eq:restitution1}
A_{n+1} = f(D_n),
\end{align}
where $A_{n+1}$ and $D_n$ are the APD and diastolic interval (DI) at beats $n+1$ and $n$, respectively. At the tissue scale, the diffusive coupling between cells influences the dynamics through the conduction velocity (CV) restitution relation, which describes how the depolarization wave speed depends on DI, defined here by the function $cv(D)$. CV restitution causes the activation interval  $T_n=A_n+D_n$ (the interval between the arrival of the $n^{th}$ and $n^{th}+1$ stimuli)  to vary along the cable, thereby coupling the maps (\ref{eq:restitution1}) in a non-local fashion as first shown in an analysis of the alternans bifurcation in a ring geometry~\cite{Courtemanche1993PRL}. Diffusive coupling also influences the repolarization dynamics. Starting from Eq.~(\ref{eq:restitution2}), Echebarria and Karma (EK)~\cite{Echebarria2002PRL,Echebarria2007PRE} showed that this effect can be captured by 
 a non-local spatial coupling between maps of the form
\begin{align}\label{eq:restitution2}
A_{n+1}(x) = \int_0^LG(x,x')f[D_n(x')]dx',  
\end{align}
where $A_{n+1}(x)$ and $D_n(x)$ are the APD and DI at beats $n+1$ and $n$, respectively, at location $x$ along the cable, and $G$ is a Green's function that encompasses the non-local electrotonic coupling along the cable due to the diffusion of $V_m$. For a simple choice of ionic model, EK  derived an analytical expression for the Green's function $G$. Furthermore, they carried out a weakly nonlinear multiscale expansion of the system of spatially coupled maps close to the alternans bifurcation [i.e., the maps (\ref{eq:restitution1}) at each point along the cable coupled non-locally by CV restitution and the Green's function]. This expansion of the form $A_n=A^*+(-1)^na+\dots$ assumes that $a$ is small close to the bifurcation point, where $A^*$ is the fixed point value of the APD at this point. Secondly, it exploits the fact that the alternans amplitude $a$ varies slowly in space on the diffusive scale $\xi \sim \sqrt{D_VA^*}$ that characterizes the spatial range of the Green's function $G$. The spatial scale of the variation of $a$ is generally characterized by the wavelength $\lambda_s$ of SDA equal to twice the spacing between nodes. Exploiting the fact that $a$ is small ($a\ll A^*$) and $\xi\ll \lambda_s$, EK reduced the system of spatially coupled maps to the integro-partial-differential equation~\cite{Echebarria2002PRL,Echebarria2007PRE}
\begin{equation}
\tau_{bcl}\partial_ta=\sigma a -\chi a^3+\xi^2\partial_x^2a-w\partial_xa -\frac{1}{\Lambda}\int_0^x a(x') dx',\label{EKVm}
\end{equation}
where $\Lambda$ is a lengthscale related to the slope of the CV restitution curve, $w$ is a short lengthscale $\sim D_V/cv(D^*)$, $\tau_{bcl}$ is the pacing period or basic cycle length (BCL), and $\sigma$ and $\chi$ can be expressed in terms of derivatives of the APD restitution curve evaluated at the fixed point, and $\sigma$ measures the distance from the bifurcation point.  

Analysis of this amplitude equation has yielded a fundamental understanding of the formation of SDA in terms of a linear instability forming periodic wave patterns~\cite{Echebarria2002PRL,Echebarria2007PRE,Dai2008SIAM,Dai2010ESAIM}. Depending on the relative magnitude of $\xi$, $w$ and $\Lambda$, SDA formation is associated with a bifurcation to standing waves with stationary nodes and a wavelength $\lambda_s \sim (w\Lambda)^{1/2}$ or traveling waves with moving nodes and $\lambda_s \sim (\xi^2\Lambda)^{1/3}$~\cite{Echebarria2002PRL,Echebarria2007PRE}. The assumption $\xi\ll \lambda_s$ under which Eq.~(\ref{EKVm}) is derived holds in both cases owing to the fact that both $w$ and $\xi$ are much smaller than $\Lambda$ for typical physiological parameters. This equation has also been instrumental for developing methods of controlling and suppressing alternans~\cite{Echebarria2002Chaos,Jordan2004JCE,Christini2006PRL,KroghMadsen2010PRE}.

\subsection{Calcium-driven alternans}    

While Eq.~(\ref{EKVm}) is derived under the assumption that alternans is voltage-driven, both laboratory and numerical experiments have shown that alternans can also be calcium-driven, i.e., mediated by an instability in the intracellular [Ca$^{2+}$]$_i$ dynamics~\cite{Chudin1999BiophysJ,Shiferaw2003BiophysJ,Pruvot2004CircRes,Bien2006BiophysJ,Picht2006,Qu2007,Restrepo2008BiophysJ,Rovetti2010,Alvarez2012,Entcheva2012BiophysJ,Sato2013PLOS}. Calcium alternans drive repolarization alternans owing to the well-known property that V$_m$ and [Ca$^{2+}$]$_i$ dynamics are bi-directionally coupled~\cite{Bers2001}. Membrane depolarization activates the L-type calcium current $I_{Ca}$ and Ca$^{2+}$ entry into the cell triggers Ca$^{2+}$ release from the sarcoplasmic reticulum (an intracellular calcium store). The transient rise of [Ca$^{2+}$]$_i$ known as the calcium transient (CaT) in turn influences calcium-sensitive membrane currents. The increase of [Ca$^{2+}$]$_i$ tends to inactivate $I_{Ca}$, thereby shortening the APD, but drives the sodium-calcium exchanger $I_{NCX}$ current into a forward mode of Ca$^{2+}$ extrusion that is depolarizing (i.e., three Na$^{+}$ ions are exchanged with one Ca$^{2+}$ ion across the membrane), thereby prolonging the APD. Consequently, depending on the balance of those two currents, the net effect of the CaT can be to prolong or shorten the APD and, concomitantly, produce Ca alternans that are in phase or out of phase with $V_m$ alternans. The condition leading to in phase (out of phase) Ca and $V_m$ alternans at a cellular level has been identified as positive (negative) Ca-to-$V_m$ coupling, which is associated with dominance of $I_{NCX}$ ($I_{Ca}$)~\cite{Shiferaw2005PRE,Restrepo2009Chaos,Wan2012}. 

Simulations of ionic models have demonstrated that calcium-driven alternans can exhibit more complex spatiotemporal behaviors on a tissue scale than voltage-driven alternans~\cite{Sato2006CircRes,Sato2007BiophysJ,Zhao2008PRE,Christini2012,Karma2013AR}. A qualitatively distinguishing feature of the calcium-driven case is that the amplitude and phase of CaT alternans can jump discontinuously in space. Such jumps are possible because Ca$^{2+}$ diffusion is several orders of magnitude smaller than $V_m$ diffusion, both intracellularly and across cells. Hence [Ca$^{2+}$]$_i$ varies rapidly over a scale of a few microns, thereby allowing CaT alternans to become spatially discordant in two neighboring cells or even in two regions of the same cell. For negative Ca-to-$V_m$ coupling, those subcellular discordant CaT alternans have been shown to be promoted by a Turing-like instability mediated by $V_m$ and Ca$^{2+}$ diffusion~\cite{Shiferaw2006,Gaeta2009,Restrepo2009Chaos}. This instability makes patterns of alternans on a tissue scale quite complex~\cite{Sato2006CircRes,Zhao2008PRE,Sato2013PLOS}. Patterns can display several jumps of CaT amplitude, with a wide range of spacings between jumps. Furthermore, pattern formation can be strongly history-dependent~\cite{Zhao2008PRE,Sato2013PLOS}.

While negative Ca-to-$V_m$ coupling can be induced by feedback control of the pacing interval~\cite{Gaeta2009} or pharmacologically~\cite{Wan2012}, and may occur in certain pathologies such as heart failure, positive Ca-to-$V_m$ coupling is more often seen in experiments and is believed to be more prevalent. For this case, the CaT alternans amplitude can also become spatially discontinuous even in the absence of a Turing instability~\cite{Sato2006CircRes,Sato2007BiophysJ}. The spatial gradient of CaT amplitude has been found in one-dimensional ionic model simulations to become steeper in the nodal region with increasing strength of calcium-driven instability~\cite{Sato2006CircRes}. Based on this observation, it was proposed that the steepness of the CaT amplitude profile in the nodal region could be used to distinguish between cases where alternans is voltage-driven and calcium-driven~\cite{Sato2006CircRes}. In this qualitative picture, a clear signature of the calcium-driven case would be the observation of a CaT alternans profile that is significantly steeper than the APD alternans profile or even discontinuous. 

At a more quantitative level, the formation and dynamical consequences of spatial discontinuities in the CaT alternans profile remains poorly understood. From a theoretical standpoint, it would be desirable to generalize the amplitude equation approach to develop a basic understanding of calcium-driven SDA patterns for positive Ca-to-$V_m$ coupling. This extension is in principle straightforward close to the alternans bifurcation, where the amplitudes of APD and CaT alternans [$a$ and $c$, respectively, where the amplitude of the CaT at beat $n$ is expanded in the form $C_n=C^*+(-1)^n c+\dots$] vary smoothly, i.e. both $a$ and $c$ vary on a scale larger than the range $\xi$ of the $V_m$ diffusive coupling. We indeed confirm in the present work by a linear stability analysis that, close to the bifurcation, the wavelength of smooth SDA patterns is governed by the same scaling laws for calcium- and voltage-driven alternans, consistent with the expectation that Eq.~(\ref{EKVm}) provides a universal description of those patterns near onset.
However, the divergence of the spatial gradient of $c$ further away from the bifurcation renders the multi-scale expansion leading to Eq.~(\ref{EKVm}) invalid. In particular, $c$ can vary on a scale shorter than $\xi$ or even become discontinuous and $a$ can vary on a scale comparable to $\xi$, but not much larger than $\xi$.   
 
In a previous report~\cite{Skardal2012PRL} we showed that, even in the absence of a multi-scale expansion, analytical insights into the formation of SDA can be obtained by investigating a reduced system of coupled integro-difference equations. This system describes beat-to-beat variations of the amplitudes of voltage and calcium alternans and handles discontinuous jumps in Ca alternans amplitude. It is derived by assuming a simple generic form for the iterative maps of the local bi-directionally coupled $V_m$-$[Ca]_i$ dynamics and couples those maps spatially using the CV-restitution relation and the diffusive Green's function defined by Eq.~(\ref{eq:restitution2}). We showed previously~\cite{Skardal2012PRL} that this system reproduces the transition from smooth to discontinuous SDA patterns with increasing strength of calcium-driven instability and highlighted a range of novel dynamical behavior including a hysteresis of node motion related to the novel phenomenon of unidirectional pinning. In this paper we develop this theory further, providing a more complete picture of the spatiotemporal dynamics of calcium-driven alternans for the case of positive Ca-to-$V_m$ coupling.

\subsection{Outline}

The rest of this paper is organized as follows. In Sec.~\ref{sec2} we present the full derivation of the reduced system describing the amplitude of calcium and voltage alternans. We assume here that alternans is mediated by an instability in the [Ca$^{2+}$]$_i$ dynamics and account for bi-directional coupling between [Ca$^{2+}$]$_i$ and V$_m$ dynamics. In Sec.~\ref{sec3} we present a numerical survey of the reduced system and a description of its phase space. We also present evidence that the dynamics of the reduced system robustly captures the dynamics exhibited by a detailed ionic model. In Sec.~\ref{sec4} we use a linear stability analysis to quantify the bifurcation characterizing the onset of alternans, as well as the spatial properties of solutions in the smooth regime that appears immediately after onset. We find that the wavelength $\lambda_s$ of stationary and traveling SDA patterns obeys the same scalings as predicted by Eq.~(\ref{EKVm}) for the case of voltage-driven alternans. This finding is consistent with the expectation that Eq.~(\ref{EKVm}) provides a universal description of alternans dynamics close to the alternans bifurcation. In Sec.~\ref{sec5} we continue our analysis by studying the strongly nonlinear regime where discontinuous patterns form. This analysis includes a description of the unique hysteresis found for large degrees of instability. In Sec.~\ref{sec6} we present several numerical experiments of a detailed ionic model. First, we show that the novel phenomena described by our reduced model can be observed in ionic models. Second, we use results from our reduced model to quantitatively predict dynamics in ionic models. Finally, in Sec.~\ref{sec7} we close with a discussion of spatially discordant alternans, physiological implications of our work, and other conclusions.

\section{Derivation of the Amplitude Equations}\label{sec2}

We now present a detailed derivation of a reduced system of integro-difference equations governing the dynamics of the amplitude of calcium and voltage alternans along a one-dimensional cable assuming a calcium-mediated instability. In this paper we will restrict our analysis to a one-dimensional cable of tissue. This case has experimental relevance~\cite{Boyden2000CircRes,Christini2006PRL} and can be later generalized to two dimensions. As in Refs.~\cite{Echebarria2002PRL,Echebarria2007PRE,Restrepo2008BiophysJ} we will assume that the cable has length $L$ with ends located at $x=0$ and $x=L$, and that the cable is periodically paced at the $x=0$ end. We add to Eq.~(\ref{eq:restitution2}) an equation describing the evolution of [Ca$^{2+}$]$_i$ dynamics as well as account for the bi-directional coupling between [Ca$^{2+}$]$_i$ and V$_m$. Thus, we begin from the coupled system of equations
\begin{align}
C_{n+1}(x) &= f_c\left[C_n(x),D_n(x)\right], \label{eq:CaGeneral} \\
A_{n+1}(x) &= \int_0^L G(x,x')f_a\left[D_n(x'),C_{n+1}(x')\right]dx', \label{eq:APDGeneral}
\end{align}
where $C_{n}(x)$ gives the peak calcium concentration Ca at beat $n$ at location $x$ along the cable. We note that the calcium dynamics in Eq.~(\ref{eq:CaGeneral}) are not spatially coupled due to the fact that diffusion of calcium occurs on a timescale that is much slower than the diffusion of voltage, and is therefore negligible. Equations~(\ref{eq:CaGeneral}) and (\ref{eq:APDGeneral}) state that Ca depends on a local combination of Ca and DI of the previous beat, while APD depends on a non-local combination of DI at the previous beat and Ca at the current beat, weighted by the Green's function $G$. We note that APD depends on the current value of Ca for the physiological reason that the V$_m$ action potential is buoyed by the influx of Ca$^{2+}$ ions via the L-type calcium current and thus influenced by the current [Ca$^{2+}$]$_i$ dynamics. 

For a paced cable with no-flux boundary conditions (i.e. $dV_m/dx=0$ at both ends of the cable), the non-local Green's function in Eq.~(\ref{eq:APDGeneral}) is given by
\begin{equation}
G(x,x')=G(x'-x)+G(x'+x)+G(2L-x'-x),\label{nofluxKernel}
\end{equation}
where
\begin{align}\label{eq:Green}
G(x) = \frac{1}{\sqrt{2\pi\xi^2}}e^{-x^2/2\xi^2}\left[1+\frac{wx}{2\xi^2}\left(1-\frac{x^2}{\xi^2}\right)\right] 
\end{align}
is an asymmetric Gaussian derived for a simple ionic model in Appendix~B of Ref.~\cite{Echebarria2007PRE}.  Importantly, $G$ has two intrinsic length scales $\xi$ and $w$: $\xi$ describes the length scale of electrotonic coupling due to voltage diffusion and is given by $\xi=\sqrt{2D_V A^*}$, where $A^*$ is the critical APD taken at the onset of alternans, and $w$ describes the symmetry-breaking effect that results from a pulse traveling in the positive direction and is given by $w=2D_V/cv^*$, where $cv^*$ is the critical conduction velocity value taken at the onset of alternans. We note that typically $w\ll\xi$~\cite{Echebarria2007PRE}, so in order to understand the generic behavior of calcium-driven alternans it is sufficient to consider the limit of small $w$. 

In order to quantify the amplitude of calcium and voltage alternans, we now introduce the quantities 
\begin{align}
c_n(x) &= [C_n(x)-C^*]/C^*, \label{eq:ampc}\\
a_n(x) &= [A_n(x)-A^*]/A^*, \label{eq:ampa}\\
d_n(x) &= [D_n(x)-D^*]/D^*, \label{eq:ampd}
\end{align}
which describe a suitably scaled difference in Ca, APD, and DI at beat $n$ from the critical Ca, APD, and DI value, respectively, at the onset of alternans. Thus, $c_n$, $a_n$, and $d_n$ measure a non-dimensional amplitude of alternans in Ca, APD, and DI. Next, in order to study the generic dynamics of calcium-driven alternans, we choose the following forms of the functions $f_c$ and $f_a$
\begin{align}
f_c(C_n,D_n)/C^* &= 1\overbrace{-rc_n+f(c_n)}^{I}+\overbrace{\alpha d_n(x)}^{II},\label{eq:fc}\\
f_a(D_n,C_{n+1})/A^* &= 1+\underbrace{\beta d_n}_{III} + \underbrace{\gamma c_{n+1}}_{IV},\label{eq:fa}
\end{align}
where terms labelled I-IV are chosen to model different aspects of the dynamics of calcium and voltage alternans. In term I we require that the function $f(c)$ is odd and captures the nonlinearity of local calcium dynamics. In this paper we will assume that $f$ is strictly cubic, i.e., $f(c)=c^3$, for convenience and in order to connect with Refs.~\cite{Echebarria2002PRL,Echebarria2007PRE}. Thus, term I models local calcium dynamics with local degree of instability $r$. In the absence of term II, the period-one solution $c_n=0$ is stable for $0\le r<1$ and loses stability at $r=1$, at which point a period-doubling bifurcation occurs and gives rise to stable period-two solutions $c_{n+1}=-c_n=\pm\sqrt{r-1}$ (which are stable for $1<r<2$).
 
Term III describes the dependence of APD on DI, capturing the effect of APD restitution. Here $\beta$ describes the slope of the APD restitution curve, which we assume to be positive and less than one to ensure that alternans are calcium-driven. Finally, terms II and IV describe the voltage-to-calcium and calcium-to-voltage coupling, respectively. In this paper we will consider the typical case of positive voltage-to-calcium and positive calcium-to-voltage coupling, and therefore assume $\alpha,\gamma\ge0$. As a measure of the total bi-directional coupling, we find it useful to define the parameter $\eta=\alpha\gamma$.
 
In order to obtain a closed system, we seek to eliminate $d_n$ in Eqs.~(\ref{eq:fc}) and (\ref{eq:fa}) in favor of $a_n$. To this end, following Refs.~\cite{Echebarria2002PRL,Echebarria2007PRE}, we note that the effect of CV restitution causes the activation interval $T_n(x)$, which describes the time between the $n^{th}$ and $(n-1)^{th}$ depolarizations at point $x$, to vary along the cable. By definition we have that
 \begin{align}
 T_n(x)=A_n(x) + D_n(x).\label{eq:T1}
 \end{align}
 On the other hand, $T_n(x)$ is also given by the pacing period $\tau_{bcl}$ at the pacing site $x=0$ plus the difference in times taken for the $n^{th}$ and $(n-1)^{th}$ stimuli to arrive at point $x$. Thus, in terms of CV restitution, we have that
 \begin{align}
 T_n(x) = \tau_{bcl} + \int_0^x\frac{dx'}{cv[DI_n(x')]} -  \int_0^x\frac{dx'}{cv[DI_{n-1}(x')]}.\label{eq:T2}
 \end{align}
Setting the right-hand sides of Eqs.~(\ref{eq:T1}) and (\ref{eq:T2}) equal and linearizing about $A^*$ and $D^*$ yields
 \begin{align}
 a_n(x)+d_n(x) = -\frac{1}{\Lambda}\int_0^x[d_n(x')-d_{n-1}(x')]/2dx',\label{eq:ad}
 \end{align}
where $\Lambda$ is given by $(cv^*)^2/2cv'^*$. We now note that the amplitude of alternans evolves very slowly near the onset of alternans, as well as far from onset in many ionic models, including the model we use here (e.g., in typical ionic model simulations several thousands of beats are required to bypass transient dynamics in the cable equation), so that $d_{n-1}(x)\approx-d_n(x)$. We then insert this approximation into Eq.~(\ref{eq:ad}), define $b_n(x)=a_n(x)+d_n(x)$ and take a derivative with respect to $x$, which yields the following linear, non-homogeneous ordinary differential equation
 \begin{align}
 b_n'(x)=-[b_n(x)-a_n(x)]/\Lambda,\label{eq:b}
 \end{align}
which can be solved analytically for $b_n(x)$ in terms of $a_n(x)$. After inserting back $b_n(x)=a_n(x)+d_n(x)$ and noting that at the pacing site $x=0$ the pacing rate remains constant, giving the initial condition $a_n(0)+d_n(0)=0$, we find that
\begin{align}
d_n(x) = -a_n(x) + \frac{1}{\Lambda}\int_0^xe^{(x'-x)/\Lambda}a_n(x')dx'.\label{eq:d}
\end{align}

Before presenting the closed system of equations, we make a few remarks about Eq.~(\ref{eq:d}). First, since $\Lambda$ is inversely proportional to the derivative of $cv$, $\Lambda$ is typically very large due to the flatness of typical CV restitution curves (in Sec.~\ref{sec5} we will explicitly calculate the CV restitution curve and $\Lambda$ for a detailed ionic model). Thus, $\Lambda^{-1}$ will be treated as a small parameter in much of the analysis that follows. Next, we will see that in steady-state, solutions to our reduced model remain stationary or have a very small velocity, so that our approximation $d_{n-1}(x)=d_n(x)$ remains valid away from the bifurcation. Finally, in Refs.~\cite{Echebarria2002PRL,Echebarria2007PRE} voltage-driven alternans profiles were found to have a spatial wavelength that scaled like $\lambda_s\sim(w\Lambda)^{1/2}$ or $\lambda_s\sim(\xi^2\Lambda)^{1/3}$, in either case $\lambda_s\ll\Lambda$, so the exponential $e^{(x'-x)/\Lambda}$ in Eq.~(\ref{eq:d}) could be approximated by one. In contrast, we will find that for calcium-driven alternans a certain regime of solutions yields the scaling $\lambda_s\sim\Lambda$, and therefore we will keep the full form of Eq.~(\ref{eq:d}).

We finally close the dynamics of the amplitude of Ca and APD alternans by inserting Eq.~(\ref{eq:d}) along with Eqs.~(\ref{eq:fc}) and (\ref{eq:fa}) into Eqs.~(\ref{eq:CaGeneral}) and (\ref{eq:APDGeneral}). This yields the following reduced system:
\begin{widetext}
\begin{align}
c_{n+1}(x) &= -rc_n(x) + c_n^3(x) -\alpha a_n(x) + \frac{\alpha}{\Lambda}\int_0^xe^{(x'-x)/\Lambda}a_n(x')dx',\label{eq:mapc}\\
a_{n+1}(x) &= \int_0^L G(x,x')\left[-\beta a_n(x') + \frac{\beta}{\Lambda}\int_0^{x'}e^{(y-x')/\Lambda}a_n(y)dy+\gamma c_{n+1}(x')\right]dx'.\label{eq:mapa}
\end{align}
\end{widetext}
Equations~(\ref{eq:mapc}) and (\ref{eq:mapa}) contain various parameters, but we will primarily be concerned with the dynamical effects that the degree of local calcium instability $r$ and length scale of CV restitution $\Lambda$ have on steady-state solutions. Thus, we will typically consider the APD restitution parameter $\beta$, the bi-directional coupling parameters $\alpha$ and $\gamma$, as well as the electrotonic coupling length scale parameters $\xi$ and $w$ [which appear in the Green's function $G$ in Eq.~(\ref{eq:Green})] to be given and fixed. Finally, we note that although Eqs.~(\ref{eq:mapc}) and (\ref{eq:mapa}) are in principle valid near the onset of alternans, we find that they also describe the features of alternans in the strongly nonlinear regime.

In the rest of this paper, we will study the dynamics of the reduced system in Eqs.~(\ref{eq:mapc}) and (\ref{eq:mapa}). In Sec.~\ref{sec3} we will present a numerical survey and describe the phase space of the reduced system. In Secs.~\ref{sec4} and \ref{sec5} we present detailed analyses of the reduced system. Finally, in In Sec.~\ref{sec6} we will use the reduced system to predict qualitatively and quantitatively the dynamics of a detailed ionic model. 

\section{Numerical Survey and Phase Space}\label{sec3}

\begin{figure}[t]
\centering
\epsfig{file =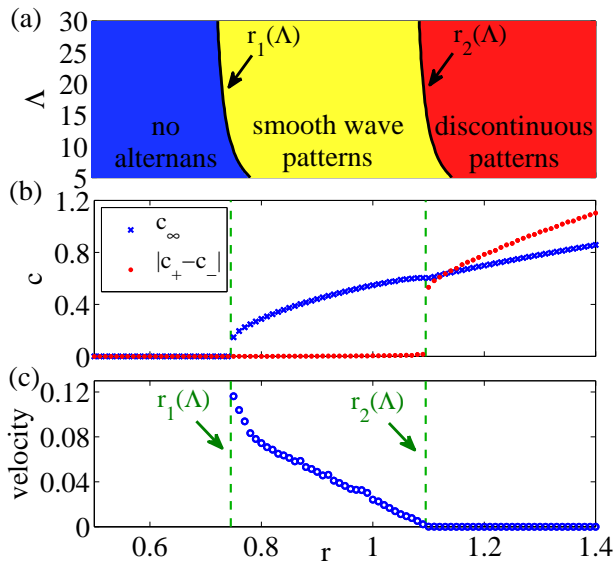, clip =,width=1.0\linewidth }
\caption{(Color online) (a) Phase space of Eqs.~(\ref{eq:mapc}) and (\ref{eq:mapa}) describing three solution regimes. As $r$ increases (left to right) these regimes are no alternans (blue), smooth wave patterns (yellow), and discontinuous patterns (red), each separated by bifurcations $r_1(\Lambda)$ and $r_2(\Lambda)$. For $\Lambda=15$ (b) the maximal alternans amplitude $c_\infty$, jump amplitude $|c_+-c_-|$, and (c) velocity for a range of $r$ values. Other parameter values are $\alpha,\gamma=\sqrt{0.3}$, $\beta=0$, $\xi=1$, and $w=0$ with $L=30$ and $\Delta x=0.005$.}\label{fig:PhaseSpace}
\end{figure}

We now present a numerical survey of the reduced system given by Eqs.~(\ref{eq:mapc}) and (\ref{eq:mapa}) on a cable of length $L=30$ with spatial discretization $\Delta x=0.005$, using $\alpha,\gamma=\sqrt{0.3}$, $\beta=0$, $\xi=1$, and $w=0$. It should be emphasized that $\xi$ is a dimensional length estimated to be in the range of a few millimeters~\cite{Echebarria2002PRL,Echebarria2007PRE}. We are reporting all amplitude equation results with lengths in units of $\xi$. Therefore the choice $\xi=1$ is equivalent to defining dimensionless length variables $\tilde x=x/\xi$, $\tilde \Lambda =\Lambda/\xi$, etc, and dropping the tilde symbol for convenience. The choice $\beta=0$ and $w=0$ is also made for convenience since we find that nonzero values result in qualitatively similar behavior. (Theoretical results for nonzero values of $\beta$ and $w$ that complement those presented in the following section are presented in Appendix~\ref{appA}.) In our numerical simulations of Eqs.~(\ref{eq:mapc}) and (\ref{eq:mapa}), we use a discretization of the spatial coordinate, $x=0,\Delta x,2\Delta x,\dots,L$. The discretization length $\Delta x$ is always chosen between $\Delta x=0.05$ (the cell size) and $\Delta x = 0.005$ (the distance that calcium diffuses during one beat). Depending on the computational requirements of a given numerical experiment (e.g., the necessity to simulate long cables, or to resolve small spatial scales), a value of $\Delta x$ in this range was chosen, with larger $\Delta x$ used for those experiments that require more computational resources and smaller $\Delta x$ used when small spatial scales need to be resolved. (Given a cable discretized with $N=L/\Delta x+1$ points, $\mathcal{O}(N^2)$ operations are required to evolve Eqs.~(\ref{eq:mapc}) and (\ref{eq:mapa}) forward a single beat using  the trapezoidal rule for numerical integration.) Importantly, we verified that our conclusions do not depend on the particular choice of $\Delta x$. The only difference is a slightly different convergence rate to steady state after a change of parameters. The value of $\Delta x$ used for each simulation will be always specified in the text and figure captions. In Appendix~\ref{appB} we report results that describe the effect that using different spatial discretizations $\Delta x$ has on the transient dynamics of Eqs.~(\ref{eq:mapc}) and (\ref{eq:mapa}).

We summarize the results from varying the degree of calcium instability $r$ and CV length scale $\Lambda$ in Fig.~\ref{fig:PhaseSpace}. In Fig.~\ref{fig:PhaseSpace} (a) we show the phase space which consists of three different regimes of solutions separated by two bifurcations. The left-most regime is colored blue and corresponds to relatively small values of $r$, where we find that the only stable solutions are identically zero, i.e., $c_n(x)\equiv0$ and $a_n(x)\equiv0$. We refer to this as the no alternans regime. As $r$ is increased, we next cross the first bifurcation $r_1(\Lambda)$ and enter the middle regime, which is colored yellow and corresponds to slightly larger values of $r$. We find that at $r=r_1(\Lambda)$ the identically zero solutions corresponding to no alternans lose stability and give rise to solutions that form smooth wave patterns. Therefore, the bifurcation $r_1(\Lambda)$ corresponds to the onset of alternans. Furthermore, we find that these smooth wave patterns can either travel with some finite velocity towards the pacing site at $x=0$, or remain stationary. Typically, stationary solutions only form if the asymmetry length scale $w$ is large enough, as in the voltage-mediated instability studied in Refs.~\cite{Echebarria2002PRL,Echebarria2007PRE}. Finally, as $r$ is increased further, we cross a second bifurcation $r_2(\Lambda)$ and enter the right-most regime, which is colored red and corresponds to larger values of $r$. As $r$ crosses $r_2(\Lambda)$ we find that the smooth patterns of $c_n(x)$ found in the middle regime develop a jump discontinuity at each node, while the amplitude of voltage alternans $a_n(x)$ remains smooth. Furthermore, regardless of whether smooth solutions had a finite velocity or were stationary, these discontinuous patterns are always stationary. 

\begin{figure}[b]
\centering
\epsfig{file =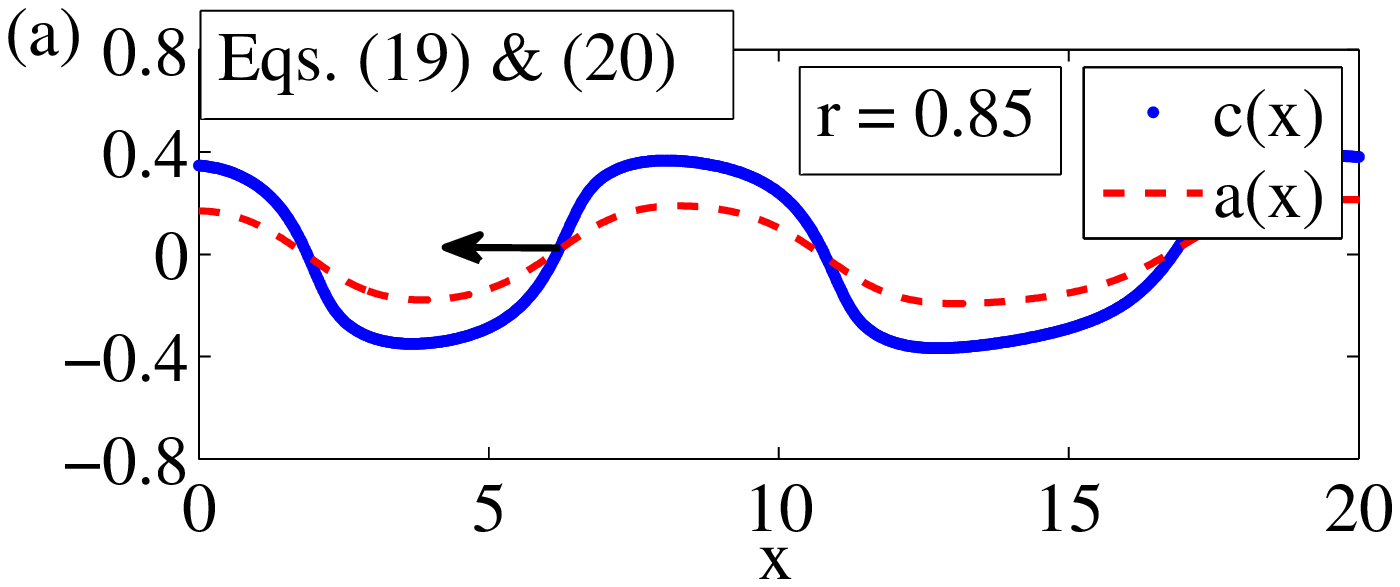, clip =,width=1.0\linewidth } \\
\epsfig{file =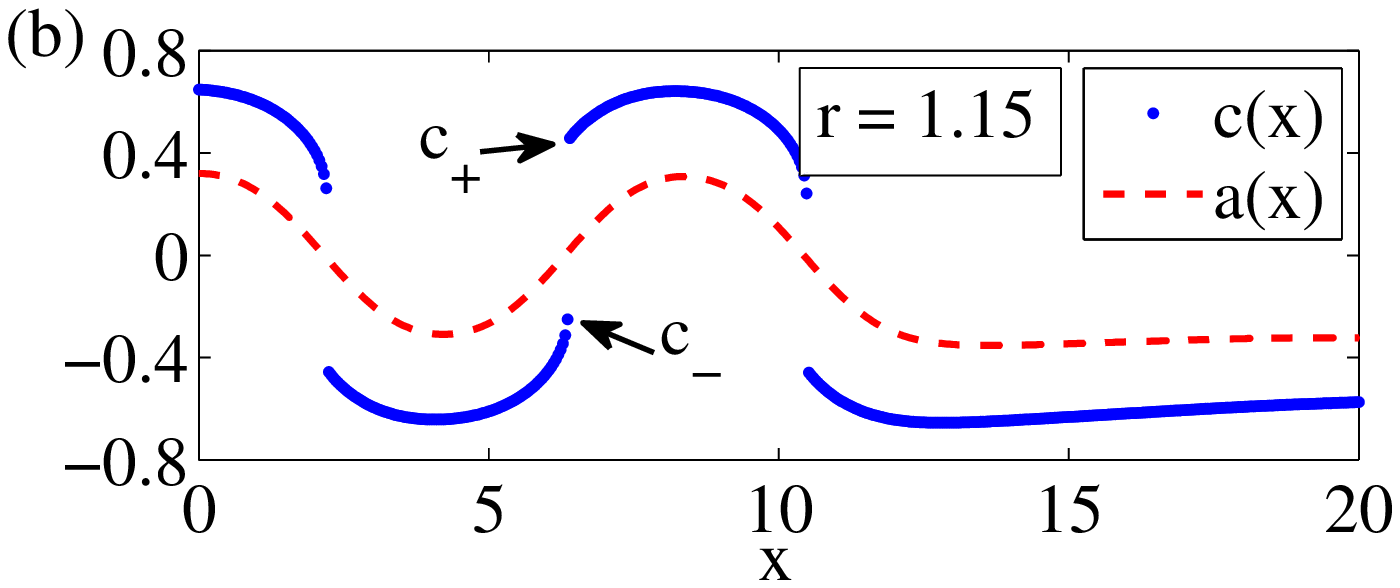, clip =,width=1.0\linewidth }
\caption{(Color online) Examples of solutions of Eqs.~(\ref{eq:mapc}) and (\ref{eq:mapa}) for (a) $r=0.85$ and (b) $r=1.15$. Calcium $c_n(x)$ and voltage $a_n(x)$ profiles are plotted as blue dots and red curves, respectively. Other parameter values are $\Lambda=15$, $\alpha,\gamma=\sqrt{0.3}$, $\beta=0$, $\xi=1$, and $w=0$ with $L=20$ and $\Delta x=0.04$. With the choice $\xi=1$, $x$  and $\Lambda$ are in units of $\xi$ (see text). }\label{fig:Examples}
\end{figure}

In Figs.~\ref{fig:PhaseSpace} (b) and (c) and we describe the nature of solutions in the three regions in more detail by setting the CV restitution length scale parameter $\Lambda=15$ and plotting several quantities as the degree of calcium instability $r$ is increased. In Fig.~\ref{fig:PhaseSpace} (b), we plot the maximal amplitude of calcium alternans $c_\infty=\max_{0\le x\le L}|c(x)|$ in blue crosses. In particular, we note that at the bifurcation $r_1(\Lambda)$ (indicated by the first vertical dashed green line) $c_\infty=0$ transitions from zero to non-zero values, indicating the onset of alternans, and $c_\infty$ continues to increase as $r$ increases. Next, if a node exists at $x=x_0$ (i.e., $c(x_0)=0$) we define the left and right limiting values $c_-$ and $c_+$ as the values of $c(x)$ just to the left and right of $x_0$ [see Fig.~\ref{fig:Examples} (b)]. We compute the maximal difference $|c_+-c_-|$ and plot it in red dots in Fig.~\ref{fig:PhaseSpace} (b). We note that $|c_+-c_-|$ remains zero for $r<r_2(\Lambda)$, indicating solutions are continuous, but at $r_2(\Lambda)$ (indicated by the second vertical dashed green line) $|c_+-c_-|$ jumps to a finite positive value, indicating that solutions develop discontinuities at the nodes. Finally, we calculate the velocity of solutions, which we plot in Fig.~\ref{fig:PhaseSpace} (c). In the no alternans regime no such velocity exists because solutions are identically zero. In the smooth regime we find positive velocities that diminish with increasing $r$, until at $r_2(\Lambda)$ the velocity vanishes and remains zero in the  discontinuous regime. As we will show in Sec.~\ref{sec4}, a linear stability analysis allows us to predict the velocity at the onset of alternans. Fig.~\ref{fig:PhaseSpace} presents a detailed picture of the bifurcations that we find in the system. In the next sections we analyze the bifurcation at $r_1(\Lambda)$ and $r_2(\Lambda)$ as well as the properties of both smooth and discontinuous solutions.

\begin{figure}[t]
\centering
\epsfig{file =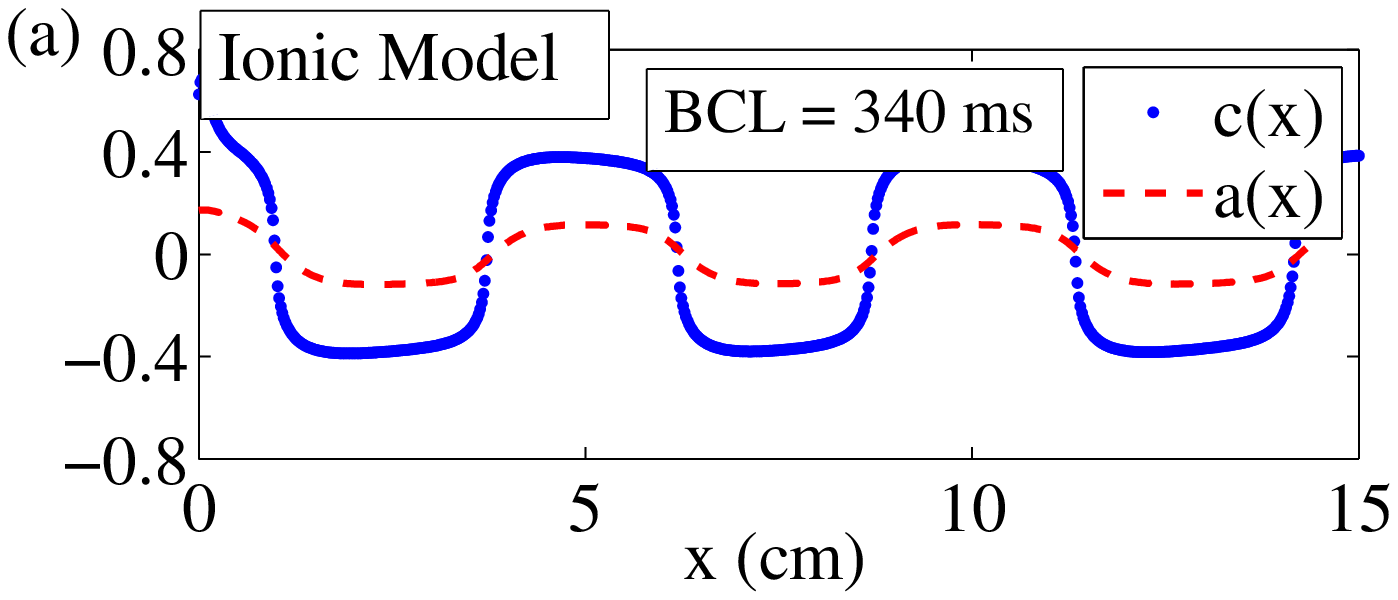, clip =,width=1.0\linewidth } \\
\epsfig{file =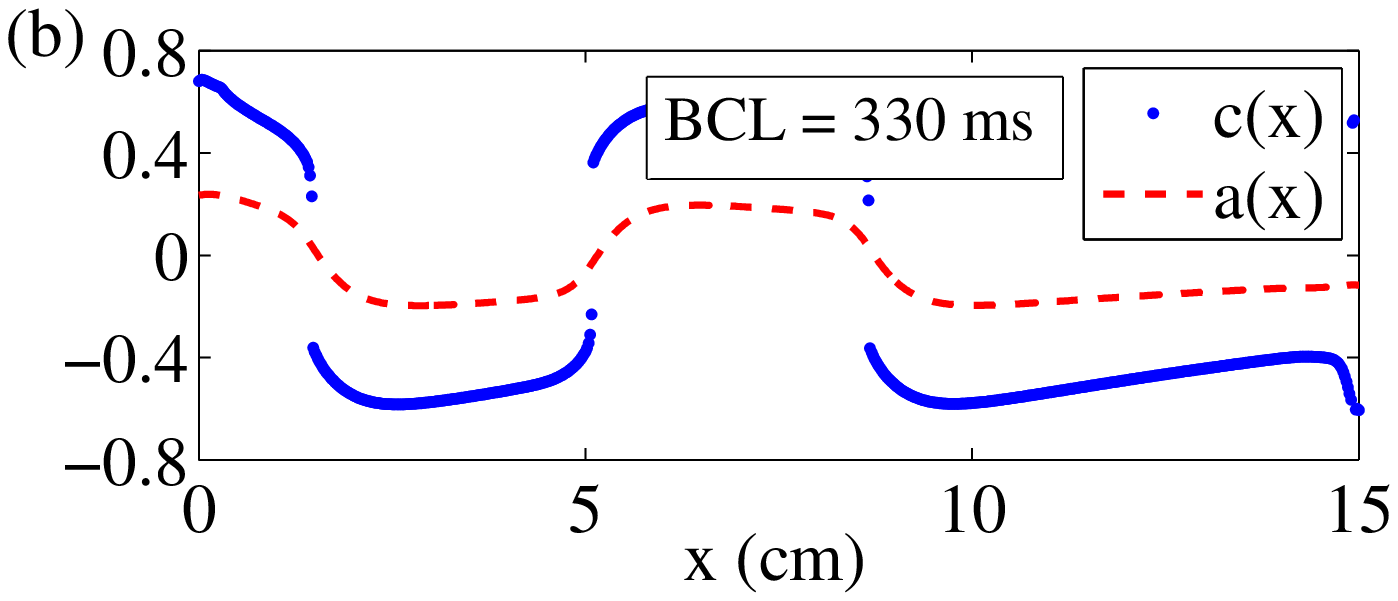, clip =,width=1.0\linewidth }
\caption{(Color online) Example solutions of the cable equation with the Shiferaw-Fox model for (a) $BCL=340$ ms and (b) $BCL=330$ ms. The amplitude of calcium alternans $c_n(x)$ and voltage alternans $a_n(x)$ are plotted as blue dots and red curves, respectively.}\label{fig:ExamplesIonic}
\end{figure}

We close this section by  presenting plots of alternans profiles obtained from Eqs.~(\ref{eq:mapc}) and (\ref{eq:mapa}) and comparing them with profiles obtained from the ionic model. In Figs.~\ref{fig:Examples} (a) and (b) we plot representative $c_n(x)$ and $a_n(x)$ profiles on a cable of length $L=20$ with spatial discretization $\Delta x=0.04$ from both the smooth and discontinuous regime for $r=0.85$ and $1.15$, respectively, for fixed $\Lambda=15$. Calcium  and voltage profiles $c_n(x)$ and $a_n(x)$ are plotted in blue dots and dashed red, respectively. Since we have used $w=0$, the solutions in the smooth regime were found to have a finite velocity in the direction of the pacing site at $x=0$, indicated by the arrow. We also denote $c_-$ and $c_+$ for the discontinuous pattern in panel (b).

To complement the alternans profiles obtained from the reduced system in Eqs.~(\ref{eq:mapc}) and (\ref{eq:mapa}), we compute alternans profiles from the cable equation~(\ref{eq:cable}) with a detailed ionic model. In particular, we perform simulations of Eq.~(\ref{eq:cable}) with $D_V=5\cdot10^{-4}$ cm$^{2}$/ms and $C_m=1$ $\mu$F/cm$^2$, and the ion currents are given by the so-called Shiferaw-Fox model, i.e., we use the ionic currents of Fox et al.~\cite{Fox2001AJPHC} coupled with the calcium-cycling dynamics of Shiferaw et al.~\cite{Shiferaw2003BiophysJ}. We describe in Sec.~\ref{sec6} in detail specifics about the parameters used, as well as particular parameter values, but note here that we choose parameters to ensure that alternans is calcium-driven. 

As with most ionic models, beat-to-beat dynamics of the Shiferaw-Fox model evolve slowly, so to reach steady-state we simulate through a transient of $12,000$ beats. Once steady-state is reached, we extract the amplitudes corresponding to those in Eqs.~(\ref{eq:ampc}) and (\ref{eq:ampa}) by approximating $C^*=(C_n+C_{n-1})/2$ and $A^*=(A_n+A_{n-1})/2$, yielding 
\begin{align}
c_n(x) &= [C_n(x)-C_{n-1}(x)]/[C_n(x)+C_{n-1}(x)],\label{eq:ampc2}\\
a_n(x) &= [A_n(x)-A_{n-1}(x)]/[A_n(x)+A_{n-1}(x)].\label{eq:ampa2}
\end{align}
In Figs.~\ref{fig:ExamplesIonic} (a) and (b) we plot the steady-state amplitudes $c_n(x)$ and $a_n(x)$ given by Eqs.~(\ref{eq:ampc2}) and (\ref{eq:ampa2}) along a cable of length $L=15$ cm with spatial discretization $\Delta x=0.02$ cm, for pacing periods of $BCL=340$ ms and $330$ ms, respectively. For $BCL=340$ ms both the amplitude of calcium and voltage alternans form smooth wave patterns, analogous to the smooth solutions from the reduced model, e.g., from Fig.~\ref{fig:Examples} (a). For $BCL=330$ ms, while the amplitude of voltage alternans remains smooth, the amplitude of calcium alternans develops a discontinuity, and resembles the discontinuous solutions from the reduced model, e.g., from Fig.~\ref{fig:Examples} (b). 

\section{Linear Stability Analysis: Onset of Alternans and Smooth Wave Patterns}\label{sec4}

We now present a linear stability analysis of the reduced model given by Eqs.~(\ref{eq:mapc}) and (\ref{eq:mapa}). We begin by studying the onset of alternans, described by the bifurcation at $r_1(\Lambda)$ (see Fig.~\ref{fig:PhaseSpace}) and the properties of smooth solutions that arise immediately after the onset of alternans [see Fig.~\ref{fig:Examples} (a)]. In particular, because the bifurcation $r_1(\Lambda)$ describes a transition from solutions $c_n(x)$ and $a_n(x)$ that are identically zero to non-zero, we study the dynamics of perturbations to solutions $c_n(x)\equiv0$ and $a_n(x)\equiv0$. For this linear stability analysis we will consider the limit of long cable length $L$, but note that our predictions are also accurate for cables of more realistic lengths as well. Furthermore, since the length scales of electronic coupling and CV tend to satisfy $\xi\ll\Lambda$, we will take the non dimensional parameter $\xi\Lambda^{-1}$ to be a small parameter.

For the sake of simplicity, in the analysis presented below we will consider the limit of no asymmetry in the Greens function in Eq.~(\ref{eq:Green}), i.e., $w=0$. Furthermore, setting the APD restitution parameter $\beta$ to zero yields a much simpler set of equations to study, and therefore the analysis below will be for the case $\beta=0$. In Appendix~\ref{appA} we will present the complementary theoretical results for both $\beta\ne0$ and $w\ne0$. In subsections~\ref{sec4subA} and \ref{sec4subB} we study the onset of alternans and spatial properties of smooth solutions, respectively, and in subsection~\ref{sec4subC} we compare and contrast our findings for calcium-driven alternans to those of voltage-driven alternans.

For the purpose of carrying out the linear stability analysis, we consider a semi-infinite cable paced at $x=0$. The Green's function $G(x,x')$ defined by Eq.~(\ref{nofluxKernel}) reduces for such a cable to $G(x,x')=G(x'-x)+G(x'+x)$. In the present section, we treat the case $w\approx 0$ leading to traveling waves. As for the voltage-driven case~\cite{Echebarria2002PRL,Echebarria2007PRE}, we find that the amplitude of the traveling waves grows exponentially with distance away from the pacing site. This makes the traveling wave eigenmode insensitive to boundary effects at the paced end of the cable. This allows us to self-consistently carry out the linear stability analysis with the Green's function for an infinite cable, i.e.  $G(x,x')=G(x'-x)$. The same turns out to be true for the case of standing waves treated in Appendix~\ref{appA}. In this case, the zero flux boundary condition at the paced end is essential so that the form $G(x,x')=G(x'-x)+G(x'+x)$ should be used in principle for the linear stability analysis. However, since the standing wave eigenmode is $\sim \cos kx$, the eigenvalue equation obtained by analyzing this mode over the semi-infinite domain $x>0$ with $G(x,x')=G(x'-x)+G(x'+x)$ is identical to the eigenvalue equation obtained by analyzing the mode $\sim \exp(ikx)$ over the infinite domain $-\infty < x <+\infty$ with $G(x,x')=G(x'-x)$. 

\subsection{Onset of alternans}\label{sec4subA}

We consider a perturbation to the solution $c_n(x)\equiv0$ and $a_n(x)\equiv0$ of the form $\delta c_n(x)=c\lambda^ne^{ikx}$ and $\delta a_n(x)=a\lambda^ne^{ikx}$ for constants $c,a\ll1$. Thus, perturbations are described by the growth parameter $\lambda$ and the wave number $k$. Since we are interested in the onset of alternans, we will search for a growth parameter of unit magnitude, i.e., $|\lambda|=1$. Inserting these perturbations into Eqs.~(\ref{eq:mapc}) and (\ref{eq:mapa}), we find after neglecting a term of order $\mathcal{O}(c^3)$ that
\begin{align}
(\lambda+r)c(ik+\Lambda^{-1}) &= -\alpha ika, \label{eq:linstab1}\\
a &=\gamma ce^{-k^2\xi^2/2}.\label{eq:linstab2}
\end{align}
By combining Eqs.~(\ref{eq:linstab1}) and (\ref{eq:linstab2}) we find the dispersion relation
\begin{align}
\lambda = -r-\eta\frac{ik}{ik+\Lambda^{-1}}e^{-k^2\xi^2/2},\label{eq:linstab3}
\end{align}
which gives the growth parameter $\lambda$ of a perturbation with given wave number $k$.

We now recall that ultimately we are interested in a cable of finite length, and therefore we will restrict our attention to perturbations that grow with zero group velocity, i.e., possibly traveling solutions that grow inside of a stationary envelope. This is refered to as an {\it absolute instability} and is characterized by the condition $\partial\lambda/\partial k=0$~\cite{Sandstede2002}. On the other hand, perturbations that grow with some finite group velocity, i.e., that grow inside of a moving envelope, will vanish from a finite domain in finite time, and therefore we do not consider these {\it convective instabilities}, which are characterized by $\partial\lambda/\partial k\ne0$.

Enforcing the absolute instability condition $\partial\lambda/\partial k=0$ on Eq.~(\ref{eq:linstab3}), we find that
\begin{align}
\Lambda^{-1}=k^2\xi^2(ik+\Lambda^{-1}),\label{eq:linstab4}
\end{align}
which describes the wave number $k$ corresponding to the absolute instability we are looking for. We next solve for $k$ in Eq.~(\ref{eq:linstab4}) perturbatively, finding
\begin{align}
\Lambda k=\left(\frac{\Lambda}{\xi}\right)^{2/3}i^{-1/2} + \frac{i}{3} + \mathcal{O}\left[\left(\frac{\xi}{\Lambda}\right)^{2/3}\right].\label{eq:linstab5}
\end{align}
After inserting Eq.~(\ref{eq:linstab5}) into Eq.~(\ref{eq:linstab3}), we have that the growth parameter is given by
\begin{align}
\lambda=-r&-\eta+\frac{3\eta i^{-2/3}}{2}\left(\frac{\xi}{\Lambda}\right)^{2/3} \nonumber \\ &-\frac{13\eta i^{-4/3}}{8}\left(\frac{\xi}{\Lambda}\right)^{4/3} + \mathcal{O}\left[\left(\frac{\xi}{\Lambda}\right)^2\right],\label{eq:linstab6}
\end{align}
Finally, by setting $|\lambda|=1$ we find that the critical bifurcation value describing the onset of alternans is given by
\begin{align}
r_1(\Lambda) = 1-\eta+\frac{3\eta}{4}\left(\frac{\xi}{\Lambda}\right)^{2/3}+\mathcal{O}\left[\left(\frac{\xi}{\Lambda}\right)^{4/3}\right].\label{eq:linstab7}
\end{align}

\begin{figure}[t]
\centering
\epsfig{file =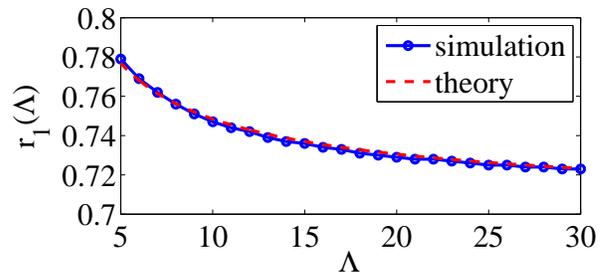, clip =,width=1.0\linewidth }
\caption{(Color online) The critical value $r_1(\Lambda)$ describing the onset of alternans computed directly from numerical simulations of Eqs.~(\ref{eq:mapc}) and (\ref{eq:mapa}) (blue circles) compared to our theoretical prediction given by Eq.~(\ref{eq:linstab7}) (dashed red). Other parameters are $\alpha,\gamma=\sqrt{0.3}$, $\beta=0$, $\xi=1$, and $w=0$ with $L=100$ and $\Delta x=0.05$.}\label{fig:Onset1}
\end{figure}

To verify this result, we perform numerical simulations of Eqs.~(\ref{eq:mapc}) and (\ref{eq:mapa}) to find the onset of alternans. On a long cable ($L=100$) with spatial discretization $\Delta x=0.05$, over a range of $\Lambda$, we start from a small $r$ value and perturb the zero solution $c_n(x)\equiv0$ and $a_n(x)\equiv0$, evolving the system forward in time to see if perturbations grow or decay. If perturbations decay, we increase $r$ slightly, perturb the zero solution again and repeat. If perturbations grow, we continue evolving the system to steady-state to confirm that alternans form and store the observed $r_1(\Lambda)$ value. In Fig.~\ref{fig:Onset1} we plot the results from simulation in blue circles as well as our theoretical prediction from Eq.~(\ref{eq:linstab7}) in dashed red. Other parameters are $\alpha,\gamma=\sqrt{0.3}$, $\beta=0$, $\xi=1$, and $w=0$. We label the region $r<r_1(\Lambda)$ and $r>r_1(\Lambda)$ ``no alternans'' and ``alternans'', respectively, since the solution $c_n(x)\equiv0$ and $a_n(x)\equiv0$ is stable and unstable to perturbation in the respective regions. We note that the agreement between onset as observed from numerical simulations and our theoretical prediction is excellent.

We note that for a different form of nonlinearity in Eq.~(\ref{eq:mapc}), i.e., a more general odd form of $f(c_n)$, we recover the same results. This follows from the fact that in our linear stability analysis we neglected all nonlinear terms, eventually obtaining Eq.~(\ref{eq:linstab1}). In the case of non-zero $\beta$ or $w$, which we treat separately in Appendix~\ref{appA}, we find that the critical onset value is given by Eqs.~(\ref{eq:appA5}) and (\ref{eq:appA12}), respectively. In particular, Eq.~(\ref{eq:appA5}) for $\beta\ne0$ gives the critical onset value implicitly.

\subsection{Spatial properties of smooth solutions}\label{sec4subB}

We now turn our attention to the properties of solutions that form in the smooth regime, i.e., immediately after the onset of alternans at $r=r_1(\Lambda)$. In particular, given the profiles we observe [see Fig.~\ref{fig:Examples} (a)], we will focus on the spatial wavelength $\lambda_s$ and velocity $v$ of steady-state solutions. We find that we can quantify both by considering smooth solutions near the onset of alternans. However, we note that the velocity decreases quickly as $r$ is increased past $r_1(\Lambda)$ [see Fig.~\ref{fig:PhaseSpace} (c)].

In the analysis above, we considered perturbations characterized by the growth parameter $\lambda$ and the wave number $k$, i.e., $c_n(x),a_n(x)\propto \lambda^ne^{ikx}$. The spatial wavelength of solutions is given by $\lambda_s=2\pi/k_{Re}$, where $k_{Re}$ is the real part of the wave number $k$. Using the wave number in Eq.~(\ref{eq:linstab5}), we find that to leading order
\begin{align}
\lambda_s=\xi\frac{4\pi}{\sqrt{3}}\left(\frac{\Lambda}{\xi}\right)^{1/3}. \label{eq:linstab8}
\end{align}

Next, we note that near the onset of alternans the growth parameter $\lambda$ has approximately unit magnitude, so that $\lambda\approx-e^{i\Omega}$ for some $\Omega\in\mathbb{R}$, where we have included the negative sign to account for periodic flipping of solutions. Thus, with $k=k_{Re}+ik_{Im}$, we can express solutions as $c_n(x),a_n(x)\propto e^{-k_{Im}x}e^{ik_{Re}(x+n\Omega/k_{Re})}$. Note that from Eq.~(\ref{eq:linstab5}) we have that $k\approx(\sqrt{3}-i)/2(\xi^2\Lambda)^{1/3}$, which implies that $k_{Im}<0$ such that the modes grow exponentially with distance away from the pacing site as announced earlier. The velocity of solutions in the direction of the pacing site is given by $v=\Omega/k_{Re}$. When $\Omega$ is small, as in our case, it can be approximated in terms of the imaginary part of $\lambda$, i.e., $\Omega\approx-\lambda_{Im}$. Using the growth parameter and wave number from Eqs.~(\ref{eq:linstab5}) and (\ref{eq:linstab6}), we find that to second order the velocity is given by
\begin{align}
v=\frac{3\eta\xi}{2}\left(\frac{\xi}{\Lambda}\right)^{1/3}-\frac{13\eta\xi}{8}\left(\frac{\xi}{\Lambda}\right), \label{eq:linstab9}
\end{align}
where we have included the second-order term to increase the precision for more moderate values of $\Lambda$.

\begin{figure}[t]
\centering
\epsfig{file =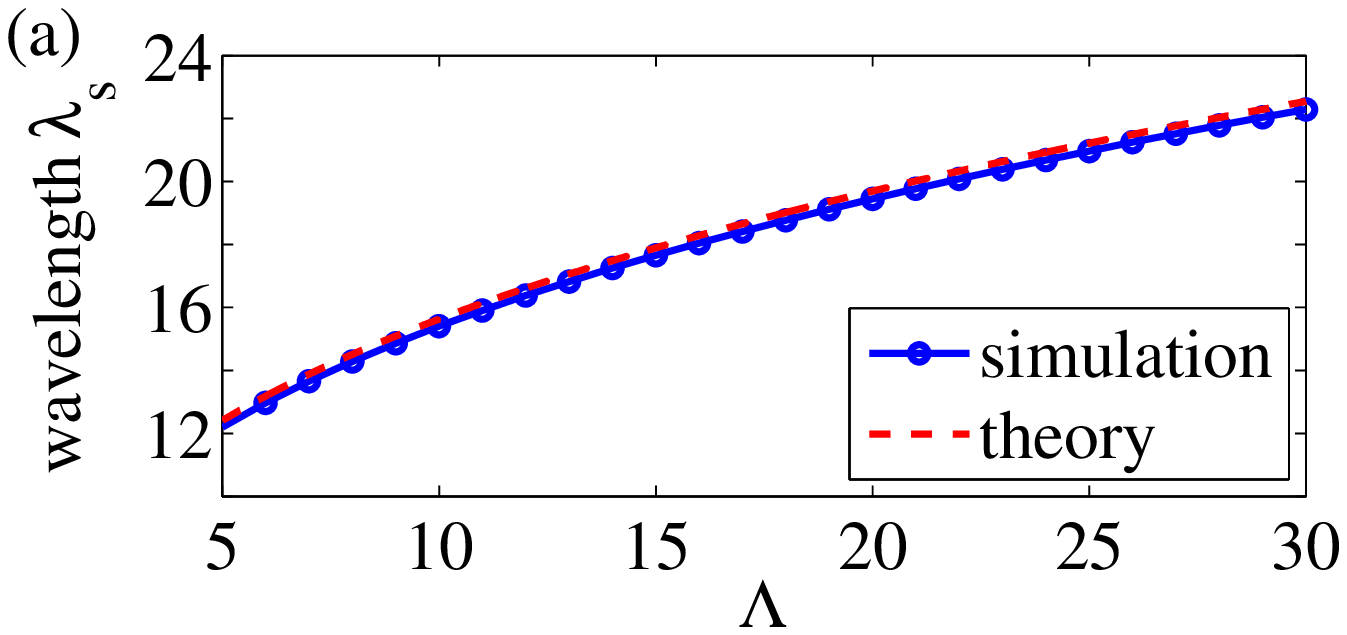, clip =,width=1.0\linewidth } \\
\epsfig{file =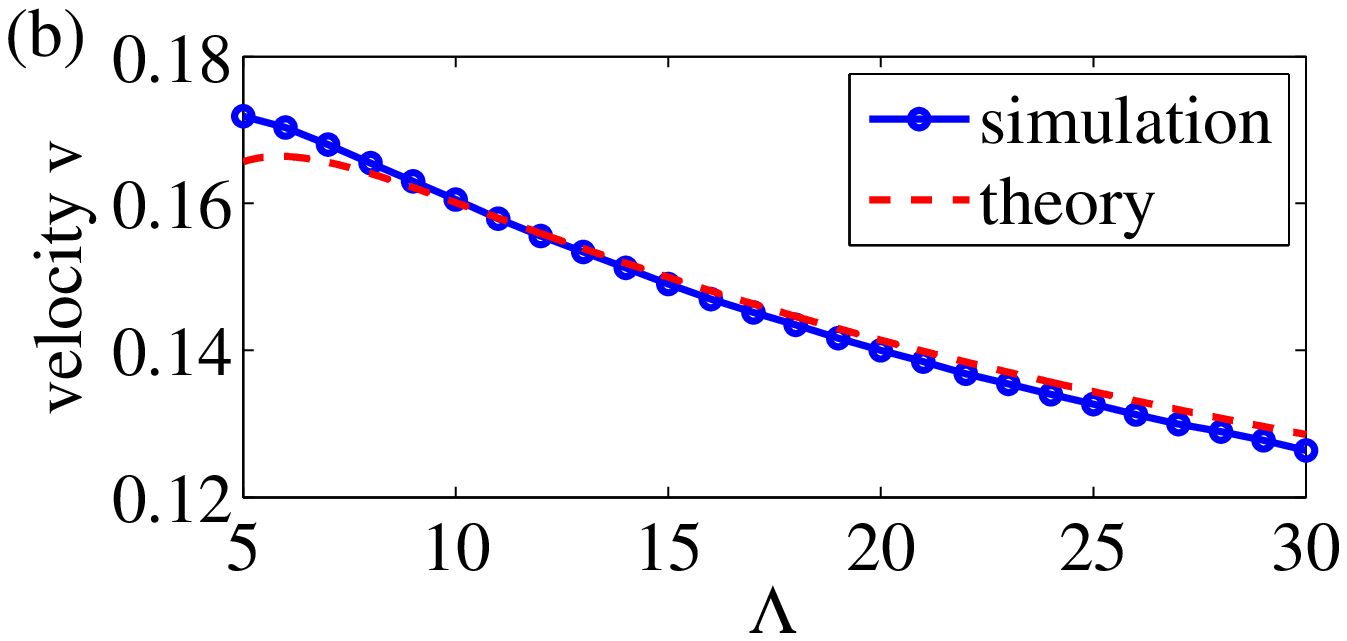, clip =,width=1.0\linewidth }
\caption{(Color online) (a) The spatial wavelength $\lambda_s$ and (b) velocity $v$ of smooth solutions near the onset of alternans computed directly from numerical simulations of Eqs.~(\ref{eq:mapc}) and (\ref{eq:mapa}) (blue circles) compared to our theoretical predictions given by Eqs.~(\ref{eq:linstab8}) and (\ref{eq:linstab9}) (dashed red). Other parameters are $\alpha,\gamma=\sqrt{0.3}$, $\beta=0$, $\xi=1$, and $w=0$ with $L=100$ and $\Delta x=0.05$.}\label{fig:OnsetAB}
\end{figure}

To verify this result, we  calculate the spatial wavelength $\lambda_s$ and velocity $v$ directly from numerical simulations of Eqs.~(\ref{eq:mapc}) and (\ref{eq:mapa}) immediately after the onset of alternans. In Fig.~\ref{fig:OnsetAB} we plot the spatial wavelength $\lambda_s$ and velocity $v$ in panels (a) and (b), respectively, from direct simulation in blue circles as well as our theoretical predictions from Eqs.~(\ref{eq:linstab8}) and (\ref{eq:linstab9}) in dashed red. As previously, simulations are done on a long cable ($L=100$) with spatial discretization $\Delta x=0.05$ with $\alpha,\gamma=\sqrt{0.3}$, $\beta=0$, $\xi=1$, and $w=0$. We note that the agreement between both the spatial wavelength and velocity as observed from numerical simulations and our theoretical prediction is excellent.

We note that for the case of non-zero $\beta$ or $w$, the spatial wavelength at onset is given by Eqs.~(\ref{eq:linstab8}) and (\ref{eq:appA13}), respectively. Note that the inclusion of non-zero $\beta$ leaves the spatial wavelength at onset unchanged from the $\beta=0$ case, but as discussed in Appendix~\ref{appA}, a sufficiently large $w$ changes the scaling entirely from $\lambda_s\sim(\xi^2\Lambda)^{1/3}$ to $\lambda_s\sim(w\Lambda)^{1/2}$. Furthermore, the velocity for non-zero $\beta$ is given by Eq.~(\ref{eq:appA6}) while sufficiently large $w$ yields stationary solutions.

\subsection{Comparison to voltage-driven alternans}\label{sec4subC}

We conclude this section with a brief comparison of the properties of calcium-driven alternans governed by Eqs.~(\ref{eq:mapc}) and (\ref{eq:mapa}) and voltage-driven alternans governed by Eq.~(\ref{EKVm})~\cite{Echebarria2002PRL,Echebarria2007PRE} near onset. We find remarkable similarities between the dynamics, suggesting that the dynamics near onset are universal. In particular, both calcium- and voltage-driven alternans admit two classes of solutions after onset that depends on the asymmetry parameter $w$: traveling and stationary wave patterns. For both traveling and stationary solutions, the scaling of the spatial wavelength is equivalent for calcium- and voltage-driven alternans.

In contrast, the critical onset value and velocity of traveling wave patterns of calcium-driven alternans is not precisely equivalent to the voltage-driven case. In particular, the model for calcium-driven alternans incorporates the bi-directional coupling parameters $\alpha$ and $\gamma$, which do not appear in the model for voltage-driven alternans. Here these bi-directional coupling parameters surface in the expressions for $r_1(\Lambda)$ and $v$ [Eqs.~(\ref{eq:linstab7}) and (\ref{eq:linstab9}) for $\beta=0$ and $w=0$, or Eqs.~(\ref{eq:appA5}), (\ref{eq:appA6}), (\ref{eq:appA12}) otherwise] as $\eta=\alpha\gamma$. However, these expressions are equivalent to those for the voltage-driven case under a shift and scaling by $\eta$.

\section{Strongly Nonlinear Regime: Dynamics of Discontinuous Patterns}\label{sec5}

We now consider the strongly non-linear regime where discontinuous calcium profiles form [see Fig.~\ref{fig:Examples} (b)], and the properties of these solutions. We emphasize that due to the electrotonic coupling of voltage dynamics due to diffusion, these solutions are non-physical for voltage-driven alternans, and thus only observed when alternans is calcium-driven. Throughout this section we will primarily be interested in the steady-state solutions of Eqs.~(\ref{eq:mapc}) and (\ref{eq:mapa}). We note, however, that the transient dynamics of Eqs.~(\ref{eq:mapc}) and (\ref{eq:mapa}) are interesting in their own right, and therefore we present results from numerical investigation of transient behavior in Appendix~\ref{appB}.

We begin this section by studying the nature of the discontinuities that form and the {\it jumping points} at each discontinuity in subsection~\ref{sec5subA}. In subsection~\ref{sec5subB} we show the hysteretic behavior inherent in the discontinuous regime, first describing the symmetrizing of jumping points, then describing unidirectional pinning. In subsection~\ref{sec5subC} we present a framework for understanding the hysteretic behavior described previously. In the subsection~\ref{sec5subD} we study the scaling properties of the spatial wavelength of solutions. Finally, in subsection~\ref{sec5subE} we address the combined effects of random fluctuation of SDA under node dynamics induced by changes of parameters.

\subsection{Discontinuities}\label{sec5subA}

\begin{figure}[b]
\centering
\epsfig{file =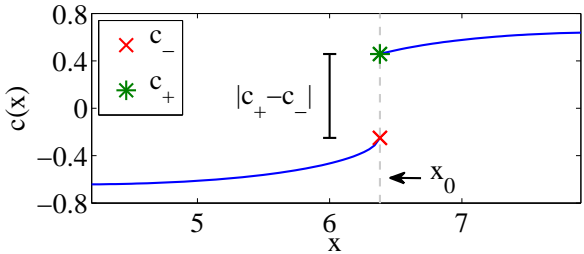, clip =,width=1.0\linewidth } 
\caption{(Color online) Illustration of jumping points $c_-$ and $c_+$ and the jump amplitude $|c_+-c_-|$ at a node $x_0$ for a discontinuous calcium profile $c(x)$ using parameters $r=1.15$, $\Lambda=15$, $\alpha,\gamma=\sqrt{0.3}$, $\beta=0$, $\xi=1$, and $w=0$ with $L=20$ and $\Delta x=0.005$.}\label{fig:Excmcp}
\end{figure}

Due to the importance of nodes in spatially discordant alternans, we will begin by studying the shape of the phase reversals that form in the discontinuous regime, and later study their locations. Recall that given a steady-state solution that has developed a discontinuity at $x=x_0$ the left and right jumping points, respectively, are defined by
\begin{align}
c_-=\lim_{x\to x_0^-}c(x),\hskip2ex\text{ and }\hskip2ex c_+ = \lim_{x\to x_0^+}c(x),\label{eq:nlin1}
\end{align}
and the total jump amplitude of the discontinuity is then given by $|c_+-c_-|$ (see Fig.~\ref{fig:Excmcp}).

\begin{figure}[t]
\centering
\epsfig{file =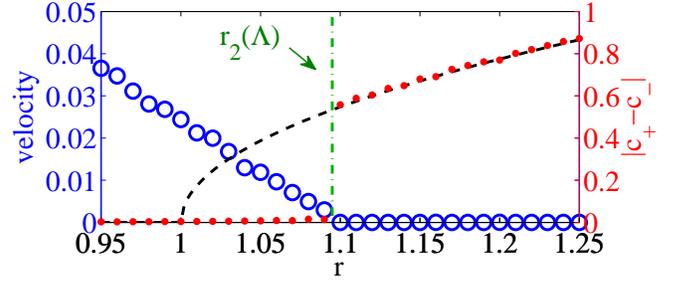, clip =,width=1.0\linewidth } 
\caption{(Color online) The velocity and jump amplitude $|c_+-c_-|$ of steady-state solutions, plotted in blue circles and red dots, respectively, over a range of $r$ values for fixed $\Lambda=15$. The theoretical prediction for the jump amplitude of normal jumps is plotted in dashed black and the bifurcation value $r_1(\Lambda)$ describing the transition from smooth to discontinuous solutions is denoted by the vertical green dot-dashed line. Other parameters are $\alpha,\gamma=\sqrt{0.3}$, $\beta=0$, $\xi=1$, and $w=0$ with $L=30$ and $\Delta x=0.005$.}\label{fig:VelocityJump}
\end{figure}

To gain some insight into the transition from the smooth regime to the discontinuous regime, we fix $\Lambda$ and simulate Eqs.~(\ref{eq:mapc}) and (\ref{eq:mapa}) over a range of $r$ values. In Fig.~\ref{fig:VelocityJump} we plot the resulting velocity and jump amplitude $|c_+-c_-|$ simultaneously in blue circles and red dots, respectively, for fixed $\Lambda=15$. Simulation were performed on a cable of length $L=30$ with spatial discretization $\Delta x=0.005$, and other parameters are $\alpha,\gamma=\sqrt{0.3}$, $\beta=0$, $\xi=1$, and $w=0$. We note that for smaller $r$ values the velocity is finite while the jump amplitude is approximately zero, implying that solutions remain smooth. As $r$ increases the velocity decays until, seemingly at the same time, the velocity vanishes and the jump amplitude jumps to a finite value, implying that the corresponding solutions are in fact discontinuous at each node. This finite value is plotted in dashed black and can be predicted analytically, as we show below. Thus, the bifurcation $r_2(\Lambda)$ describing the transition from smooth to discontinuous solutions is given by this point where $|c_+-c_-|$ jumps, and is denoted by the vertical green dot-dashed line.

To gain some more insight into the jumping points $c_-$ and $c_+$ and the jump amplitude $|c_+-c_-|$, we now consider stationary period-two solutions of Eq.~(\ref{eq:mapc}). Assuming solutions of the form $-c_{n+1}(x)=c_n(x)=c(x)$, we take a derivative of Eq.~(\ref{eq:mapc}) with respect to space and find that away from each discontinuity solutions satisfy
\begin{align}
\Lambda c'(x)=\frac{-\alpha\Lambda a'(x) - (r-1)c(x)+c^3(x)}{r-1-3c^2(x)}.\label{eq:nlin2}
\end{align}
Thus, we see that when $3c^2(x)=r-1$ the denominator on the right-hand-side of Eq.~(\ref{eq:nlin2}) vanishes, causing the derivative $c'(x)$ to diverge and the profile $c(x)$ to develop a jump discontinuity. Thus, upon formation of discontinuities, the left jumping point is given by $c_-=\pm\sqrt{(r-1)/3}$. To find the right jumping point we note that stationary solutions satisfy the cubic equation
\begin{align}
(r-1)c(x)-c^3(x)=A(x),\label{eq:nlin3}
\end{align}
where $A(x)=-\alpha a(x)+\frac{\alpha}{\Lambda}\int_0^xe^{(x'-x)/\Lambda}a(x')dx'$. Since $a(x)$ is smoothed by the Green's function at each iteration, the quantity $A(x)$ remains smooth through the discontinuity in $c(x)$. The right jumping point $c_+$ is given by the other root of Eq.~(\ref{eq:nlin3}) at $x=x_0$, where $A(x_0)=(r-1)c_-+c_-^3=\pm2(r-1)^{3/2}/3\sqrt{3}$, yielding $c_+=\mp2\sqrt{(r-1)/3}$. Finally, the total jump amplitude is given by $|c_+-c_-|=\sqrt{3(r-1)}$. In Fig.~\ref{fig:VelocityJump} we plot this theoretical prediction of $|c_+-c_-|$ in dashed black, noting that the agreement with numerical simulations is excellent. 

We emphasize here that upon formation of discontinuities, the left and right jumping points take the values described above. We will refer to these as {\it normal jumps}. As we will see below, upon changes of parameters, the values of $c_-$ and $c_+$ can potentially change. However, we will see below that normal jumps play an important role in the dynamics in the discontinuous regime.

To further understand the bifurcation $r_2(\Lambda)$ that characterizes the transition from the smooth regime to the discontinuous regime (see Fig.~\ref{fig:VelocityJump}), we now study how discontinuities form in profiles as $r$ approaches $r_2(\Lambda)$ from below and surpasses it. In particular, we consider the length scale of the phase reversal corresponding to a given node. For solutions found in the smooth regime [see Fig.~\ref{fig:Examples} (a)] we expect the length scale of phase reversals to be finite. However, for patterns in the discontinuous regime [see Fig.~\ref{fig:Examples} (b)], we expect the length scale of the phase reversal to be comparable with the numerical discretization length scale.

\begin{figure}[t]
\centering
\epsfig{file =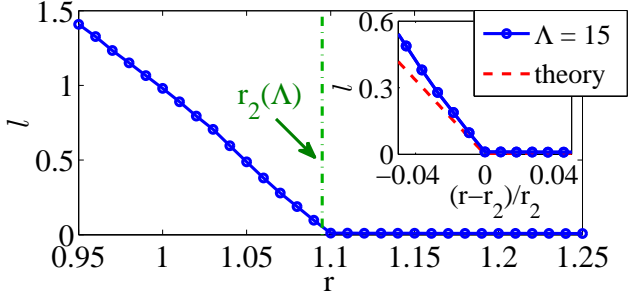, clip =,width=1.0\linewidth } 
\caption{(Color online) The length scale $l$ of phase reversals defined by Eq.~(\ref{eq:phasereversal}) over a range of $r$ values given $\Lambda=15$. Note that as $r$ approaches $r_2(\Lambda)$ (denoted by the vertical dot-dashed green line) from below, $l$ approaches zero. Other parameters are $\alpha,\gamma=\sqrt{0.3}$, $\beta=0$, $\xi=1$, and $w=0$ with $L=30$ and $\Delta x=0.005$. Inset: Comparison to the analytical expression for the flat CV limit.}\label{fig:PhaseReversal}
\end{figure}

Given a node at $x=x_0$, the corresponding length scale $l$ of the phase reversal, assuming a non-zero smooth profile, can be defined as
\begin{equation}\label{eq:phasereversal}
l=\frac{2c_\infty}{|c'(x_0)|},
\end{equation}
where $c_\infty=\max_{0\le x\le L}|c(x)|$ is the maximum value taken by $c(x)$ on the cable. In the main panel of Fig.~\ref{fig:PhaseReversal} we plot the length scale $l$ of phase reversals for the same parameter values as those from Fig.~\ref{fig:VelocityJump}. We note that in the smooth regime $l$ takes on finite values, and $l$ approaches zero as $r$ approaches $r_2(\Lambda)\approx1.1$. In particular, as $r$ approaches $r_2(\Lambda)$ from below the derivative $c'(x_0)$ increases, yielding a sharper phase reversal, until at $r=r_2(\Lambda)$ the derivative diverges, giving way to discontinuities at each node. In the inset of Fig.~\ref{fig:PhaseReversal} we compare these results to the theoretical prediction of $l$ for the limit of flat CV restitution where $\Lambda\to\infty$ and for $w=0$ by plotting $l$ vs $(r-r_2)/r_2$. In this limit it can be shown that as $r$ approaches $r_2$ the length scale of the phase reversal is given by
\begin{equation}
l\approx\frac{(1-r)\sqrt{2\pi\xi^2}}{\eta\text{Li}_{1/2}(\beta)/\beta},\label{eq:phasereversalFlatCV}
\end{equation}
where $\text{Li}_{s}(\beta)=\sum_{j=1}^\infty\beta^j/j^s$ is the polylogarithm function. We present the full derivation of Eq.~(\ref{eq:phasereversalFlatCV}) together with other properties of the flat CV case in Appendix~\ref{appC}. This theoretical prediction agrees well with simulations, indicating that the scaling of the phase reversal length scale near the critical value $r_2(\Lambda)$ for finite $\Lambda$ is similar to that for flat CV restitution.

\subsection{Hysteresis and unidirectional pinning}\label{sec5subB}

We will now consider the effects that changes of parameters have on solutions once steady-state is reached in the discontinuous regime. We will find below that two unique types of hysteresis are inherent to the discontinuous solutions. First, we find that if $r$ or $\Lambda$ are increased, then the location of each node remains unchanged. However, the shape of each node (e.g., the values of the jumping points $c_-$ and $c_+$) changes, yielding non-normal jumps. More specifically, while for normal jumps we have $|c_+|=2|c_-|$, we find that after increasing $r$ or $\Lambda$ that $c_-$ and $c_+$ change in such a way that $|c_+|$ and $|c_-|$ decreases and increases, respectively, approaching one another for the case of a symmetric Green's function, i.e., $w=0$. Thus, increasing $r$ or $\Lambda$ has a symmetrizing effect on the shape of $c(x)$ about each node. To quantify this, we introduce a measure of asymmetry defined by
\begin{align}
\Delta=\frac{|c_+|-|c_-|}{|c_{+}^0|-|c_{-}^0|}=\frac{|c_+|-|c_-|}{\sqrt{(r-1)/3}},\label{eq:nlin6}
\end{align}
where $c_{-}^0$ and $c_{+}^0$ are the left and right jumping point values for a normal jump. This normalization is made so that normal jumps yield an asymmetry of $\Delta=1$ regardless of the value of $r$. Furthermore, $\Delta$ approaching zero corresponds to $c_+$ and $c_-$ approaching one another in magnitude, i.e., a symmetrizing of the shape of profiles about each node. 

\begin{figure}[b]
\centering
\epsfig{file =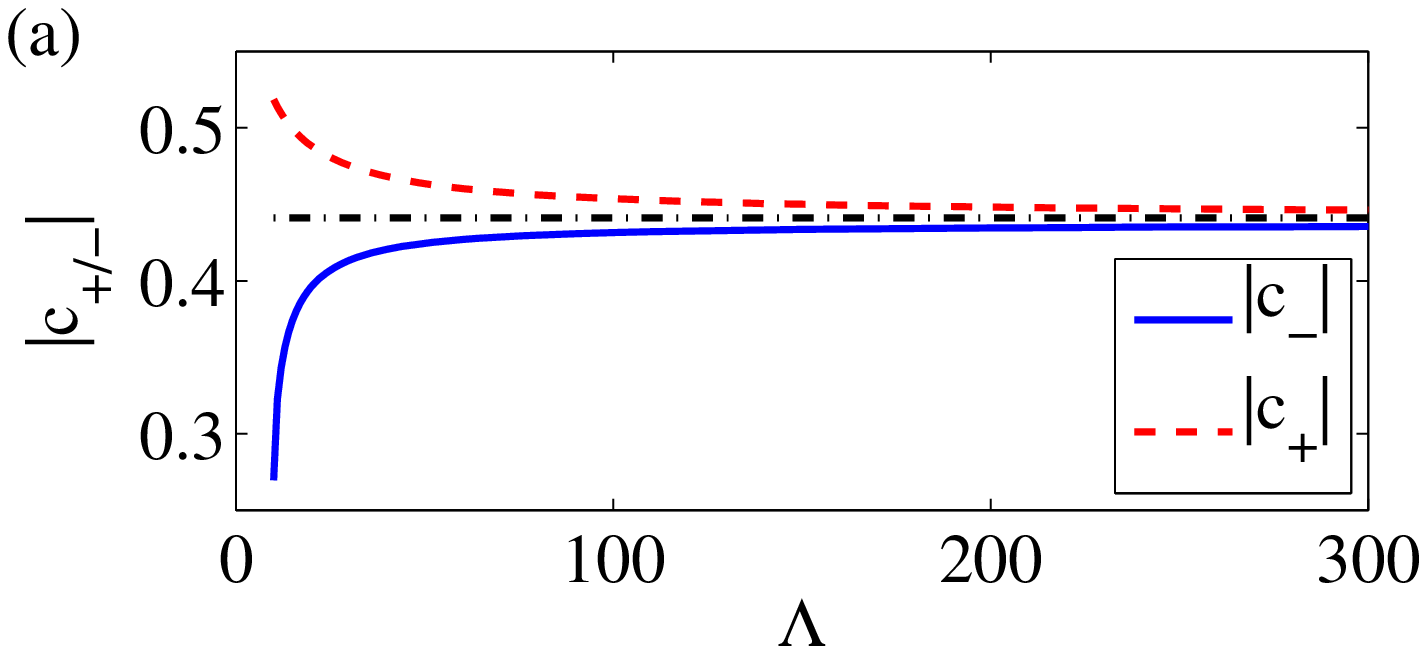, clip =,width=1.0\linewidth } \\
\epsfig{file =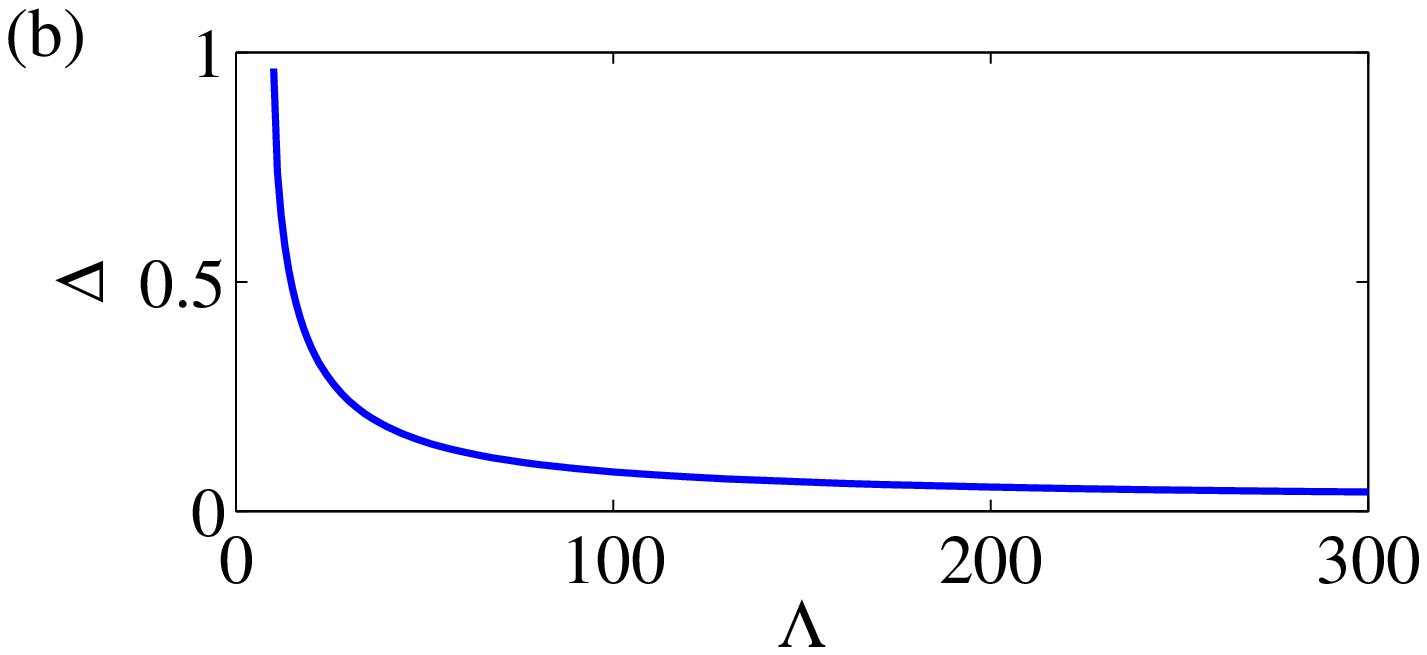, clip =,width=1.0\linewidth }
\caption{(Color online) (a) Jumping points values $|c_-|$ and $|c_+|$ and (b) asymmetry of nodes $\Delta$ as $\Lambda$ is increased from a steady-state profile with normal jumps with $r=1.2$ and $\Lambda=10$, plotted in solid blue and dashed red. Other parameters are $\alpha,\gamma=\sqrt{0.3}$, $\beta=0$, $\xi=1$, and $w=0$ with $L=30$ and $\Delta x=0.005$.}\label{fig:IncLambda}
\end{figure}

In Fig.~\ref{fig:IncLambda} we illustrate this phenomenon by obtaining a discontinuous pattern at $r=1.2$ and $\Lambda=10$, then increasing $\Lambda$ to $300$. In panel (a) we plot the magnitude of the jumping points $|c_-|$ and $|c_+|$ in solid blue and dashed red, respectively, and in panel (b) we plot the asymmetry $\Delta$ as a function of $\Lambda$. Simulations were performed on a cable of length $L=30$ with a spatial discretization of $\Delta x=0.005$, and other parameters are $\alpha,\gamma=\sqrt{0.3}$, $\beta=0$, $\xi=1$, and $w=0$. We find that $c_-$ and $c_+$ approach one another in absolute value as $\Lambda$ is increased. In fact, it can be shown by studying the large $\Lambda$ limit of Eqs.~(\ref{eq:mapc}) and (\ref{eq:mapa}) with $w=0$ that $|c_-|$ and $|c_+|$ approach the value $\sqrt{r-1}$ as $\Lambda\to\infty$, which is denoted in dot-dashed black. This result also follows from the analysis presented in Appendix~\ref{appC}. We see explicitly in panel (b) that as $\Lambda$ increases, $\Delta$ approaches zero. Furthermore, if we restore $\Lambda$ to its original value after increasing it, the profile recovers its original shape and previous jumping point values. Finally, we note that if the symmetry of the Green's function is broken with $w>0$, it can be shown that as $\Lambda$ is increased, the magnitude of the left jumping point $c_-$ eventually surpasses the magnitude of the right jumping point $c_+$, yielding a negative value for the asymmetry $\Delta$.

Next, we consider the effect that decreasing $r$ or $\Lambda$ has on discontinuous solutions. Interestingly, the effect is somewhat the opposite of what was described above: the jumping points $c_-$ and $c_+$ remain unchanged and the node locations move towards the pacing site at $x=0$. Furthermore, if $r$ or $\Lambda$ are restored to their original (larger) value, we find that the profile does {\it not} recover its original shape. Instead, the node remains pinned to the location closer to the pacing site and the shape of the node symmetrizes as described above. We refer to this phenomenon as {\it unidirectional pinning}.

\begin{figure}[t]
\centering
\epsfig{file =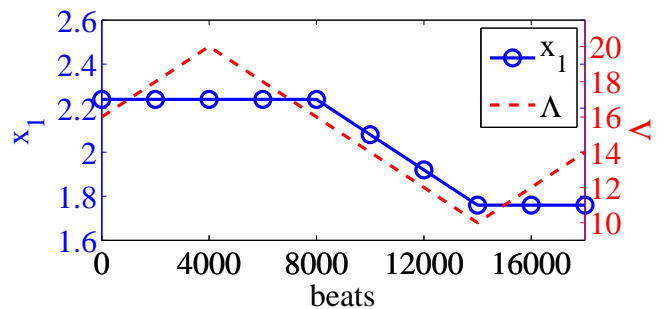, clip =,width=1.0\linewidth } 
\caption{(Color online) First node location $x_1$ and $\Lambda$, plotted in blue circles and dashed red, as $\Lambda$ is ``zig-zagged'' starting from a steady-state profile with normal jumps with $r=1.2$ and $\Lambda=16$. Other parameters are $\alpha,\gamma=\sqrt{0.3}$, $\beta=0$, $\xi=1$, and $w=0$ with $L=20$ and $\Delta x=0.02$.}\label{fig:ZigZagLambda}
\end{figure}

In Fig.~\ref{fig:ZigZagLambda} we illustrate the phenomenon of unidirectional pinning by plotting the location of the first node $x_1$ as we slowly ``zig-zag'' $\Lambda$ after obtaining a steady-state discontinuous solution at $r=1.2$ and $\Lambda=16$. Simulations were performed on a cable of length $L=20$ with a spatial discretization of $\Delta x=0.02$, and other parameters are $\alpha,\gamma=\sqrt{0.3}$, $\beta=0$, $\xi=1$, and $w=0$. Starting at $\Lambda=16$, we first increase it slowly to $20$, then decrease it slowly to $10$, and finally increase it slowly to $14$. As we begin initially increasing $\Lambda$, we note that the first node location $x_1$ (blue circles) remains pinned in its original location, and only decreases when $\Lambda$ is decreased past its previous minimum, at which point it appears to decrease linearly with $\Lambda$. Finally, when $\Lambda$ is increased again  $x_1$ remains pinned in its location nearest the pacing site. We find that all other node locations along the cable move in the same way as the first node location.

\begin{figure}[t]
\centering
\epsfig{file =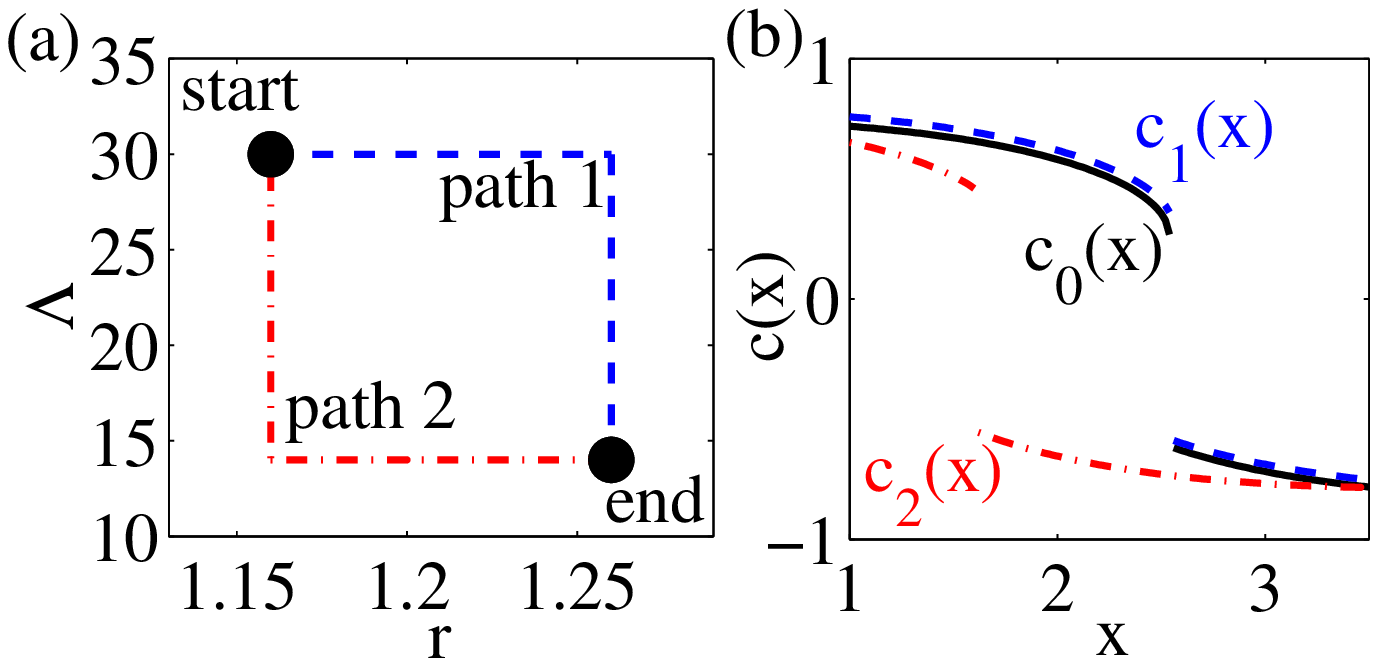, clip =,width=1.0\linewidth } \\
\epsfig{file =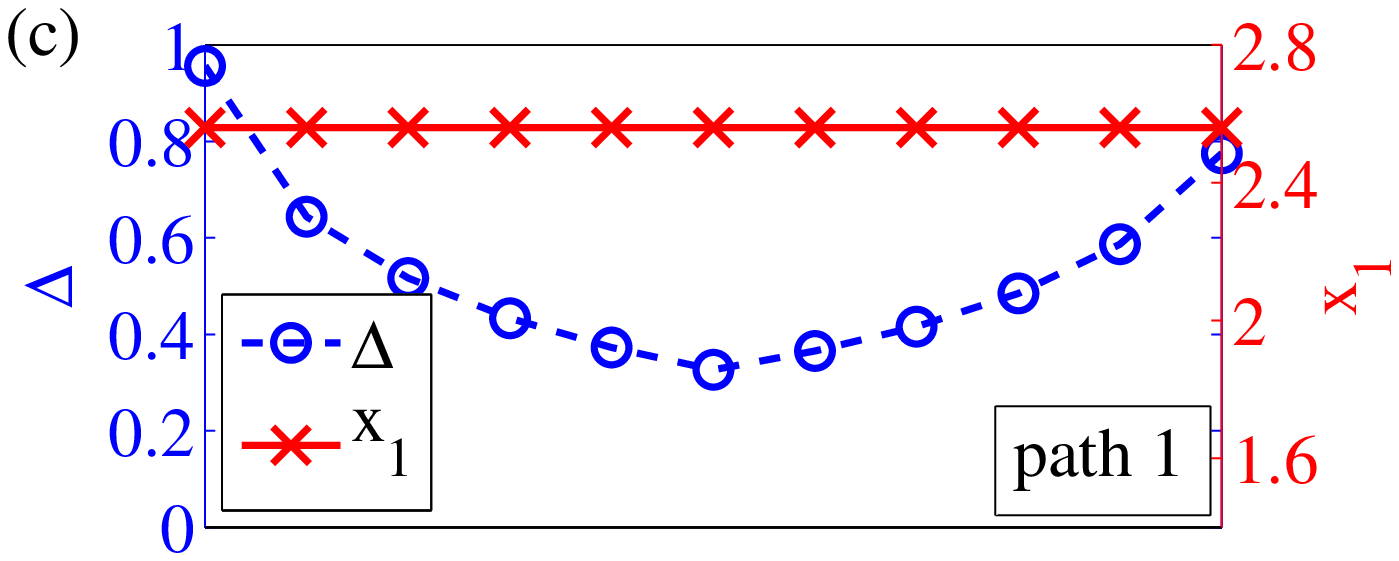, clip =,width=1.0\linewidth } \\
\epsfig{file =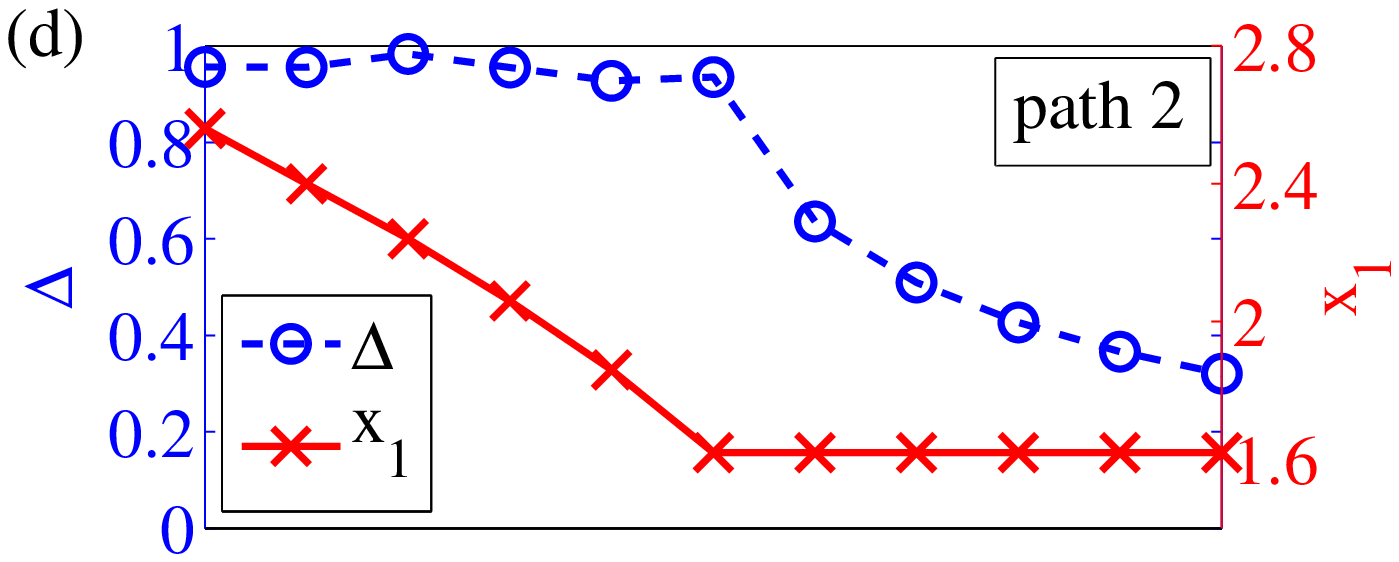, clip =,width=1.0\linewidth } 
\caption{(Color online) (a) Two paths plotted in dashed blue and dot-dashed red connecting $(r_1,\Lambda_1)=(1.16,30)$ and $(r_2,\Lambda_2)=(1.26,14)$. (b) Zoomed-in view of the first node of the initial profile $c_0(x)$ at $(r_1,\Lambda)$ and resulting profiles $c_1(x)$ and $c_2(x)$ after moving along paths one and two plotted in dashed blue and dot-dashed red. (c), (d) Asymmetry $\Delta$ (blue circles) and first node location $x_1$ (red crosses) along paths one and two. Other parameters are $\alpha,\gamma=\sqrt{0.3}$, $\beta=0$, $\xi=1$, and $w=0$ with $L=20$ and $\Delta x=0.02$.}\label{fig:rLambda}
\end{figure}

We conclude this subsection by presenting an example where, through changing both $r$ and $\Lambda$, we observe both symmetrizing of the profile near the node, as well as unidirectional pinning. To highlight the hysteretic behavior we see, we choose two parameter pairs $(r_1,\Lambda_1)=(1.16,30)$ and $(r_2,\Lambda_2)=(1.26,14)$ and construct two different paths that connect $(r_1,\Lambda_1)$ to $(r_2,\Lambda_2)$. Path one is traversed by first increasing $r$ from $r_1$ to $r_2$ while leaving $\Lambda=\Lambda_1$, then decreasing $\Lambda$ from $\Lambda_1$ to $\Lambda_2$ while keeping $r=r_2$, and path two is traversed by first decreasing $\Lambda$ from $\Lambda_1$ to $\Lambda_2$ while keeping $r=r_1$, then increasing $r$ from $r_1$ to $r_2$ while leaving $\Lambda=\Lambda_2$ [see Fig.~\ref{fig:rLambda} (a)]. Next, we perform two simulations where, after reaching steady-state at $(r_1,\Lambda_1)$, we move slowly along paths one and two.

In Fig.~\ref{fig:rLambda} we plot the results of these simulations. In panel (a) we plot the trajectories of paths one and two in the $(r,\Lambda)$ plane in dashed blue and dot-dashed red, respectively. In panel (b) we plot a zoomed-in view of the first node location for the initial profile $c_0(x)$ taken at $(r_1,\Lambda_1)$ in solid black, as well as the final profiles $c_1(x)$ and $c_2(x)$ obtained after moving along paths one and two in dashed blue and dot-dashed red, respectively. From these profiles, we can see that after moving along path one, the profile $c_1(x)$ is very similar to the initial profile $c_0(x)$, both with respect to the shape and location of the first node. However, the profile resulting from moving along path two, $c_2(x)$ is very different from $c_1(x)$, despite having the same parameters. In particular, the first node of $c_2(x)$ is much closer to the pacing site than that of $c_1(x)$. In panels (c) and (d), we explore the dynamics further by plotting the asymmetry $\Delta$ and the first node location $x_1$ in blue circles and red crosses, respectively, along paths one and two. Along the first half of path one, as $r$ increases, we see the node location remains constant and the asymmetry decreases, until the second half, where $\Lambda$ decreases, and the asymmetry is almost recovered. Along the first half of path two, however, as $\Lambda$ is decreased the asymmetry remains nearly constant at $\Delta\approx1$ while the node location decreases, until the second half where the node location remains constant and the asymmetry decreases as $r$ is increased. In particular, we note that at the end, where both simulations have the same parameter values $(r_2,\Lambda_2)$, both the asymmetry $\Delta$ and first node location $x_1$ are very different, depending on the path taken.

\subsection{Node dynamics: a framework for understanding hysteresis}\label{sec5subC}

Given the novel dynamics presented in the previous subsection, we will now present a framework for understanding the hysteretic behavior described there. To do so we will study the dynamics of $c(x)$ at a node $x_0$. Our goal is to show that in response to a change in parameters, and depending on which direction we change parameters in, then the following is true. First, starting at a normal jump, the absolute value of jumping points $|c_-|$ and $|c_+|$ can approach one another, but not move away from one another. Second, nodes move towards the pacing site at $x=0$, but not away. We note that movement towards the pacing site corresponds to the point $c(x_0)=c_-$ transitioning to $c_+$.

We begin by noting that with the definition $A_n(x) = -\alpha a_n(x)+\frac{\alpha}{\Lambda}\int_0^xe^{(x'-x)/\Lambda}a_n(x')dx'$, Eq.~(\ref{eq:mapc}) can be rewritten as
\begin{align}
c_{n+1}(x)=-rc_n(x) + c_n^3(x) + A_n(x).\label{eq:nlin7}
\end{align}
Importantly, $A_n(x)$ encompasses all the non-local effects contributing to the local evolution of $c_{n+1}(x)$. Dropping subscripts, we note that $A$ is explicitly a function of $x$ and $a(x)$, as well as the parameters $\alpha$ and $\lambda$, but it is also indirectly a function of $c(x)$ and the other parameters $r$, $\gamma$, $\beta$, $\xi$, and $w$, so in principle we have that $A=A(x,c,a,\bm{p})$, where $\bm{p}$ is a vector containing all system parameters. We saw previously that for a normal jump the right and left jumping points $c_-$ and $c_+$ are given by two roots of the cubic polynomial
\begin{align}
F(c)=(r-1)c-c^3-A(x_0)=0,\label{eq:nlin8}
\end{align}
where $A(x_0)=\pm2(r-1)^{3/2}/3\sqrt{3}$, yielding $c_-=\pm\sqrt{(r-1)/3}$ and $c_+=\mp2\sqrt{(r-1)/3}$. We note here that $c_-$ turns out to be a double root of Eq.~(\ref{eq:nlin8}).

In order to understand the node dynamics, one can study the initial local dynamics of $c=c(x_0)$ in response to small changes in $A$ induced by small changes in either the parameters or global profiles $a(x)$ or $c(x)$. More specifically, upon a change of parameters we consider how $A$ changes through its explicit dependence on these parameters, while neglecting the change in $A$ due to the implicit dependence of $c(x)$ and $a(x)$ on the modified parameters. We argue that this approach is valid because the local dynamics of $c(x_0)$ occur very quickly in comparison to the global dynamics of $c(x)$. To study the dynamics of $c(x)$ at $x=x_0$, we introduce the quantity $\Delta c_n = (-c_{n+1})-c_n$, which describes, after reflecting the $n+1^{st}$ beat, the evolution of $c(x_0)$ from beat $n$ to $n+1$. We also choose without any loss of generality the positive root of $A$. Choosing the negative root yields the same results for a flipped calcium profile. From Eq.~(\ref{eq:nlin7}), we have that
\begin{align}
\Delta c_n = (r-1)c_n-c_n^3-A_n(x_0)=F(c_n).\label{eq:nlin9}
\end{align}
Now, as we noted in Sec.~\ref{sec2}, beat-to-beat dynamics of alternans amplitudes evolve slowly, so $\Delta c_n$ can be treated as the continuous-time derivative $dc/dt$, where the time variable $t$ is in units of beats. Thus, the dynamics of Eq.~(\ref{eq:nlin9}) can be approximated by the ODE
\begin{align}
\frac{dc}{dt}=-\frac{\partial V}{\partial c},\label{eq:nlin10}
\end{align}
where $V(c)$ represents the energy potential given by
\begin{align}
V(c)=c^4/4-(r-1)c^2/2+A(x_0)c.\label{eq:nlin11}
\end{align}

\begin{figure}[t]
\centering
\epsfig{file =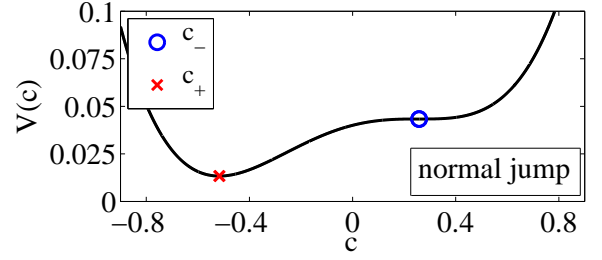, clip =,width=1.0\linewidth } 
\caption{(Color online) Potential well $V(c)$ for a normal jump with $r=1.2$ whose equilibria give the jumping points $c_-$ (blue circle) and $c_+$ (red cross).}\label{fig:Well1}
\end{figure}

In Fig.~\ref{fig:Well1} we illustrate the potential well given by $V(c)$ for a normal jump, i.e., $A(x_0)=2(r-1)^{3/2}/3\sqrt{3}$, for $r=1.2$. The steady-state jumping points $c_-$ and $c_+$ are thus represented by the equilibria of Eq.~(\ref{eq:nlin10}), plotted as a blue circles and red crosses, respectively. The equilibrium representing $c_+$ is a true minimum of $V(c)$ and therefore stable to perturbations in both directions. However, the equilibrium representing $c_-$ is semi-stable, i.e., only stable to perturbations in the positive direction, so that in response to a negative perturbation, $c(x_0)$ will ``roll down'' to the $c_+$ equilibrium.

\begin{figure}[b]
\centering
\epsfig{file =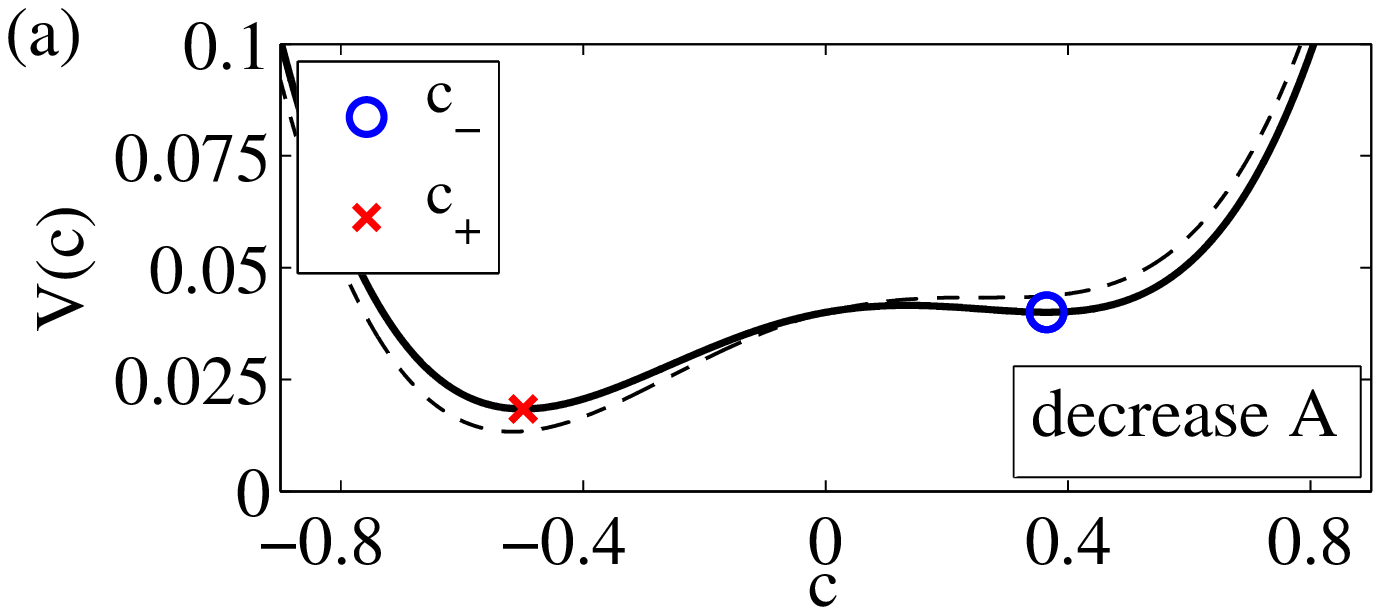, clip =,width=1.0\linewidth } \\
\epsfig{file =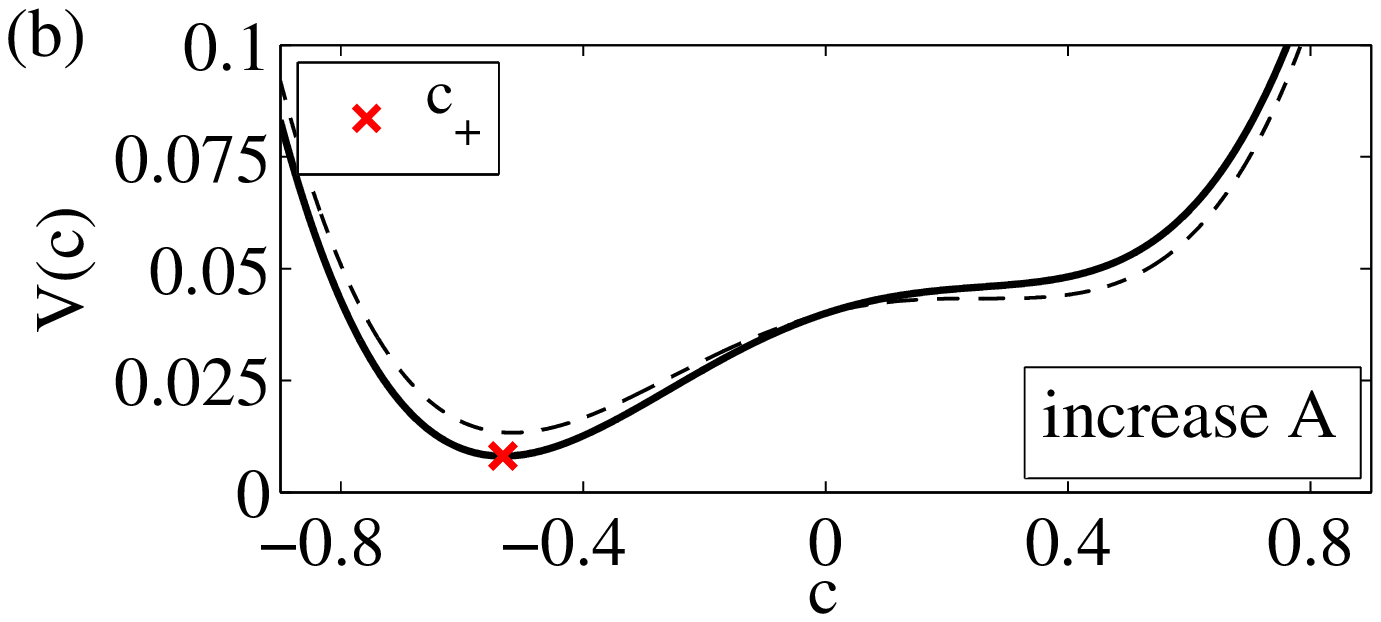, clip =,width=1.0\linewidth } 
\caption{(Color online) Potential well $V(c)$ for non-normal jump with $r=1.2$ after $A(x_0)$ is (a) decreased and (b) increased (solid lines). The original potential is shown in dashed lines. Equilibria represent the jumping points $c_-$ (blue circle) and $c_+$ (red cross).}\label{fig:Well2}
\end{figure}

We now consider how a perturbation to $A(x_0)$ changes $V(c)$. In panels (a) and (b) of Fig.~\ref{fig:Well2} we illustrate how $V(c)$ changes in response to a small decrease and increase, respectively, in $A(x_0)$ by plotting the original potential $V(c)$ as a dashed curves and the modified potential $V(c)$ as a solid curve. When $A(x_0)$ is decreased, the potential well changes in such a way that both equilibria $c_+$ and $c_-$ increase slightly, becoming more symmetric, and $c_-$ becomes a true minimum of $V(c)$ and therefore fully stable. Furthermore, since both equilibria remain, there is no switching of $c(x_0)$ corresponding to any node movement. On the other hand, when $A(x_0)$ is increased the potential well changes in such a way that the $c_+$ equilibrium remains, while the $c_-$ equilibrium vanishes. Thus, $c_+$ is the only equilibrium remaining and a point previously at $c(x_0)=c_-$ must transition to $c_+$ by ``rolling down'' the well, corresponding to movement of the node towards the pacing site. Thus, with a decrease or increase in $A(x_0)$ nodes respond by either symmetrizing their shape or moving towards the pacing site, respectively. In Appendix~\ref{appB} we illustrate in greater detail how node movement is driven by points switching from $c_-$ to $c_+$ with numerical experiments of Eqs.~(\ref{eq:mapc}) and (\ref{eq:mapa}) with different spatial discretizations. 

The analysis above provides a framework for understanding the node dynamics, i.e., symmetrizing of jumping points and unidirectional pinning, via perturbations of the nonlocal term $A(x_0)$. We now quantify how $A(x_0)$ changes in response to changes in the different parameters in the model by numerically computing the derivative of $A(x_0)$ with respect to each parameter. In particular, we are interested in the sign of each derivative. If the derivative is positive, then decreasing the parameter will cause $A(x_0)$ to decrease and the nodes to symmetrize and remain pinned, and an increase in the parameter will cause $A(x_0)$ to increase and the nodes to move towards to pacing site or asymmetrize. On the other hand, if the derivative is negative, then decreasing the parameter will cause $A(x_0)$ to increase and the nodes to move towards the pacing site or asymmetrize, and an increase in the parameter will cause $A(x_0)$ to decrease and the nodes to symmetrize and remain pinned.

\begin{table}[t]
\centering
 \caption{Numerically computed derivatives of $A(x_0)$ with respect to different parameters for $r=1.2$, $\Lambda=15$, $\alpha,\gamma=\sqrt{0.3}$, $\beta=0$, $\xi=1$, and $w=0$.}
 \label{table:dA}
\begin{tabular}{|c||c|}
\hline
  & Derivative value\\
 \hline
 \hline
 $\partial A/\partial \Lambda$ & -0.00232418\\
 \hline
 $\partial A/\partial r$& -0.0281439\\
 \hline
 \hline
 $\partial A/\partial \alpha$ & 0.462741\\
 \hline
 $\partial A/\partial \beta$ & 0.0769836\\
 \hline
 $\partial A/\partial \gamma$& 0.10796\\
 \hline
\end{tabular}
\setlength{\abovecaptionskip}{10pt}
\end{table}

In Table~\ref{table:dA} we present the results from numerically computing the derivatives of $A(x_0)$ with respect to $\Lambda$, $r$, $\alpha$, $\beta$, and $\gamma$ at $r=1.2$, $\Lambda=15$, $\alpha,\gamma=\sqrt{0.3}$, $\beta=0$, $\xi=1$, and $w=0$. We note that both $\partial A/\partial\Lambda$ and $\partial A/\partial r$ are negative, confirming that decreasing and increasing both $\Lambda$ and $r$ causes the nodes to move towards the pacing site and nodes to symmetrize, respectively, as we have shown above. Furthermore, we find that each $\partial A/\partial\alpha$, $\partial A/\partial\beta$, and $\partial A/\partial\gamma$ are positive, implying that the opposite is true, which we find to occur in numerical experiments. While the derivative values presented here are for a particular choice of parameters, further investigation suggests that the signs of the derivatives are preserved for all other relevant parameter choices, yielding qualitatively similar dynamics.

\subsection{Scaling of the spatial wavelength}\label{sec5subD}

Next we study the scaling behavior of the spatial wavelength $\lambda_s$ of solutions in the discontinuous regime. We begin by noting that in the middle portion of Fig.~\ref{fig:ZigZagLambda} the first node location $x_1$ appears to scale linearly with $\Lambda$. To investigate this further, we recall that by assuming period-two stationary solutions $-c_{n+1}(x)=c_n(x)=c(x)$ and taking a derivative of Eq.~(\ref{eq:mapc}) we obtained Eq.~(\ref{eq:nlin2}) which solutions satisfy away from discontinuities. If instead we take a derivative with respect to the scaled variable $\tilde{x}=x/\Lambda$ and redefine $\tilde{c}(\tilde{x})=c(\Lambda\tilde{x})$ and $\tilde{a}(\tilde{x})=a(\Lambda\tilde{x})$, we obtain the new ODE
\begin{align}
\tilde{c}'(\tilde{x})=\frac{-\alpha\tilde{a}'(\tilde{x})-(r-1)\tilde{c}(\tilde{x})+\tilde{c}^3(\tilde{x})}{r-1-3\tilde{c}^2(\tilde{x})},\label{eq:nlin12}
\end{align}
along with the following expression for $\tilde{a}(\tilde{x})$,
\begin{align}
\tilde{a}(\tilde{x}) = \int G_{\tilde{\xi}}(\tilde{x},\tilde{x}')[-\beta\tilde{a}(\tilde{x}')+\gamma\tilde{c}(\tilde{x}')]d\tilde{x}',\label{eq:nlin13}
\end{align}
where $\tilde{\xi}=\xi/\Lambda$. For simplicity we have assumed that $w=0$ and we have included in the notation of $G$ explicitly its length scale $\tilde{\xi}$. In the limit $\xi/\Lambda\to0$ the Green's function becomes a delta function, Eq.~(\ref{eq:nlin13}) yields $\tilde{a}(\tilde{x})=\gamma\tilde{c}(\tilde{x})/(1+\beta)$, and Eq.~(\ref{eq:nlin12}) has solutions with a wavelength $\tilde{\lambda}$ independent of $\Lambda$. Therefore we expect the wavelength of the profiles for the original system (\ref{eq:nlin2}) for large $\Lambda$ to be approximately proportional to $\Lambda$, i.e., $\lambda_s\approx\tilde{\lambda}\Lambda$, plus a small correction of order $\xi/\Lambda$.

\begin{figure}[t]
\centering
\epsfig{file =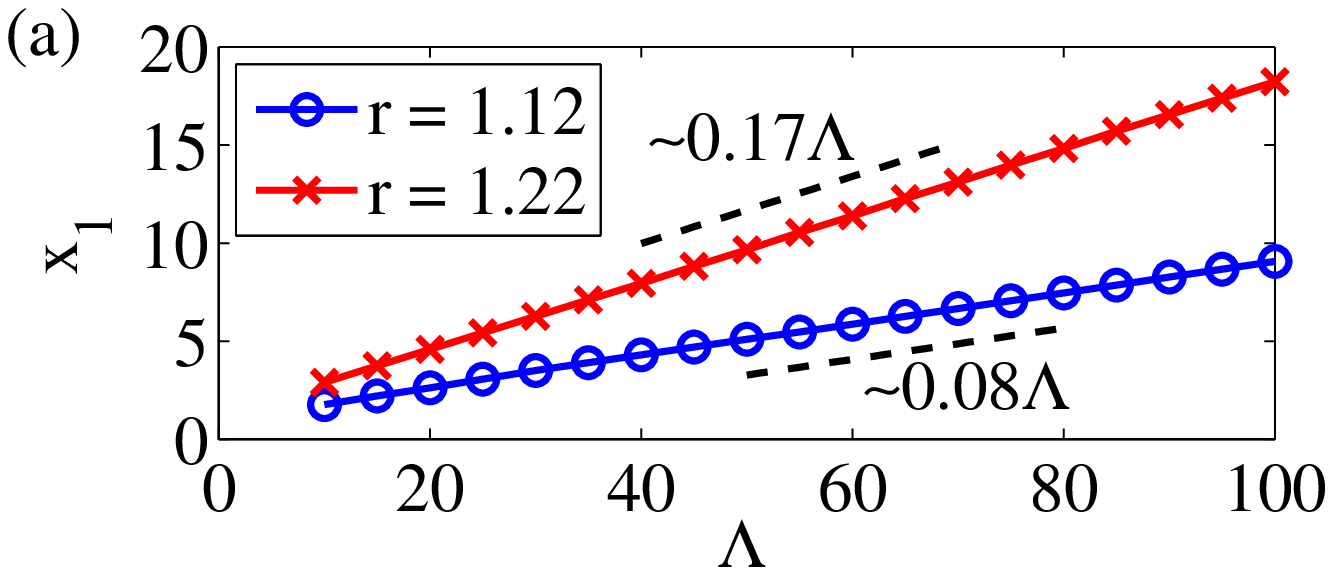, clip =,width=1.0\linewidth } \\
\epsfig{file =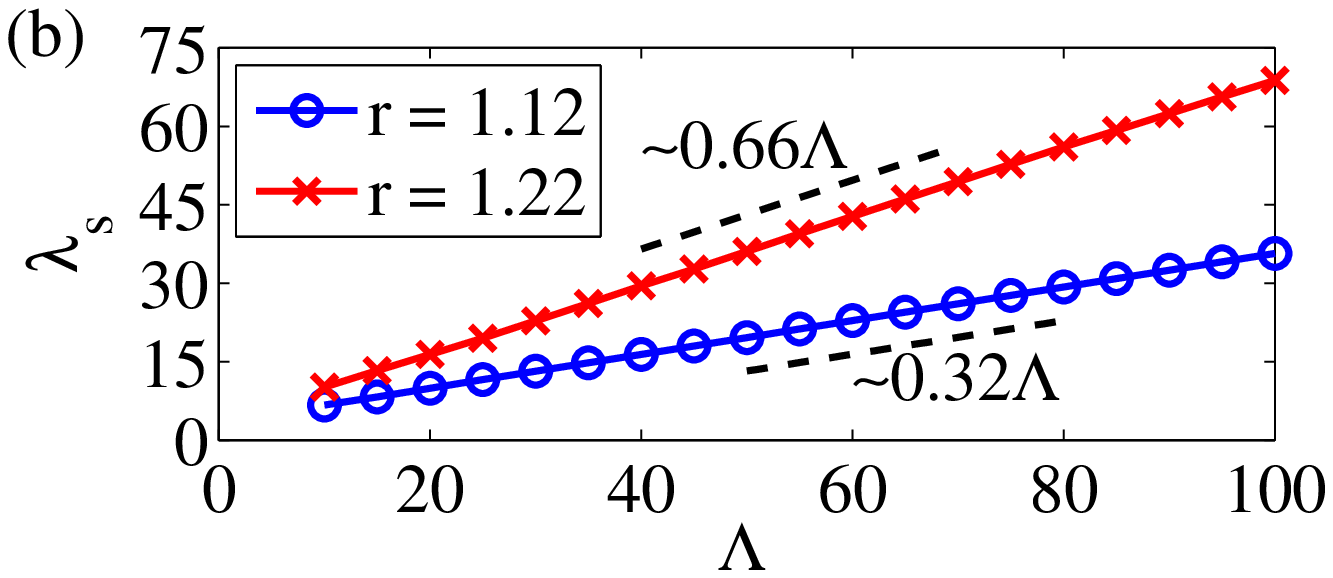, clip =,width=1.0\linewidth }
\caption{(Color online) (a) First node location $x_1$ and (b) spatial wavelength $\lambda_s$ of solutions in the discontinuous regime for $r=1.12$ and $1.22$. Results are well-approximated by $x_1=1.08+0.08\Lambda$ and $\lambda_s=3.54+0.32\Lambda$ ($r=1.12$) and $x_1=1.12+0.17\Lambda$ and $\lambda_s=3.16+0.66\Lambda$ ($r=1.22$). Other parameters are $\alpha,\gamma=\sqrt{0.3}$, $\beta=0$, $\xi=1$, and $w=0$ with $L=200$ and $\Delta x=0.04$.}\label{fig:Scaling}
\end{figure}

To test this hypothesis, we simulate Eqs.~(\ref{eq:mapc}) and (\ref{eq:mapa}) until we obtain a steady-state discontinuous solution, then slowly decrease $\Lambda$ and observe the spatial scaling behavior of solutions. In Figs.~\ref{fig:Scaling} (a) and (b) we plot the first node location $x_1$ and the spatial wavelength $\lambda_s$, respectively, for $r=1.12$ (blue circles) and $1.22$ (red crosses). Simulations were performed on a long cable ($L=200$) with spatial discretization $\Delta x=0.04$, and other parameters are $\alpha,\gamma=\sqrt{0.3}$, $\beta=0$, $\xi=1$, and $w=0$. We find that both $x_1$ and $\lambda_s$ quantities scale with $\Lambda$ and are well approximated by $x_1=1.08+0.08\Lambda$ and $\lambda_s=3.54+0.32\Lambda$ for $r=1.12$ and $x_1=1.12+0.17\Lambda$ and $\lambda_s=3.16+0.66\Lambda$ for $r=1.22$, agreeing with our hypothesis. In particular, the scaling of spatial wavelengths in the discontinuous regime, i.e., $\lambda_s\sim\Lambda$, is qualitatively different than the scaling at the onset of alternans, where $\lambda_s\sim(\xi^2\Lambda)^{1/3}$ or $\lambda_s\sim(w\Lambda)^{1/2}$ [see Eqs.~(\ref{eq:linstab8}) and (\ref{eq:appA13})].

\subsection{Random fluctuations and node dynamics}\label{sec5subE}

Numerical simulations of ionic models have shown that random cell-to-cell fluctuations in the initial phase of Ca alternans give rise to {\it nodal areas}, i.e., relatively thin regions that contain several rapid, fine-scale phase reversals in Ca alternans~\cite{Sato2013PLOS}. While fine-scale phase reversals tend to be eliminated rapidly in regions of large APD alternans, they can remain in nodal regions where the APD alternans amplitude is small. It remains unclear what are the effects of changes in control parameters on those nodal areas with multiple jumps of Ca alternans amplitude. In particular, are such profiles subject to the unidirectional pinning phenomenon described above for single jumps? Here we show that multiple jumps in nodal areas do in fact display unidirectional pinning dynamics. Furthermore, we show that node dynamics tends to sharpen nodal areas into a single node (i.e. collapse multiple jumps into a single jump), effectively ``washing away'' the effect of random initial fluctuations.

\begin{figure}[b]
\centering
\epsfig{file =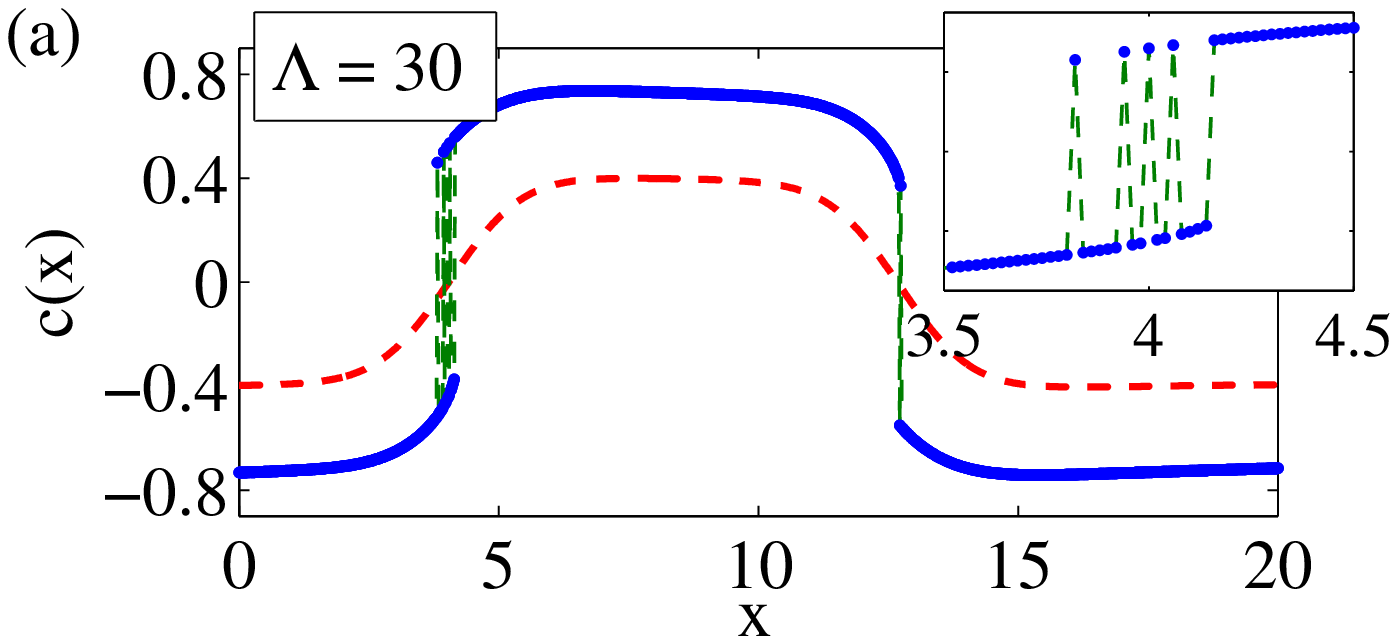, clip =,width=1.0\linewidth } \\
\epsfig{file =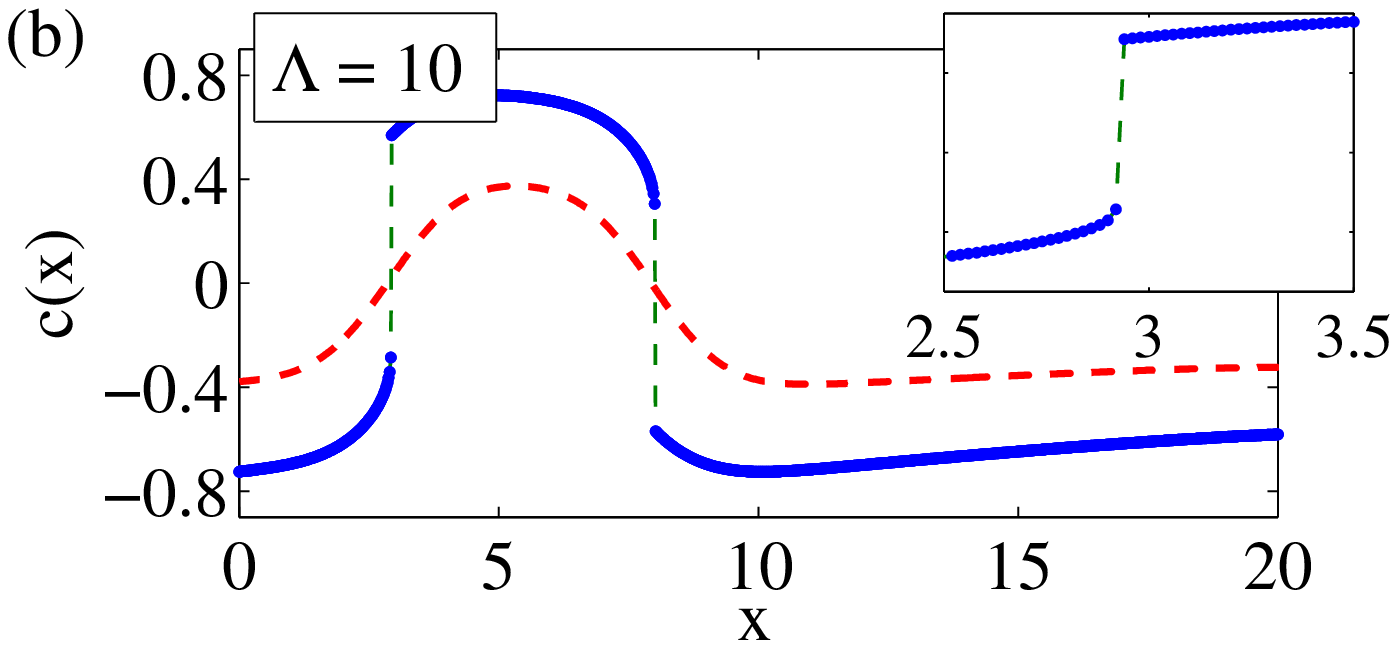, clip =,width=1.0\linewidth }
\caption{(Color online) Alternans profiles $c(x)$ (blue dots) and $a(x)$ (dashed red) for random initial conditions [each point $c(x)$ drawn independently from the uniform distribution $\mathcal{U}(-0.1,0.1)$] for (a) initial CV parameter $\Lambda=30$ and (b) after slowly decreasing $\Lambda$ to $10$. Insets: zoomed-in vein of the first nodal area. Other parameters are $r=1.2$, $\alpha,\gamma=\sqrt{0.3}$, $\beta=0$, $\xi=1$, $w=0$ with $L=20$ and $\Delta x=0.02$.}\label{fig:Random}
\end{figure}

To show this, we perform the following experiment. Beginning with a random initial calcium profile $c(x)$ where each point is drawn independently from the uniform distribution $\mathcal{U}(-0.1,0.1)$ and initial CV parameter $\Lambda=30$, we first evolve Eqs.~(\ref{eq:mapc}) and (\ref{eq:mapa}) until steady-state is reached. Simulations were performed on a cable of length $L=20$ with a spatial discretization of $\Delta x=0.02$, and other parameters are $r=1.24$, $\alpha,\gamma=\sqrt{0.3}$, $\beta=0$, $\xi=1$, and $w=0$. In Fig.~\ref{fig:Random} (a) we plot the resulting $c(x)$ and $a(x)$ profiles (blue dots and dashed red, respectively). To highlight the fine-scale variations in $c(x)$ in the nodal area we connect adjacent points of $c(x)$ with thin, dashed green lines and show a zoomed-in view of the first nodal area in the inset. Here we find near $x=4$ a nodal area containing nine rapid phase reversals. Next we slowly decrease $\Lambda$ to $10$, thereby inducing unidirectional node motion towards the pacing site, and plot the steady-state profile in Fig.~\ref{fig:Random} (b), again showing a zoomed-in view of the first nodal area in the inset. In addition to the node movement towards the pacing site, the nodal region has agglomerated into a single node near $x=3$, thus washing away the remnant effects of random initial fluctuations in $c(x)$. We conclude that unidirectional node motion collapses multiple jumps of Ca alternans amplitude accumulating in the nodal area into a single jump.

This phenomenon can be understood in terms of the amplitude equations as follows. First, note that in nodal areas (see Fig.~\ref{fig:Random}) $a(x)\approx0$. Thus, in these relatively thin regions calcium dynamics is entirely driven by local effects and CV restitution, explaining the emergence of fine-scale cell-to-cell variations in $c(x)$. Next, the agglomeration of phase reversals in a nodal area into a single node is the result of CV restitution inducing movement of the furthest phase reversal in a nodal area towards the pacing site until it collides and combines with those closer to the pacing site, eventually resulting in a single node. We note that, in contrast with our findings of Sec.~\ref{sec5subB}, this typically does not occur immediately as $\Lambda$ is decreased from its initial value, but only after CV restitution has become sufficiently steep. This is due to the fact that in the presence of fluctuations the (first) node often forms in a location closer to the pacing site than dictated by the effect of CV restitution. Thus, $\Lambda$ must be decreased by a finite amount to compensate for the initial node position before node movement towards the pacing site is induced. Importantly, if the opposite experiment is performed, i.e., $\Lambda$ is increased, then nodal areas {\it do not} agglomerate and the multiple jumps of Ca alternans amplitude originating from initial cell-to-cell fluctuations remain. Thus, nodal area agglomeration only occurs when nodal movement is induced, with this motion being always towards the pacing site due to unidirectional pinning.

\section{Ionic Model, Restitution Curves, and Numerical Experiments}\label{sec6}

Equipped with a detailed understanding of the dynamics of the amplitude equations~(\ref{eq:mapc}) and (\ref{eq:mapa}), we shift our focus to the dynamics of the cable equation [Eq.~(\ref{eq:cable})] with a detailed ionic model. In particular, we aim to show that the results we have obtained from the reduced model can be used to predict, both qualitatively and quantitatively, behavior of the cable equation with a biologically robust detailed ionic model. As we mentioned in Sec.~\ref{sec4}, we have chosen to use the Shiferaw-Fox ionic model, which combines the calcium cycling dynamics of Shiferaw et al.~\cite{Shiferaw2003BiophysJ} with the ionic current dynamics of Fox et al.~\cite{Fox2001AJPHC}. Importantly, the coupling between detailed calcium and voltage dynamics given by the Shiferaw-Fox model allows for a robust enough model to produce calcium-driven alternans for relatively large parameter ranges.

We will begin this section by describing the important parameter values of the Shiferaw-Fox model that we have chosen in subsection~\ref{sec6subA}. Next, in subsection~\ref{sec6subB} we describe and present the APD and CV restitution curves for the Shiferaw-Fox model. We will then show with numerical simulations that we can predict both qualitatively and quantitatively the behavior of the Shiferaw-Fox model using results from our reduced model. In subsection~\ref{sec6subC} we present examples of unidirectional pinning in the Shiferaw-Fox model. In subsection~\ref{sec6subD} we study the discontinuous solutions. Finally, in subsection~\ref{sec6subE} we study the scaling of spatial wavelengths.

\subsection{Details of the ionic model}\label{sec6subA}

The most important feature of the Shiferaw-Fox model for our purposes is its ability to produce calcium-driven alternans. To this end, we choose parameter values of the model to promote instabilities in the calcium dynamics while suppressing instabilities in the voltage dynamics. Voltage-driven alternans is often caused by long inactivation timescales of the voltage-dependent gating variables. These long inactivation timescales cause ionic current dynamics to take longer to equilibrate between subsequent beats, thus more easily transitioning to a regime of period-two dynamics. To suppress instabilities in the voltage dynamics we identify the voltage inactivation timescale $\tau_f$ which directly affects the voltage equation~(\ref{eq:cable}) through the L-type calcium current. In particular, we choose $\tau_f$ to be relatively small, i.e., $\tau_f=30$ ms as compared to other typical values of $\tau_f\approx40$-$60$ ms for which voltage driven alternans can be achieved relatively easily. We note that variables and parameters described in this text refer to the Shiferaw-Fox model as implemented in Ref.~\cite{KroghMadsen2007BiophysJ}.

In the calcium-cycling dynamics of the Shiferaw-Fox model, the primary mechanism for calcium ions entering the cell cytoplasm, aside from the standard L-type calcium current, is the release of stored calcium from the sarcoplasmic reticulum (SR), a network of rigid tubule-like structures that store calcium within the cell. The release of calcium from the SR occurs via a positive-feedback process in response to the activation of the L-type calcium current. In the Shiferaw-Fox model, the rate of calcium release by this mechanism is determined by a parameter $u$. Larger (smaller) choices of $u$ typically correspond to more (less) instability in the calcium cycling dynamics. Thus, to promote calcium instabilities, we choose a relatively large release parameter of $u=9$ ms$^{-1}$.

We also make other parameter choices that should be noted before moving on. First, recall that we are interested in studying calcium-driven alternans when the calcium-to-voltage (as well as the voltage-to-calcium) coupling is positive. To ensure that this coupling is positive, following~\cite{Shiferaw2005PRE}, we change the calcium-inactivation exponent $\gamma$, which affects the inactivation of the L-type calcium current. In short, $\gamma<1$ ($\gamma>1$) typically corresponds to positive (negative) calcium-to-voltage coupling. Here we set $\gamma=0.2$.

With the parameter choices described above, another parameter we can change is the BCL, i.e., the period at which the cable is paced at $x=0$. By decreasing (increasing) BCL, we allow the tissue less (more) time to equilibrate between subsequent beats, thus promoting (suppressing) instabilities. Regarding our reduced system, decreasing (increasing) BCL corresponds to increasing (decreasing) $r$. In addition to the degree of calcium instability $r$, another parameter that played a large dynamical role in the reduced system was the CV restitution length scale $\Lambda$. Thus, it will be useful to identify a parameter in the Shiferaw-Fox model that will effectively change $\Lambda$ as well. To this end, we consider the spiking behavior of the voltage dynamics that occurs at the beginning of each action potential, since, as with many other types of excitable media, the velocity with which activity propagates through the tissue depends primarily on the sharpness of the front of the propagating dynamics. The dynamics responsible for the spiking at the beginning of each action potential is primarily contained in the fast sodium current. Therefore, by controlling the timescale of the fast-sodium dynamics, as done in Ref.~\cite{Sato2006CircRes}, we can modulate the sharpness of the spike. In particular, we introduce a scaling parameter $\tau$ that scales the timescale of the fast sodium $j$-gate dynamics, i.e., $1/(\alpha_m+\beta_m)\to\tau/(\alpha_m+\beta_m)$. Increasing $\tau$ slows the dynamics of the $j$-gate, yielding a more mild spike at the beginning of each action potential. Thus, as we will see below, increasing (decreasing) $\tau$ effectively decreases (increases) CV. Below we will show more precisely how $\tau$ can be used to change the shape of the CV restitution curve and change $\Lambda$. We note that changing $\tau$ could potentially change other parameters of the reduced model. However, as increasing $\tau$ weakens the initial action potential upstroke, we expect that the primary effect is on CV restitution. We note that numerical simulation support this hypothesis. The Shiferaw-Fox model with similar parameter choices has also been used in other numerical studies of cardiac dynamics~\cite{KroghMadsen2007BiophysJ,Sato2007BiophysJ}. Unless otherwise noted, ionic model simulations are all performed on a cable of length $L=15$ cm with a spatial discretization of $\Delta x=0.02$ cm.

\subsection{Restitution curves}\label{sec6subB}

\begin{figure}[t]
\centering
\epsfig{file =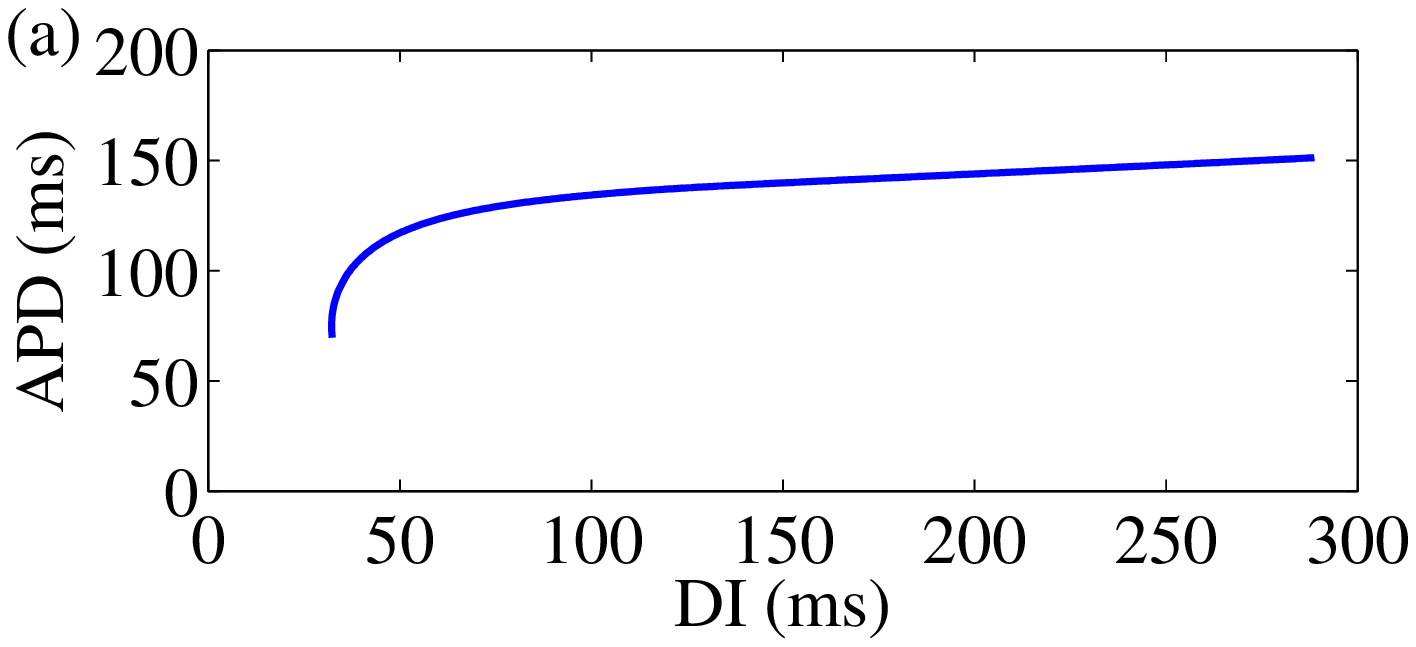, clip =,width=1.0\linewidth } \\
\epsfig{file =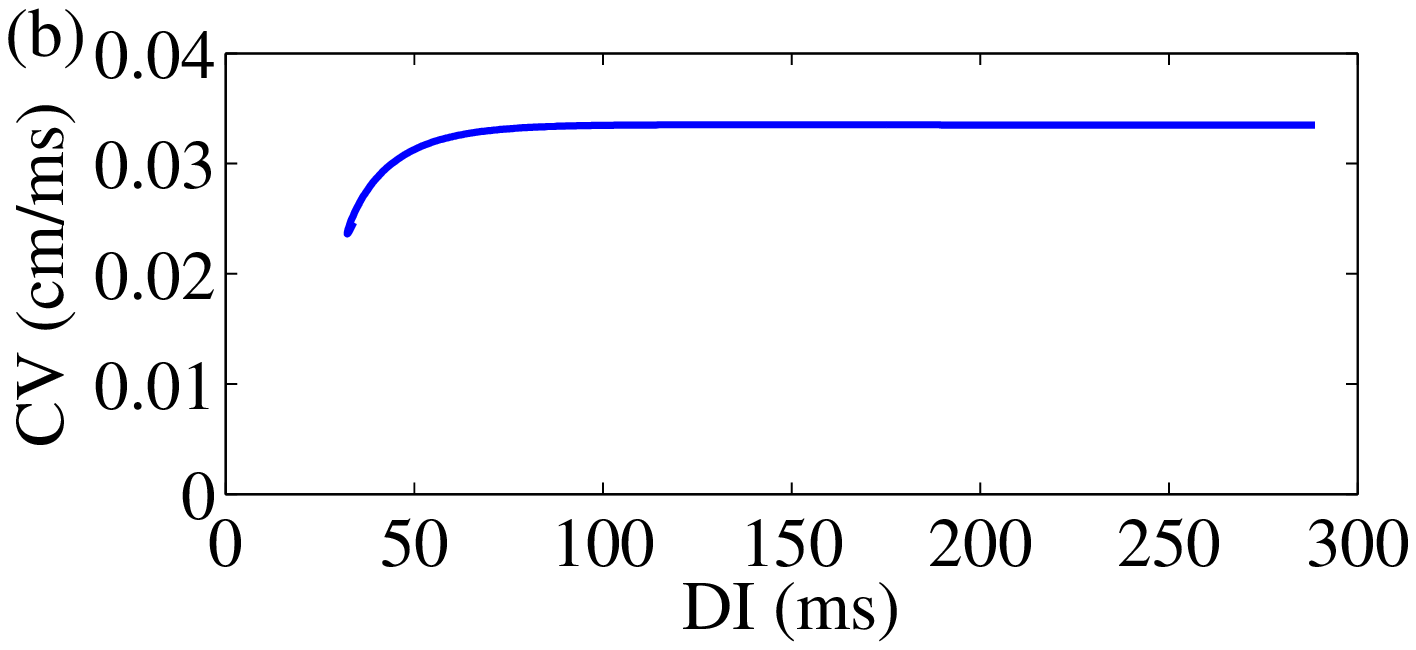, clip =,width=1.0\linewidth } 
\caption{(Color online) APD and CV restitution curves calculated for the Shiferaw-Fox model for scaling parameter value $\tau=2$.}\label{fig:Res}
\end{figure}

We now present the APD and CV restitution curves for the Shiferaw-Fox model, which describe the APD and CV, respectively, as a function of the DI at the previous beat. In particular, recall from the derivation of the reduced model in Sec.~\ref{sec2} that the CV restitution curve $cv(D)$ plays a crucial role in the dynamics as it defines the CV restitution length scale $\Lambda=(cv^*)^2/2cv'^*$. Both restitution curves are typically computed numerically by measuring the APD and CV at a point half-way through a relatively short cable. Here we take the cable to be of length $L=1$ cm. Both are calculated using the S1S2 pacing protocol, i.e., pacing a cable at a large period BCL$_1$ (taken here to be BCL$_1=440$ ms) until steady-state is reached, then decreasing the BCL to a value BCL$_2<$BCL$_1$, and storing the resulting abbreviated DI and the resulting APD and CV. This process is repeated for many values of BCL$_2$ until the APD and CV restitution curves are complete. 

\begin{figure}[t]
\centering
\epsfig{file =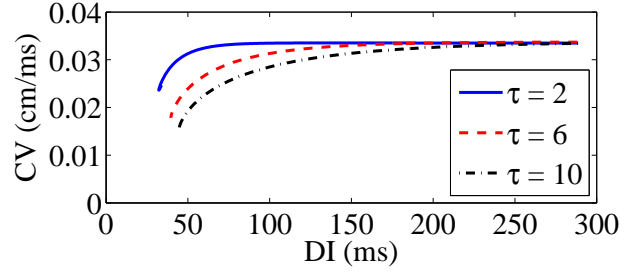, clip =,width=1.0\linewidth } 
\caption{(Color online) Effect of increasing the scaling parameter $\tau$ on the CV restitution curve for the Shiferaw-Fox model. In solid blue, dashed red, and dot-dashed black the CV restitution curves for $\tau=2$, $6$, and $10$, respectively.}\label{fig:CVRes}
\end{figure}

In Figs.~\ref{fig:Res} (a) and (b) we plot the resulting APD and CV restitution curves, respectively, obtained from the Shiferaw-Fox model using a scaling parameter value of $\tau=2$. We first note that the general shape of both curves is similar: both are monotonically increasing with a steep slope for smaller DI that becomes milder for larger DI. Recall, however, that $\Lambda$ is inversely proportional to the slope of the CV restitution curve $cv(D)$ at the onset of alternans $D=D^*$, which turns out to occur where CV restitution is very flat. Thus for smaller values of $\tau$ including $\tau=1$ (i.e., the unmodified model) and $\tau=2$, the parameter values that yield the restitution curves in Fig.~\ref{fig:Res} (b) where $cv'(D)\approx0$ yield an extremely large value of $\Lambda$. 

\begin{figure}[b]
\centering
\epsfig{file =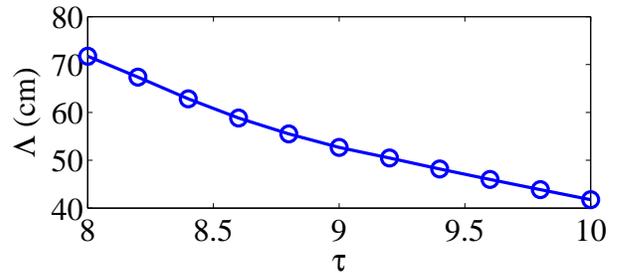, clip =,width=1.0\linewidth } 
\caption{(Color online) CV restitution length scale $\Lambda$ as computed numerically from the Shiferaw-Fox model over a range of values of the scaling parameter $\tau$.}\label{fig:TauLambda}
\end{figure}

To obtain more mild values of $\Lambda$ in our ionic model, we can modify the scaling parameter $\tau$. In fact, as observed in~\cite{Sato2006CircRes}, we find that increasing $\tau$, i.e., slowing down the fast sodium $j$-gate dynamics, tends to unflatten the CV restitution curve. In Fig.~\ref{fig:CVRes} we plot the resulting CV restitution curves for $\tau=2$, $6$, and $10$ in solid blue, dashed red, and dot-dashed black, respectively. We note in particular that as $\tau$ increases, so does the slope $cv'(D)$ along the whole curve. We note that this same technique was used in the supplemental material of Ref.~\cite{Sato2006CircRes}.

Next, we explicitly calculate $\Lambda$ for the Shiferaw-Fox model for several different values of the scaling parameter $\tau$. Since $\Lambda=(cv^*)^2/2cv'^*$, where $cv^*=cv(D^*)$ is the CV at the onset of alternans, we calculate the CV restitution curve $cv(D)$ for several values of $\tau$, from which we calculate the derivative $cv'(D)$ numerically. Next, for each value of $\tau$, we simulate a short cable while slowly decreasing BCL to find the critical onset value $D^*$. Thus, we can finally evaluate $cv$ and $cv'$ at $D^*$. In Fig.~\ref{fig:TauLambda} we plot the resulting values of $\Lambda$ as a function of the scaling parameter $\tau$ between $8$ and $10$. In particular, we note that for these parameter values, $\Lambda$ decreases monotonically with $\tau$. Therefore, via the scaling parameter $\tau$ we have a mechanism of varying the CV restitution length scale $\Lambda$ that features prominently in the dynamics of the reduced model. We also note that over this range of $\tau$ we can change $\Lambda$ by a relatively large amount, and we find that $\Lambda\gg1$, as we assumed in our analysis of the reduced model. 

\subsection{Unidirectional pinning}\label{sec6subC}

We will now present a series of numerical simulations designed to show that results from our reduced model can be used to predict behavior in the cable equation [Eq.~(\ref{eq:cable})] with the Shiferaw-Fox model. We begin with the phenomenon of unidirectional pinning. Recall that unidirectional pinning is the phenomenon observed in the discontinuous regime characterized by the fact that, by changing parameters, nodes can be moved towards, but not away from, the pacing site. 

\begin{figure}[t]
\centering
\epsfig{file =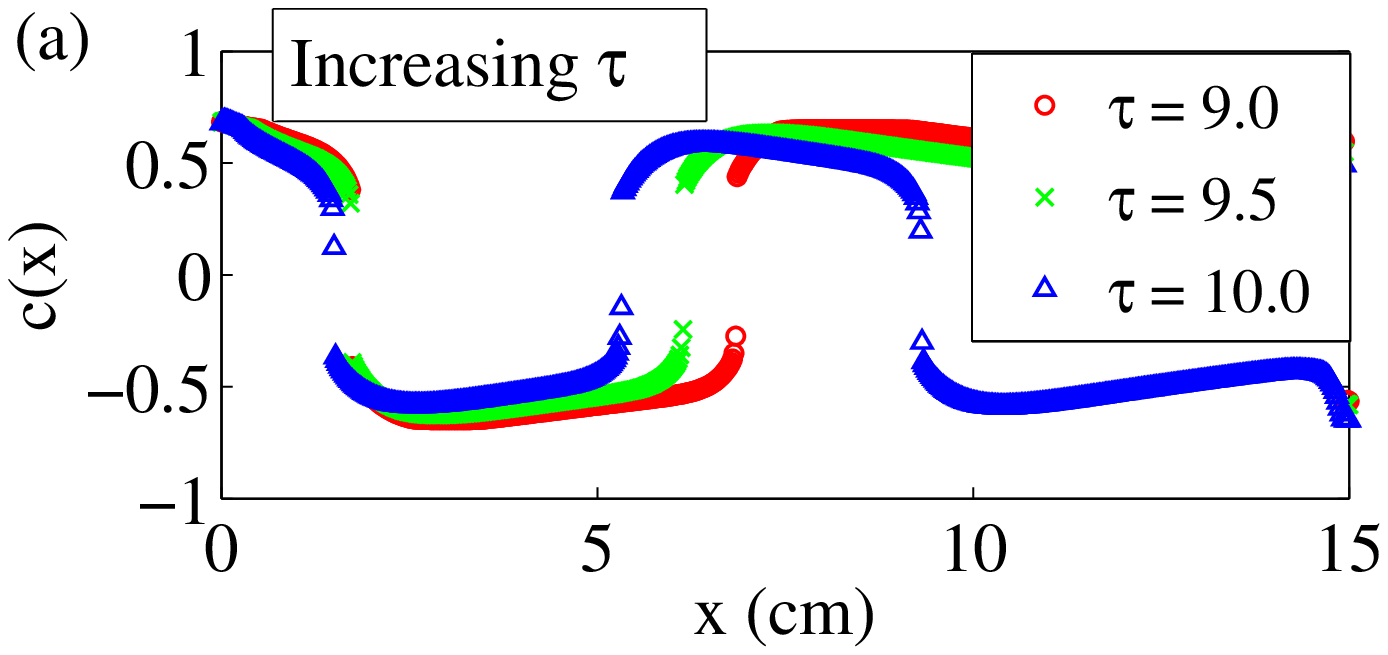, clip =,width=1.0\linewidth } \\
\epsfig{file =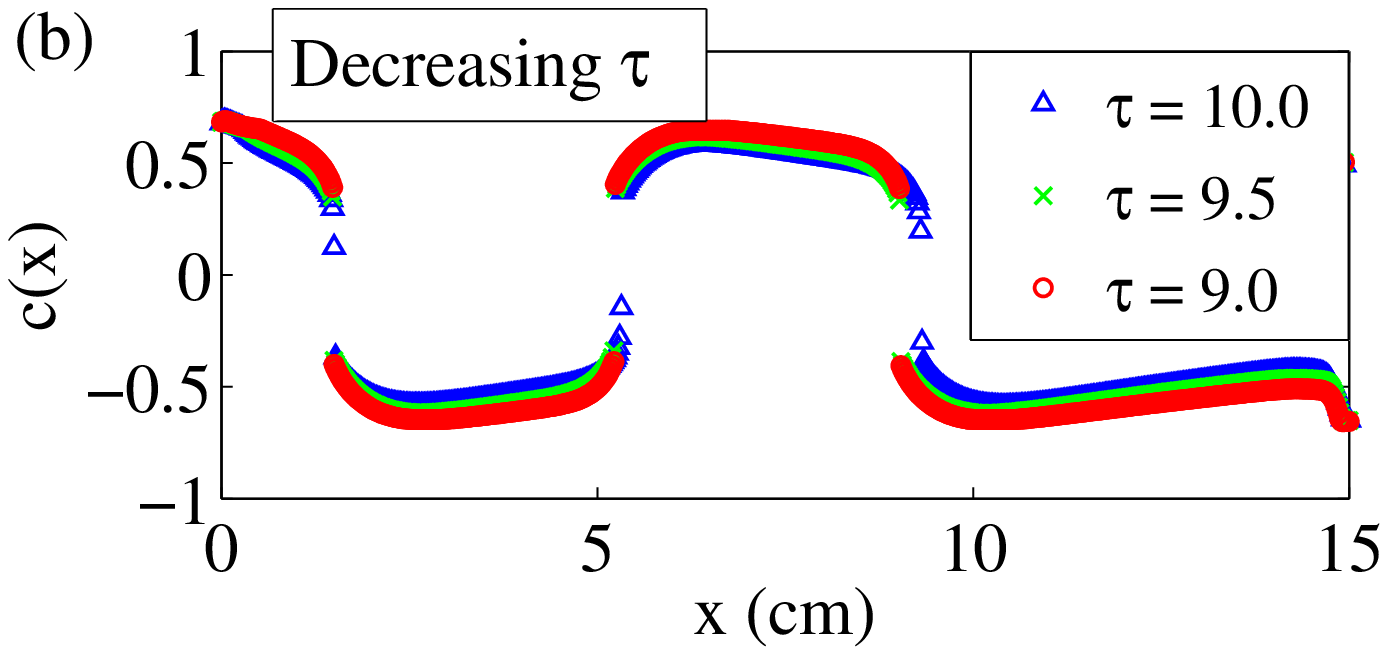, clip =,width=1.0\linewidth } \\
\epsfig{file =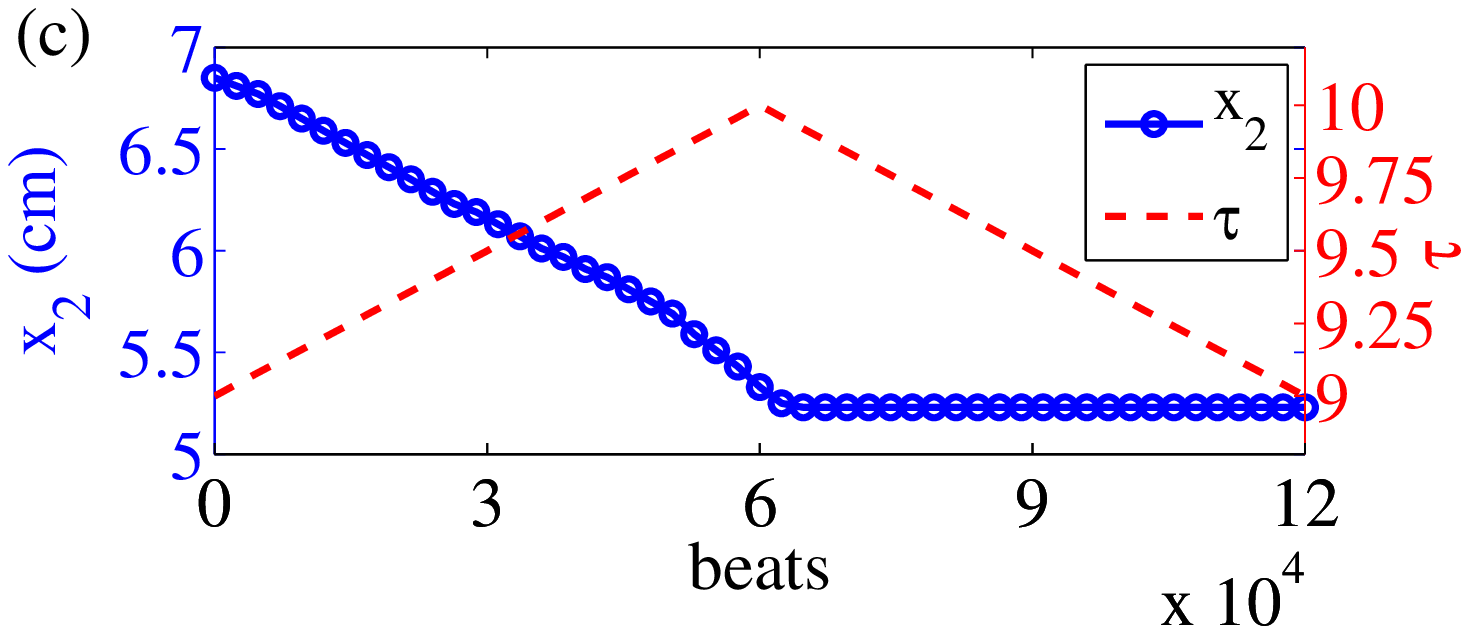, clip =,width=1.0\linewidth } 
\caption{(Color online) Unidirectional pinning in the Shiferaw-Fox model via changing $\tau$. (a) Calcium profiles $c(x)$ as $\tau$ is increased from $9$ to $10$ and (b) calcium profiles $c(x)$ as $\tau$ is restored to $9$. (c) Second node location $x_2$ (blue circles) and $\tau$ (dashed red) vs beat number.}\label{fig:IonicPinningTau}
\end{figure}

By studying the reduced model [Eqs.~(\ref{eq:mapc}) and (\ref{eq:mapa})], we have found that there are many parameters that can be changed to observe unidirectional pinning, including both the CV restitution length scale $\Lambda$ and the degree of calcium instability $r$. To test the predictions in the ionic model, we begin by modifying $\Lambda$, recalling from Fig.~\ref{fig:ZigZagLambda} that, starting from a normal jump, when $\Lambda$ is decreased the nodes move towards the pacing site, after which if we increase $\Lambda$ the nodes remain pinned in their locations close to the pacing site. Now that we have a way of changing the CV restitution length scale for the Shiferaw-Fox model, we show that unidirectional pinning can be observed in the Shiferaw-Fox model as well.

In Fig.~\ref{fig:IonicPinningTau} we plot the results from a simulation of a cable of length $L=15$ cm where we have first slowly increased $\tau$ from $9$ to $10$, then slowly decreased it from $10$ back to $9$. The pacing protocol here is to simulate the cable for $12000$ beats to achieve steady-state, then change $\tau$ by $\Delta\tau=0.02$ ms every 500 beats. Recall that by increasing (decreasing) $\tau$ we effectively decrease (increase) the CV restitution length scale $\Lambda$ (see Fig.~\ref{fig:TauLambda}). In subfigure (a) we plot the profile of the amplitude of calcium alternans $c(x)$ at $\tau=9$, $9.5$, and $10$ in red circles, green crosses, and blue triangles, respectively, as we first increase $\tau$. Note that the node locations move towards the pacing site at $x=0$ during this process, as predicted by our reduced model. Furthermore, due to the fixed finite size of the cable, an additional node forms. In subfigure (b) we plot the profile of the amplitude of calcium alternans $c(x)$ as we now decrease $\tau$, plotting profiles at $\tau=10$, $9.5$, and $9$ in blue triangles, green crosses, and red circles. Importantly, we note that as $\tau$ is restored to $9$ the nodes remain pinned in their locations close to the pacing site. To highlight this pinning, we plot in subfigure (c) the second node location, $x_2$, and $\tau$ versus the beat number in blue circles and dashed red, respectively. In this plot it is easy to see that the node first moves towards the pacing site as $\tau$ is initially increased, but remains pinned as we restore $\tau$ to its initial value. Thus, we have confirmed that unidirectional pinning is observable in detailed ionic models and is not simply an artifact of our reduced model. We also note that from (b) it is apparent that the jumping points and asymmetry of $c(x)$ about the nodes change as $\tau$ is restored to $9$, which we will study in more detail in the next subsection.

\begin{figure}[t]
\centering
\epsfig{file =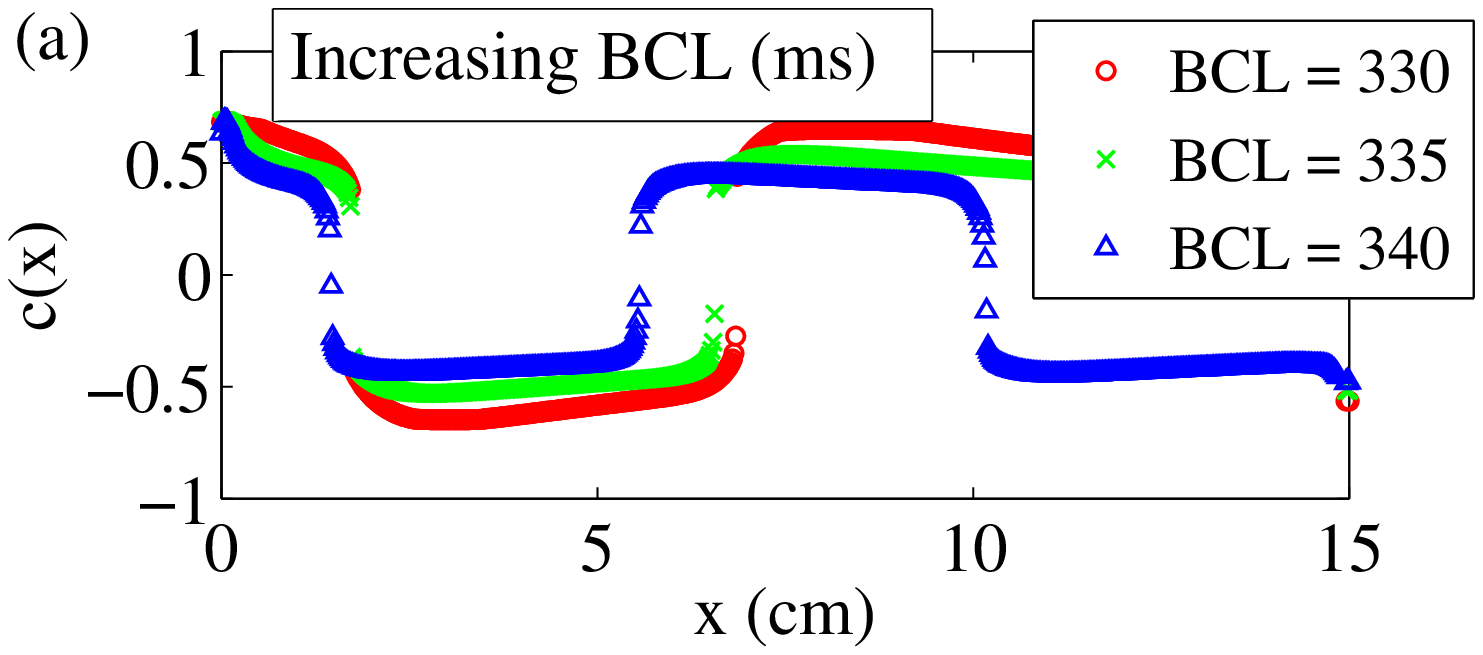, clip =,width=1.0\linewidth } \\
\epsfig{file =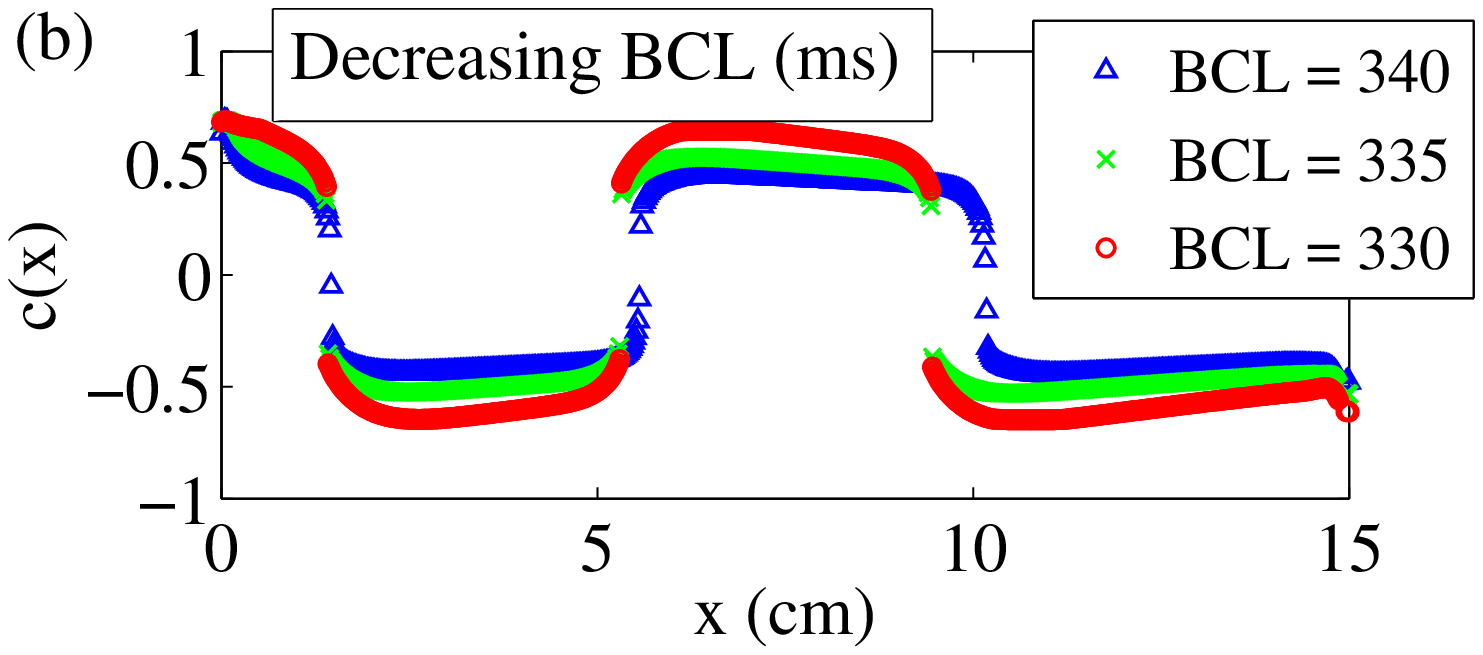, clip =,width=1.0\linewidth } \\
\epsfig{file =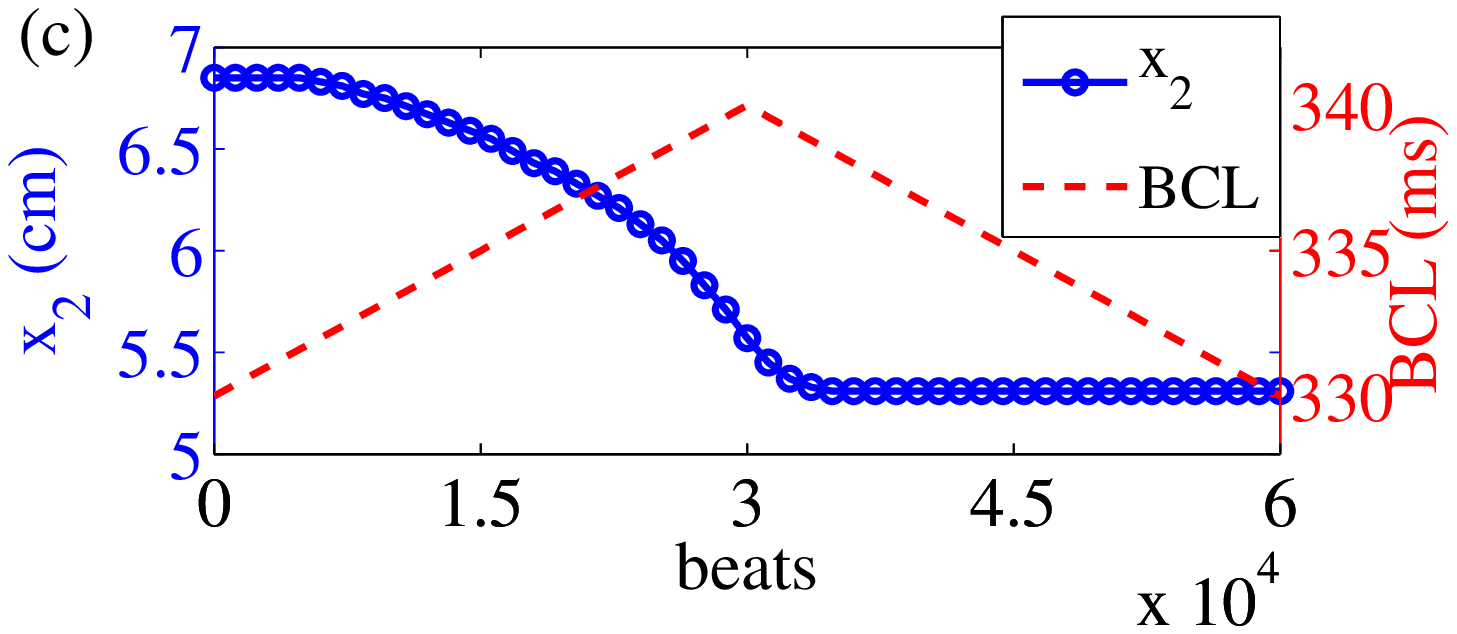, clip =,width=1.0\linewidth }
\caption{(Color online) Unidirectional pinning in the Shiferaw-Fox model via changing BCL. (a) Calcium profiles $c(x)$ as BCL is increased from $330$ ms to $340$ ms and (b) calcium profiles $c(x)$ as BCL is restored to $330$ ms. (c) Second node location $x_2$ (blue circles) and BCL (dashed red) vs beat number.}\label{fig:IonicPinningBCL}
\end{figure}

As we found in Sec.~\ref{sec5subB}, changing other parameters of the reduced model can also induce unidirectional pinning, in particular the degree of calcium instability (see Fig.~\ref{fig:rLambda}). To this end, we now show that unidirectional pinning can be achieved in the Shiferaw-Fox model by simply changing the BCL, which is easily done in experiments. Thus, we expect that the following results could be qualitatively reproduced in experiments. In Fig.~\ref{fig:IonicPinningBCL} we plot the results from a simulation of a cable of length $L=15$ cm where we have first slowly increased BCL from $330$ ms to $340$ ms, and then slowly decreased it from $340$ ms back to $330$ ms. The pacing protocol here is to simulate the cable for $12000$ beats to achieve steady-state, then change BCL by $\Delta$BCL$=0.1$ ms every 300 beats. In subfigure (a) we plot the profile of the amplitude of calcium alternans $c(x)$ at BCL$=330$ ms, $335$ ms, and $340$ ms in red circles, green crosses, and blue triangles, respectively, as we first increase BCL. Note that the node locations move towards the pacing site at $x=0$ during this process, as predicted by our reduced model, in a similar fashion to changing $\tau$ (see Fig.~\ref{fig:IonicPinningTau}). Furthermore, due to the fixed finite size of the cable, and additional node forms, as it did when $\tau$ changed. In subfigure (b) we plot the profile of the amplitude of calcium alternans $c(x)$ as we now decrease BCL, plotting profiles at BCL$=340$ ms, $335$ ms, and $330$ ms in blue triangles, green crosses, and red circles. Importantly, we note that as BCL is restored to $330$ ms the node locations remain pinned in their locations close to the pacing site. We again highlight the pinning phenomenon by plotting in subfigure (c) the second node location, $x_2$, and $\tau$ versus the beat number in blue circles and dashed red, respectively. Just as in the previous simulation where $\tau$ was modified, we see that the node first moves towards the pacing site as the BCL is initially increased, but remains pinned as we restore the BCL to its initial value. 

This confirms that unidirectional pinning can be achieved in detailed ionic models by changing only the pacing frequency. However, these results need to be interpreted carefully. In particular, it is well known that a change in BCL results in a change in CV restitution as follows~\cite{Echebarria2002PRL,Echebarria2007PRE}: a decrease (increase) in BCL yields a steeper (shallower) CV via decreasing (increasing) DI. However, a change in BCL can also affect change in the degree of calcium instability: a decrease (increase) in BCL allows the calcium dynamics less (more) time to equilibrate between beats, yielding a larger (smaller) degree of instability. Thus, changing the pacing rate yields competing effects from CV restitution and the degree of instability. Here we find that the change in CV restitution is small in comparison to the change in instability, which is dominant. Thus, node movement is induced by decreasing the degree of instability (as predicted by the reduced model and illustrated in Fig.~\ref{fig:Scaling}), i.e., by increasing BCL. In principle, however, if the change in CV restitution dominates the change in instability, we expect that node movement towards the pacing site will be induced by decreasing BCL.

\subsection{Jumping points and asymmetry}\label{sec6subD}

\begin{figure}[b]
\centering
\epsfig{file =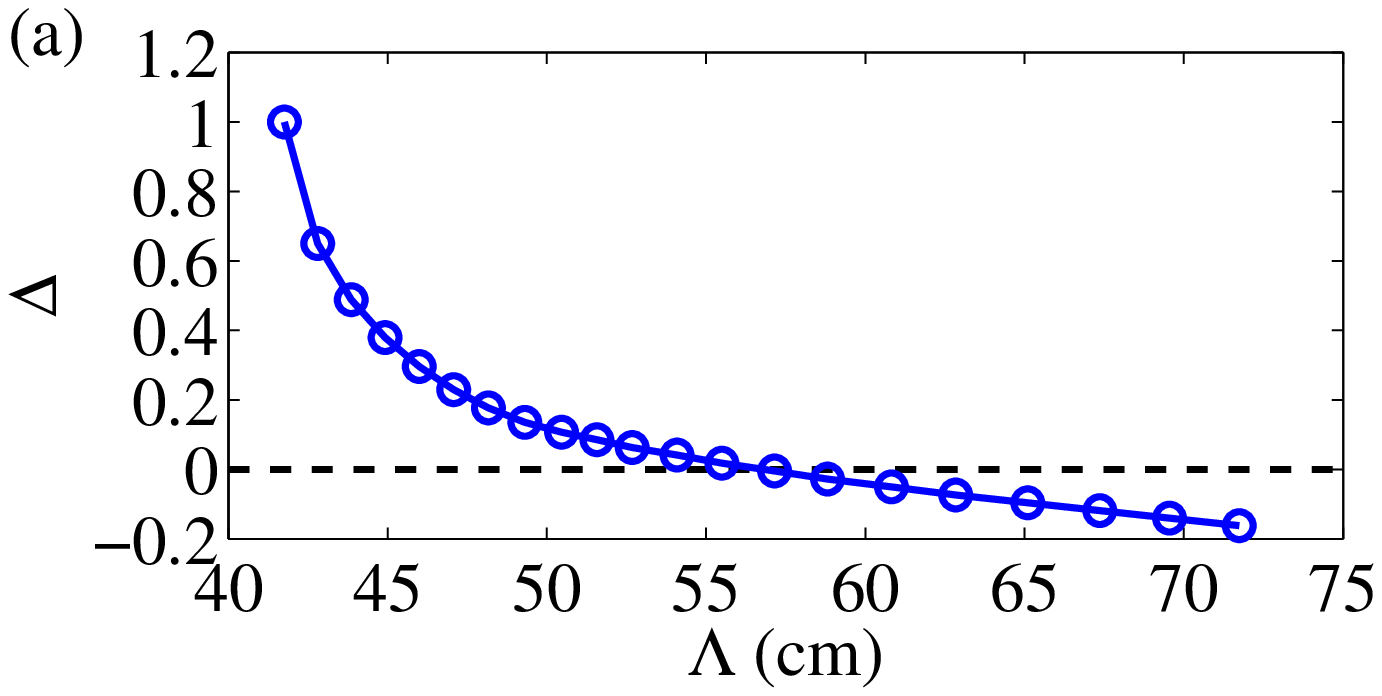, clip =,width=1.0\linewidth } \\
\epsfig{file =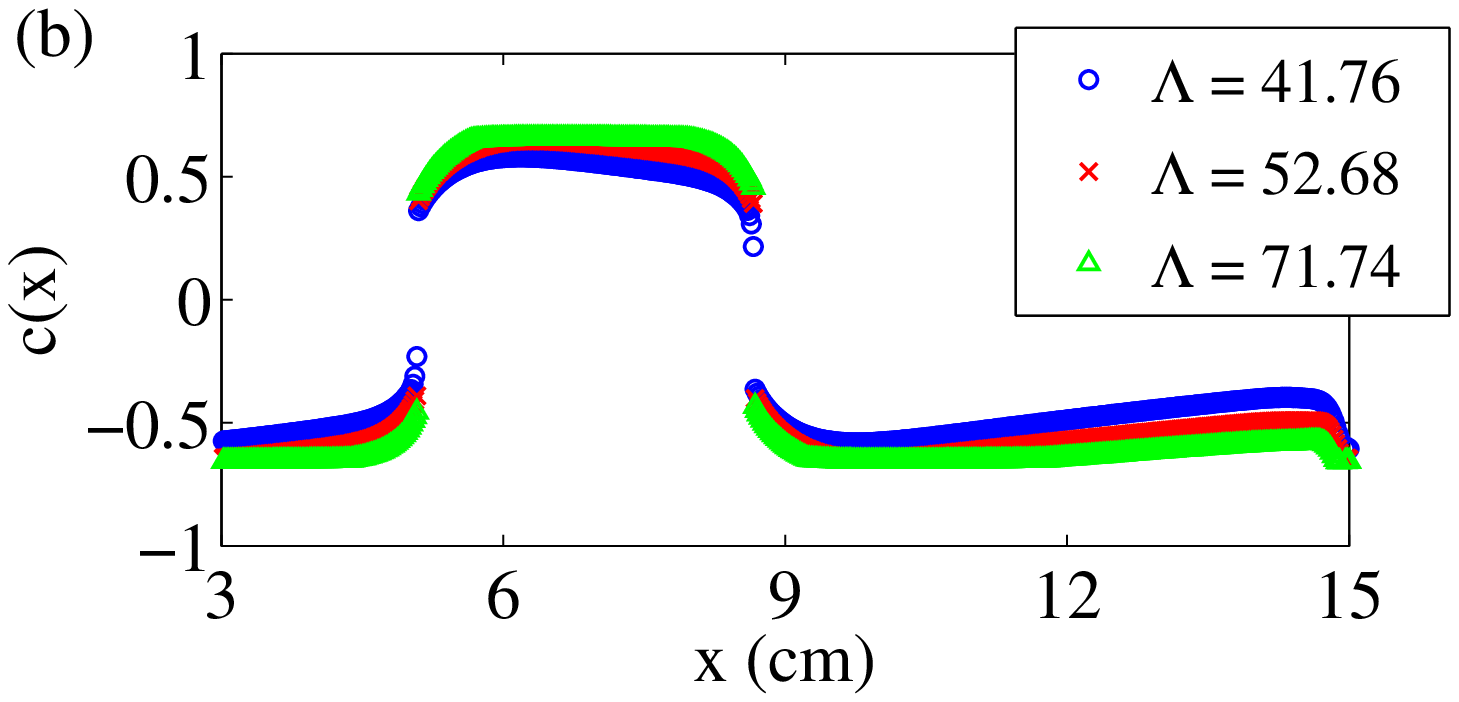, clip =,width=1.0\linewidth } 
\caption{(Color online) Asymmetry of node shapes in the Shiferaw-Fox model via changing $\tau$. (a) Asymmetry $\Delta$ in calcium profiles $c(x)$ of the Shiferaw-Fox model as $\tau$ is decreased from $10$ to $6$ and (b) representative calcium profiles $c(x)$ for three values of $\Lambda$: $\Lambda=41.76$, $52.68$, and $71.74$ (blue circles, red crosses, and green triangles).}\label{fig:Jumps}
\end{figure}

In addition to the movement and pinning of node locations in the discontinuous regime, another form of hysteresis we observed involved the shape of $c(x)$ about the nodes. In particular, we saw that increasing $\Lambda$ or $r$ in the reduced model causes a symmetrizing effect where the jumping points $c_-$ and $c_+$ would approach one another, or in terms of the asymmetry measure $\Delta$ [see Eq.~(\ref{eq:nlin6})], $\Delta$ would decrease. We now show that this phenomenon can be observed in the Shiferaw-Fox model.

In Fig.~\ref{fig:Jumps} we plot the results from a simulation of a cable of length $L=15$ cm where we have set BCL$=330$ ms and slowly decreased $\tau$ from $10$ to $6$. The pacing protocol here is to simulate the cable for $12000$ beats to achieve steady-state, then change $\tau$ by $\Delta\tau=0.1$ every 200 beats. Recall that by decreasing $\tau$ we effectively increase the CV restitution length scale $\Lambda$. In subfigure (a) we plot the asymmetry $\Delta$ vs $\Lambda$. Note these we begin at a normal jump, i.e., $\Delta=1$ after which $\Delta$ decreases. However, we see that rather than approach zero as $\Lambda\to\infty$, in the Shiferaw-Fox model $\Delta$ crosses zero at $\Lambda\approx57$, and takes on negative values, implying that $|c_-|>|c_+|$. In subfigure (b) we plot the profile of the amplitude of calcium alternans $c(x)$ at representative $\Lambda$ values, $\Lambda=41.76$, $52.68$, and $71.74$ plotted as blue circles, red crosses, and green triangles, respectively, where we can see explicitly that $|c_-|$ eventually becomes larger than $|c_+|$. Although the results of the Shiferaw-Fox model in Fig.~\ref{fig:Jumps} (a) differ from those we found with the reduced model in Fig.~\ref{fig:IncLambda} (b), in that $\Delta$ becomes negative, these results do not contradict one another. Recall that for simplicity, in our analysis of the reduced model we assumed zero asymmetry in the Green's function, i.e. $w=0$. By studying the limit of flat CV restitution, i.e., $\Lambda\to\infty$, of the reduced model in Eqs.~(\ref{eq:mapc}) and (\ref{eq:mapa}), it can be shown that, while $\Delta\to0$ as $\Lambda\to\infty$ when $w=0$, if $w>0$, $\Delta$ approaches a negative value, crossing zero at some finite value of $\Lambda$. Thus, we can infer from these results that for these particular parameters the asymmetry of the Green's function $w$ is not exactly zero.

\subsection{Scaling of spatial wavelengths}\label{sec6subE}

We close this section by investigating the scaling of spatial wavelengths found in the Shiferaw-Fox model in the smooth and discontinuous regimes. Recall that in the reduced model [Eqs.~(\ref{eq:mapc}) and (\ref{eq:mapa})] we found that the spatial wavelength $\lambda_s$ of solutions at the onset of alternans in the smooth regime scales sub-linearly with the CV restitution length scale $\Lambda$ [in particular, $\lambda_s\sim(\xi^2\Lambda)^{1/3}$ or $\lambda_s\sim(w\Lambda)^{1/2}$ for traveling or stationary patterns, respectively], while in the discontinuous regime $\lambda_s$ scales linearly with $\Lambda$.

\begin{figure}[b]
\centering
\epsfig{file =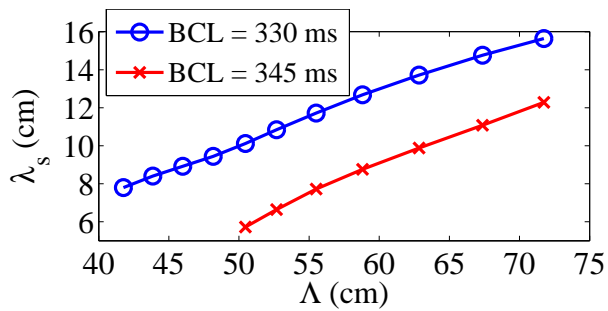, clip =,width=1.0\linewidth } 
\caption{(Color online) Scaling of the spatial wavelength $\lambda_s$ in the Shiferaw-Fox model as a function of the CV restitution length scale $\Lambda$ as $\tau$ is slowly increased from $8$ to $10$. Results in the discontinuous and smooth regimes are obtained using BCL$=330$ ms (blue circles) and $345$ ms (red crosses), respectively.}\label{fig:IonicWavelength}
\end{figure}

To track $\lambda_s$ in the different solution regimes, we set BCL$=330$ ms and $345$ ms to obtain discontinuous and smooth solutions, respectively, and track the node locations and measure $\Lambda$ as $\tau$ is slowly increased from $8$ to $10$. In Fig.~\ref{fig:IonicWavelength} we plot the results from the discontinuous regime (BCL$=330$ ms) and smooth regime (BCL$=345$ ms) in blue circles and red crosses, respectively. (We note that for BCL$=345$ ms increasing $\tau$ beyond $\approx9.5$ has a large effect on the amplitude of solutions, indicating an effective change in the effective instability and/or coupling parameters, so we ignore this data.) Although the range of $\Lambda$ values obtained is not large enough to accurately fit any scaling laws to the data, we observe that $\lambda_s$ grows approximately linearly with $\Lambda$ for BCL$=330$ ms, while a negative concavity is apparent for BCL$=345$ ms, suggesting a possibly sub-linear scaling.

The data presented here was obtained by simulations on a cable of length $L=30$ cm. We note that more accurate measurements of the spatial wavelength require both a long cable as well as a fine spatial discretization. Due to the slow transient dynamics (approximately 10,000--50,000 beats are required to obtain near steady-state behavior, depending on the cable length $L$) this makes further investigations extremely computationally expensive, which we leave open for future work.

\section{Conclusions and outlook}\label{sec7}

In this paper we have derived a system of equations that describes the spatiotemporal dynamics of voltage and calcium alternans along a one-dimensional paced cable, assuming a calcium-mediated instability. Our formulation is based on several properties including APD and CV restitution, bi-directional coupling between voltage and calcium dynamics, and diffusive cell-to-cell coupling. The resulting system is given by Eqs.~(\ref{eq:mapc}) and (\ref{eq:mapa}) and comprises two coupled integro-difference equations that describe the beat-to-beat dynamics of alternans along the cable.

By performing a linear stability analysis around the period-one solution, we found a first bifurcation corresponding to the onset of alternans. This bifurcation occurs at a degree of calcium instability that is less than that for the onset of alternans in a single cell due to the combination of the local coupling between calcium and voltage dynamics and voltage diffusion. Above onset, solutions take the form of smooth traveling or stationary wave patterns depending on the asymmetry of voltage diffusive coupling, as in the case of voltage-driven alternans. Furthermore, the spatial wavelength $\lambda_s$ of smooth SDA patterns and the velocity of traveling patterns obey similar scaling laws as for the voltage-driven case up to the introduction of the bi-directional coupling parameters [i.e., $\lambda_s\sim (\xi^2\Lambda)^{1/3}$ or $\lambda_s \sim (w\Lambda)^{1/2}$ depending on whether the pattern is traveling or stationary, respectively]. This shows that smooth SDA patterns behave similarly in the voltage- and calcium-driven cases close to the alternans bifurcation.  

A more novel aspect of the present work is the characterization of the second bifurcation from smooth to discontinuous SDA patterns at a larger degree of calcium instability. We have found numerically that the width of the spatial profile of calcium alternans varies smoothly from a finite value at the alternans bifurcation to zero at this second bifurcation. Furthermore, we have shown analytically that this width increases linearly with distance away from the second bifurcation towards the smooth side, and that the magnitude of the jump in calcium alternans amplitude increases with square-root of distance on the discontinuous side. In addition to exhibiting a jump in calcium alternans amplitude, we have shown that discontinuous SDA patterns display two key properties  distinguishing them from smooth SDA patterns. Firstly, discontinuous patterns have a wavelength that scales linearly with the CV-restitution scale (i.e., $\lambda_s\sim \Lambda$), as opposed to sublinearly [$\lambda_s\sim (\xi^2\Lambda)^{1/3}$ or $\lambda_s \sim (w\Lambda)^{1/2}$].  Secondly, node motion for those patterns is pinned unidirectionally. Namely, in response to a change of pacing frequency or physiological parameter, nodes can be induced to move towards, but not away from, the pacing site. This is in contrast to smooth patterns whose nodes can move in both directions. We have also investigated the combined effect of subcellular fluctuations of Ca$^{2+}$ cycling (highlighted by recent computational work~\cite{Sato2013PLOS}) and node dynamics. Such fluctuations typically give rise to nodal areas containing many fine-scale phase reversals instead of an isolated node. We have illustrated that inducing nodal movement causes each nodal area to agglomerate into a single node, effectively washing away the remnant effects of the Ca$^{2+}$ fluctuations.

We have validated the theoretical results with ionic model simulations. By varying certain key parameters in the ionic model, we observed both unidirectional pinning and predicted changes of alternans profile shapes. 
In particular, we show that in simulations unidirectional pinning can be observed by changing the pacing frequency or physiological parameters. This prediction could be readily tested in laboratory experiments. However, interpretation of the results requires some care. This is because the direction of the node motion can be non-trivially related to the pacing frequency that can affect both CV restitution and several parameters that characterize the degree of calcium-driven instability and the strength of the bi-directional coupling between APD and Ca alternans. Hence, while an increase of the pacing frequency tends to steepen CV restitution, which alone tends to move the node closer to the pacing site as for voltage-driven alternans~\cite{Echebarria2002PRL,Echebarria2007PRE}, this increase can also affect one or several of those parameters to cancel this effect. As a result, node motion towards the pacing site can be induced by either a decrease or increase of pacing frequency. In one example of the ionic model simulations presented here (see Fig.~\ref{fig:IonicPinningBCL}), we found that node motion towards the pacing site was induced by a decrease of pacing frequency, which reduced the degree of calcium-driven instability (reduced the amplitude of Ca alternans) while having little effect on the CV restitution slope.
This finding is consistent with the prediction of amplitude equations that decreasing the degree of calcium-driven instability pulls the node closer to the pacing site.
We have also found that the scaling of the spatial wavelength with the CV-restitution lengthscale in ionic model simulations is in good qualitative agreement with the linear behavior ($\lambda_s\sim \Lambda$) predicted by the amplitude equations.

This work extends our theoretical understanding of cardiac alternans to the common case of a calcium-driven instability with both positive voltage-to-calcium and calcium-to-voltage couplings. An important result is the understanding of the regime of discontinuous solutions that are only present when alternans is calcium-driven and displays hysteresis. It is plausible that due to the phenomenon of unidirectional pinning, which makes it difficult to expel nodes from the cable, calcium-driven alternans is potentially more arrhythmogenic than voltage-driven alternans. 

The present results further demonstrate that reduced models can help understand complex behaviors observed in ionic model simulations or experiments. However, a number of questions remain. What spatiotemporal dynamics emerges if the instability at the cellular level is both voltage- and calcium-driven, e.g. due to the combination of steep APD restitution and unstable calcium cycling? Does node motion display hysteresis for the case of negative calcium-to-voltage coupling? How is hysteresis in node motion modified by tissue heterogeneities? There is experimental evidence that anatomical heterogeneities influence the location of nodal lines in tissue~\cite{Mironov2008Circ,Ziv2009JPhys} but the combined effect of those heterogeneities and subtle dynamical effects such as unidirectional pinning remain to be explored. Finally, what is the role of hysteresis in higher dimensions where fiber anisotropy influences propagation and in the more complex setting where nodal lines form during reentry~\cite{Kim2007PNAS,Restrepo2009PRE}? We hope that this work will serve as both inspiration for more theoretical advances as well as a guide for future experiments.

\section*{ACKNOWLEDGEMENTS}

The work of P.S.S. was funded in part by the James S. McDonnell Foundation. The work of A.K. was supported by NIH/NHLBI grant R01 HL110791-01A1.

\begin{appendix}

\section{Generalizations of the linear stability analysis}\label{appA}

In this appendix we present the generalizations to the linear stability analysis presented in Sec.~\ref{sec4}. In particular we consider a non-zero APD restitution parameter $\beta$ and a non-zero asymmetry length scale $w$. For simplicity we consider each case separately, beginning with non-zero $\beta$ while still assuming $w=0$. After considering perturbations $\delta c_n(x)=c\lambda^ne^{ikx}$ and $\delta a_n(x)=a\lambda^ne^{ikx}$ to the zero solution, we recover Eq.~(\ref{eq:linstab1}) but find that Eq.~(\ref{eq:linstab2}) is replaced by
\begin{align}
\left[\lambda+\left(\beta-\frac{\beta\Lambda^{-1}}{ik+\Lambda^{-1}}\right)e^{-k^2\xi^2/2}\right]a=\gamma\lambda ce^{-k^2\xi^2/2}.\label{eq:appA1}
\end{align}
Next, we eliminate $c$ and $a$ by combining Eqs.~(\ref{eq:linstab1}) and (\ref{eq:appA1}) and find
\begin{align}
ik\eta\lambda = -(ik+\Lambda^{-1})(\lambda+r)\left(\beta-\frac{\beta\Lambda^{-1}}{ik+\Lambda^{-1}} + \lambda e^{k^2\xi^2/2}\right).\label{eq:appA2}
\end{align}
Eq.~(\ref{eq:appA2}) is quadratic in $\lambda$, and therefore has two solutions. To find the correct root, we choose that which recovers Eq.~(\ref{eq:linstab3}) in the limit, $\beta\to0^+$ which is given by 
\begin{widetext}
\begin{align}
2\lambda=-r-\frac{ik(\beta+\eta)}{ik+\Lambda^{-1}}e^{-k^2\xi^2/2}-\frac{\sqrt{r^2(ik+\Lambda^{-1})^2 - 2ie^{-k^2\xi^2/2}kr(ik+\Lambda^{-1})(\beta-\eta)-e^{-k^2\xi^2}k^2(\beta+\eta)^2}}{ik+\Lambda}.\label{eq:appA3}
\end{align}
\end{widetext}

Surprisingly, applying the absolute instability condition $\partial\lambda/\partial k=0$ to Eq.~(\ref{eq:appA3}) yields the same condition as for $\beta=0$, i.e., Eq.~(\ref{eq:linstab4}), which is solved to leading order by Eq.~(\ref{eq:linstab5}). Next, we insert Eq.~(\ref{eq:linstab4}) into Eq.~(\ref{eq:appA3}) to obtain the growth parameter
\begin{widetext}
\begin{align}
\lambda &= \frac{-r-\eta-\beta-\sqrt{(r-\beta)^2+2(r+\beta)\eta+\eta^2}}{2} + \frac{3i^{-2/3}}{4}\left(\eta+\beta+\frac{r(\eta-\beta)+(\eta+\beta)^2}{\sqrt{r^2+2r(\eta-\beta)+(\eta+\beta)^2}}\right)\left(\frac{\xi}{\Lambda}\right)^{2/3} \nonumber \\ &\hskip4ex- \frac{13i^{-4/3}}{16}\left(\frac{r^3(\eta-\beta)+3r(\eta-\beta)(\eta+\beta)^2+(\eta+\beta)^4+r^2(3\eta^2+\frac{10}{13}\eta\beta+3\beta^2)}{(r^2+2r(\eta-\beta)+(\eta+\beta)^2)^{3/2}}\right)\left(\frac{\xi}{\Lambda}\right)^{4/3}+\mathcal{O}\left[\left(\frac{\xi}{\Lambda}\right)^2\right].\label{eq:appA4}
\end{align}
\end{widetext}
Finally, the onset of alternans is found my setting $|\lambda|=1$, and is given implicitly to leading order by
\begin{widetext}
\begin{align}
1&=\frac{1}{4}\left(r+\eta+\beta+\sqrt{(r-\beta)^2+2(r+\beta)\eta+\eta^2}\right)^2 \nonumber \\&\hskip4ex+\frac{3}{8}\left(-r-\eta-\beta-\sqrt{(r-\beta)^2+2(r+\beta)\eta+\eta^2}\right)\left(\eta+\beta+\frac{r(\eta-\beta)+(\eta+\beta)^2}{\sqrt{(r-\beta)^2+2(r+\beta)\eta+\eta^2}}\right)\left(\frac{\xi}{\Lambda}\right)^{2/3}.\label{eq:appA5}
\end{align}
\end{widetext}

To find the spatial wavelength $\lambda_s$ and velocity $v$ of smooth solutions near the onset of alternans, we recall that $\lambda_s=2\pi/k_{Re}$ and $v=\Omega/k_{Re}$, and $\Omega\approx-\lambda_{Im}$. Since the wave number $k$ was found to be the same for $\beta\ne0$ as for $\beta=0$, the spatial wavelength $\lambda_s$ is also the same, given by Eq.~(\ref{eq:linstab8}). However, the velocity $v$ depends on the growth parameter, which is not the same. Combining Eq.~(\ref{eq:appA4}) with Eq.~(\ref{eq:linstab5}) yields a velocity that is to second order
\begin{widetext}
\begin{align}
v&=\frac{3\xi}{4}\left(\eta+\beta+\frac{r(\eta-\beta)+(\eta+\beta)^2}{\sqrt{r^2+2r(\eta-\beta)+(\eta+\beta)^2}}\right)\left(\frac{\xi}{\Lambda}\right)^{1/3}\nonumber \\&\hskip4ex- \frac{13\xi}{16}\left(\eta+\beta+\frac{r^3(\eta-\beta)+3r(\eta-\beta)(\eta+\beta)^2+(\eta+\beta)^4+r^2(3\eta^2+\frac{10}{13}\eta\beta)+3\beta^2}{(r^2+2r(\eta-\beta)+(\eta+\beta)^2)^{3/2}}\right)\left(\frac{\xi}{\Lambda}\right).\label{eq:appA6}
\end{align}
\end{widetext}

\begin{figure}[t]
\centering
\epsfig{file =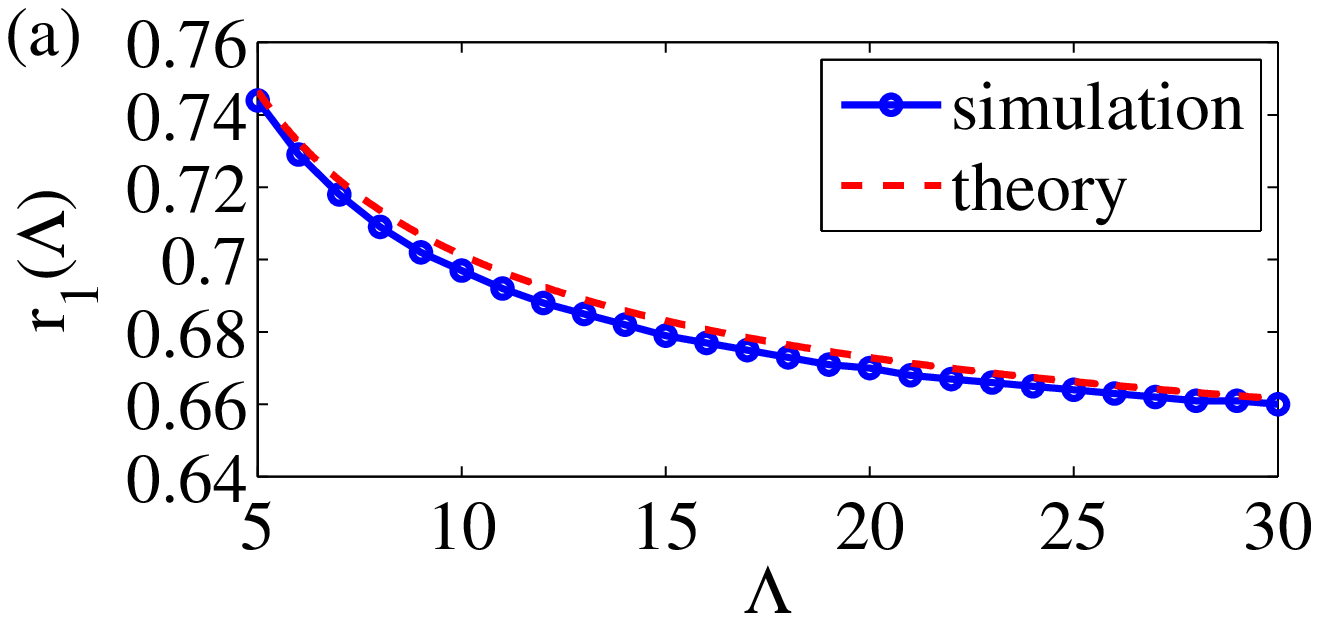, clip =,width=1.0\linewidth } \\
\epsfig{file =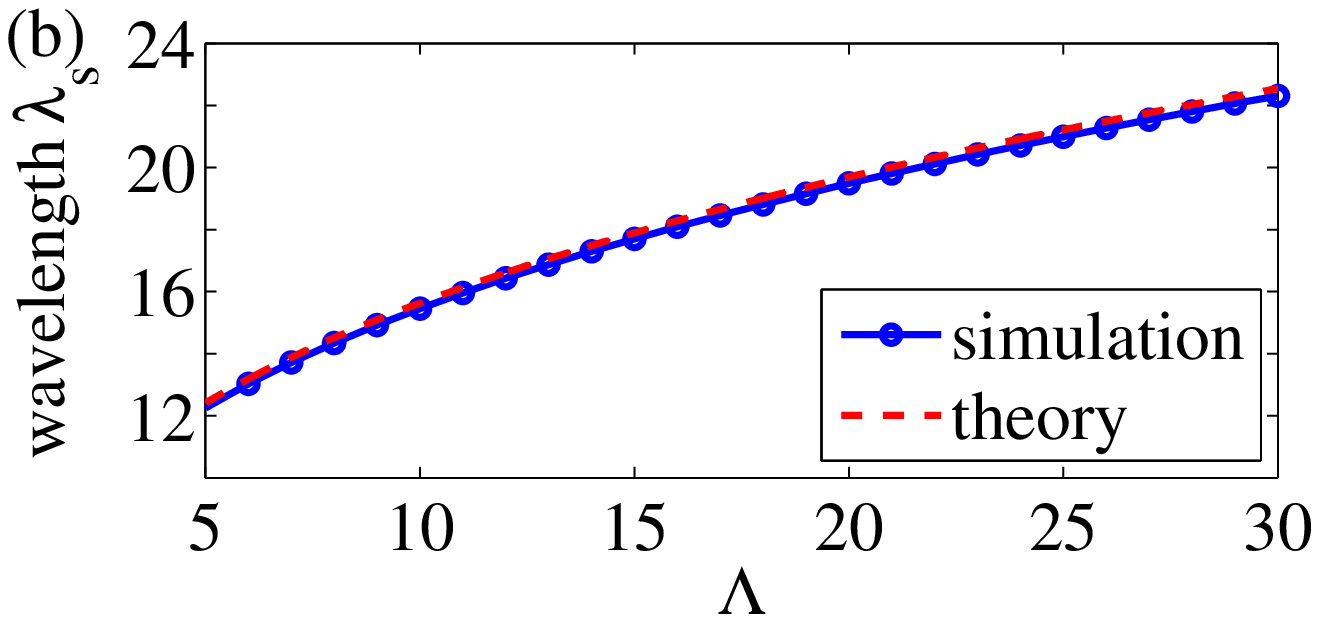, clip =,width=1.0\linewidth } \\
\epsfig{file =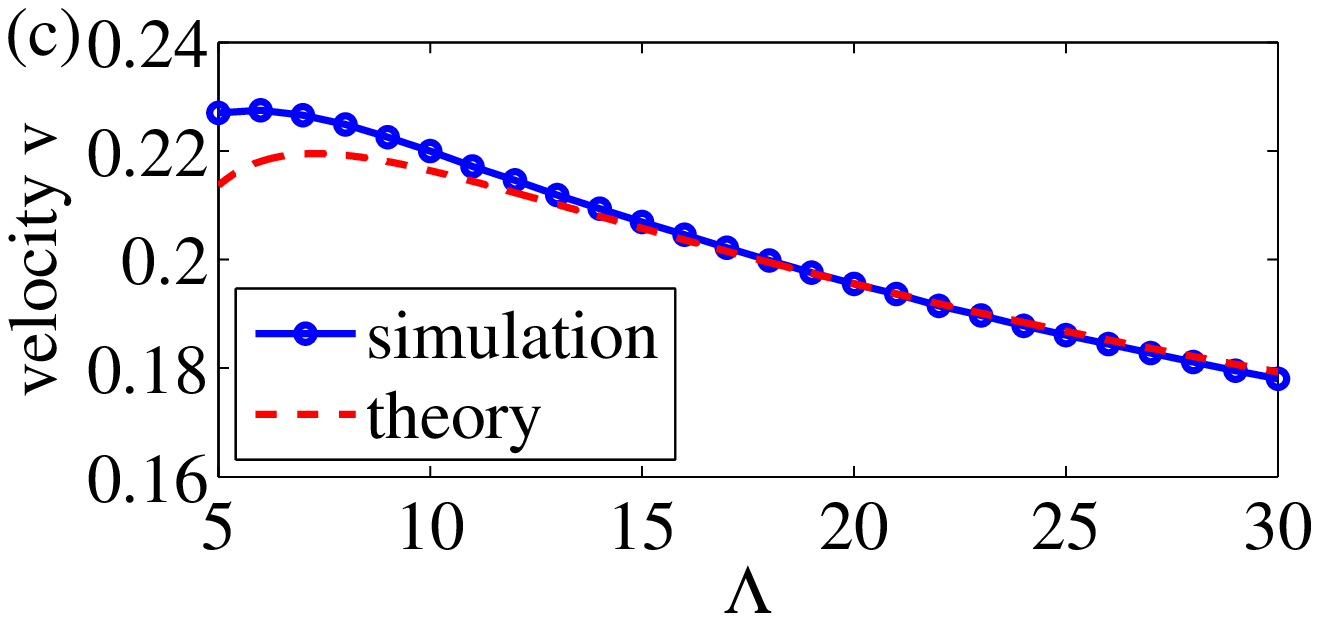, clip =,width=1.0\linewidth }
\caption{(Color online) (a) The critical onset value $r_1(\Lambda)$, (b) spatial wavelength $\lambda_s$, and (c) velocity $v$ of smooth solutions near the onset of alternans as observed from numerical simulations of Eqs.~(\ref{eq:mapc}) and (\ref{eq:mapa}) (blue circles) compared to our theoretical predictions given by Eqs.~(\ref{eq:appA5}), (\ref{eq:linstab8}), and (\ref{eq:appA6}) (dashed red) for non-zero APD restitution $\beta=0.2$. Other parameters are $\alpha,\gamma=\sqrt{0.3}$, $\xi=1$, and $w=0$.}\label{fig:Onset2}
\end{figure}

We validate these results by plotting in Fig.~\ref{fig:Onset2} (a)--(c) the critical onset value $r_1(\Lambda)$, spatial wavelength $\lambda_s$, and velocity $v$ of solutions near onset computed directly from simulations of Eqs.~(\ref{eq:mapc}) and (\ref{eq:mapa}) (blue circles) compared to the theoretical predictions given by Eqs.~(\ref{eq:appA5}), (\ref{eq:linstab8}), and (\ref{eq:appA6}) (dashed red) for $\beta=0.2$. Other parameters are $\alpha,\gamma=\sqrt{0.3}$, $\xi=1$, and $w=0$. We note that the agreement between theory and simulations is excellent.

We now consider a non-zero asymmetry length scale $w\ne0$. For simplicity, we again choose the APD restitution parameter $\beta$ to be zero. Inserting perturbations of the form $\delta c_n(x)=c\lambda^n e^{ikx}$ and $\delta a_n(x)=a\lambda^n e^{ikx}$ into Eqs.~(\ref{eq:mapc}) and (\ref{eq:mapa}), we recover Eq.~(\ref{eq:linstab1}), but find that Eq.~(\ref{eq:linstab2}) is replaced by
\begin{align}
a=\gamma ce^{-k^2\xi^2/2}\left[1-ikw\left(1-\frac{k^2\xi^2}{2}\right)\right].\label{eq:appA7}
\end{align}
Combining Eqs.~(\ref{eq:linstab1}) and (\ref{eq:appA7}), we eliminate $a$ and $c$ and find that the growth parameter satisfies
\begin{align}
\lambda &= -r-\eta\left(1-\frac{\Lambda^{-1}}{ik+\Lambda^{-1}}\right)e^{-k^2\xi^2/2}\nonumber \\ &\hskip12ex\times\left[1-ikw\left(1-\frac{k^2\xi^2}{2}\right)\right].\label{eq:appA8}
\end{align}
Next, imposing the absolute instability condition $\partial \lambda/\partial k=0$ on Eq.~(\ref{eq:appA8}) then yields the equation
\begin{align}
0&=2i\Lambda^{-1} +4kw\Lambda^{-1} + 2i(w-\Lambda^{-1}\xi^2)k^2\nonumber \\ &\hskip2ex+(2\xi^2-6w\Lambda^{-1}\xi^2)k^3-5iw\xi^2k^4\nonumber \\ &\hskip2ex+w\Lambda^{-1}\xi^4k^5+iw\xi^4k^6.\label{eq:appA9}
\end{align}
To leading order, Eq.~(\ref{eq:appA9}) is balanced by the wave number
\begin{align}
k=-1/\sqrt{w\Lambda},\label{eq:appA10}
\end{align}
which can be inserted into Eq.~(\ref{eq:appA8}) to yield to leading order the growth rate
\begin{align}
\lambda=-r-\eta + \eta\frac{\xi^2}{2w\Lambda}.\label{eq:appA11}
\end{align}
The onset of alternans can then be found by setting $\lambda=-1$ in Eq.~(\ref{eq:appA11}), yielding to leading order
\begin{align}
r_1(\Lambda)=1-\eta+\eta\frac{\xi^2}{2w\Lambda}.\label{eq:appA12}
\end{align}

Since the growth rate given in Eq.~(\ref{eq:appA11}) is purely real, solutions near onset have zero velocity $v=0$. However, the spatial wavelength $\lambda_s$ is given by $2\pi/|k_{Re}|$, which yields to leading order
\begin{align}
\lambda_s=2\pi\sqrt{w\Lambda}.\label{eq:appA13}
\end{align}

\begin{figure}[t]
\centering
\epsfig{file =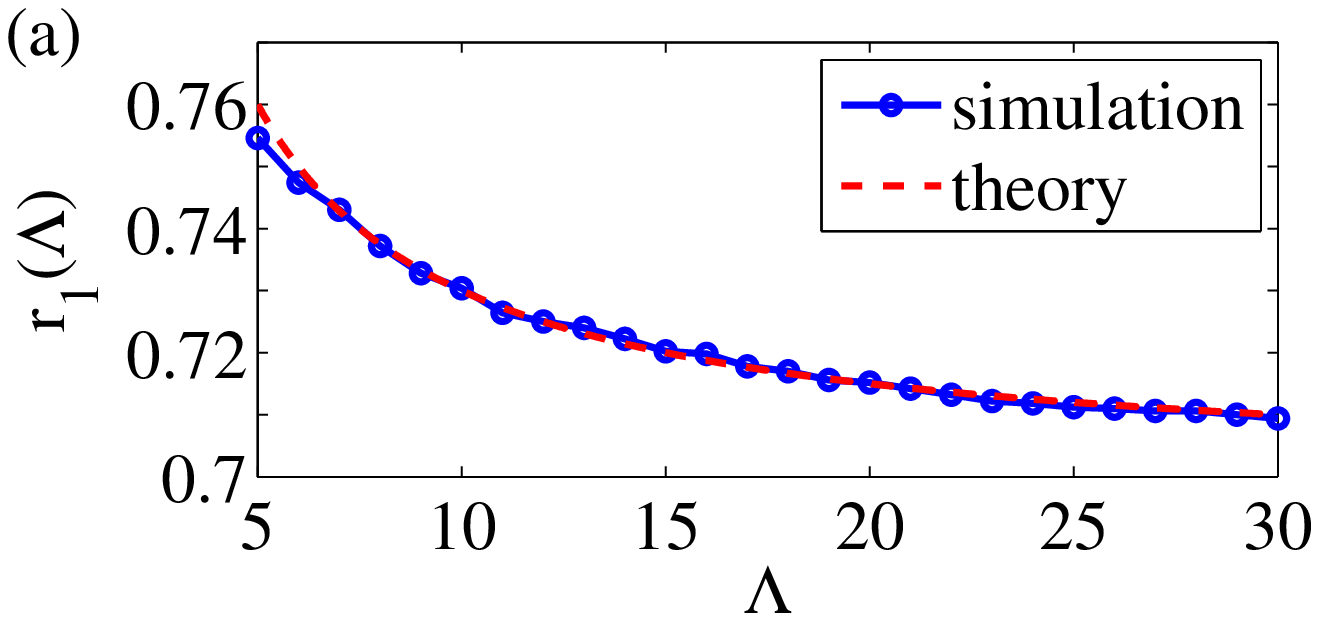, clip =,width=1.0\linewidth } \\
\epsfig{file =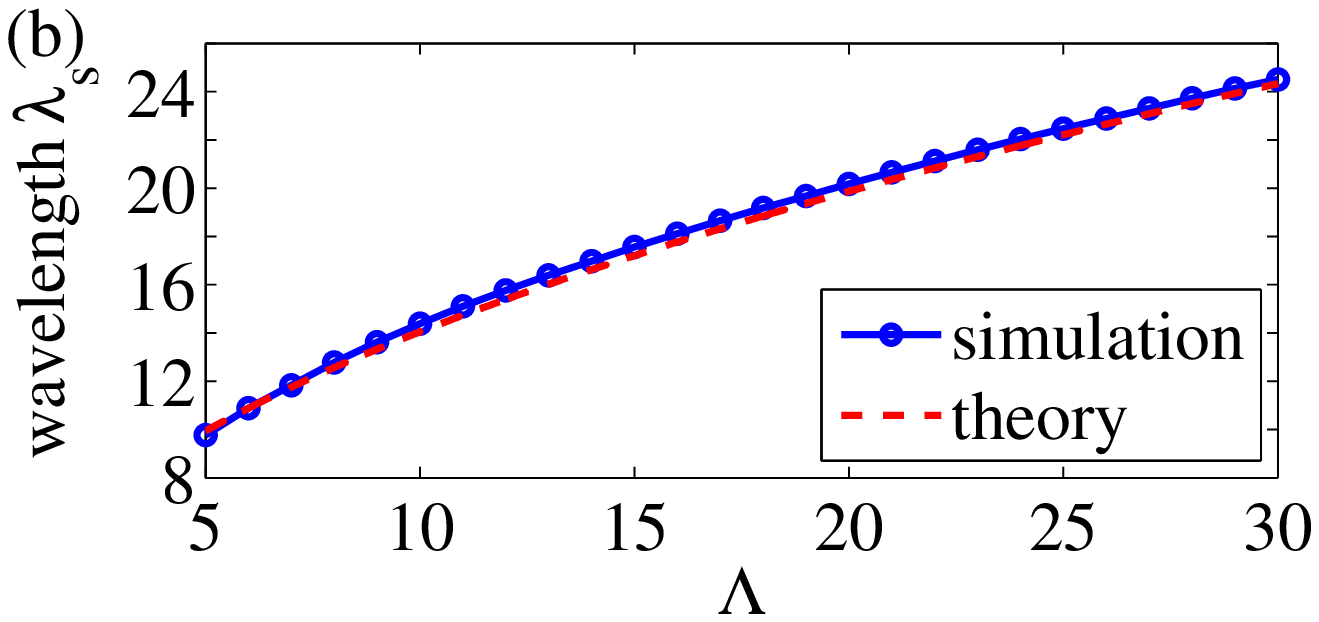, clip =,width=1.0\linewidth }
\caption{(Color online) (a) The critical onset value $r_1(\Lambda)$ and (b) spatial wavelength $\lambda_s$ of smooth solutions near the onset of alternans as observed from numerical simulations of Eqs.~(\ref{eq:mapc}) and (\ref{eq:mapa}) (blue circles) compared to our theoretical predictions given by Eqs.~(\ref{eq:appA12}) and (\ref{eq:appA13}) (dashed red) for non-zero asymmetry $w=0.4$. Other parameters are $\alpha,\gamma=\sqrt{0.3}$, $\beta=0$, and $\xi=1$.}\label{fig:Onset3}
\end{figure}

We validate these results  by plotting in Fig.~\ref{fig:Onset3} (a) and (b) the critical onset value $r_1(\Lambda)$ and spatial wavelength $\lambda_s$ of solutions near onset computed directly from simulations of Eqs.~(\ref{eq:mapc}) and (\ref{eq:mapa}) (blue circles) compared to the theoretical predictions given by Eqs.~(\ref{eq:appA12}) and (\ref{eq:appA13}) (dashed red) for $w=0.4$. Other parameters are $\alpha,\gamma=\sqrt{0.3}$, $\beta=0$ and $\xi=1$. As predicted by our theory, solutions are stationary. The agreement between theory and simulations for both $r_1(\Lambda)$ and $\lambda_s$ is excellent.

\section{Transient node dynamics}\label{appB}

In this appendix we investigate the transient node dynamics of solutions in the discontinuous regime. Specifically, we study the evolution of node locations as movement is induced towards the pacing site. We also address here the effect that using different spatial discretization $\Delta x$ has on the the transient dynamics of Eqs.~(\ref{eq:mapc}) and (\ref{eq:mapa}). 

\begin{figure}[b]
\centering
\epsfig{file =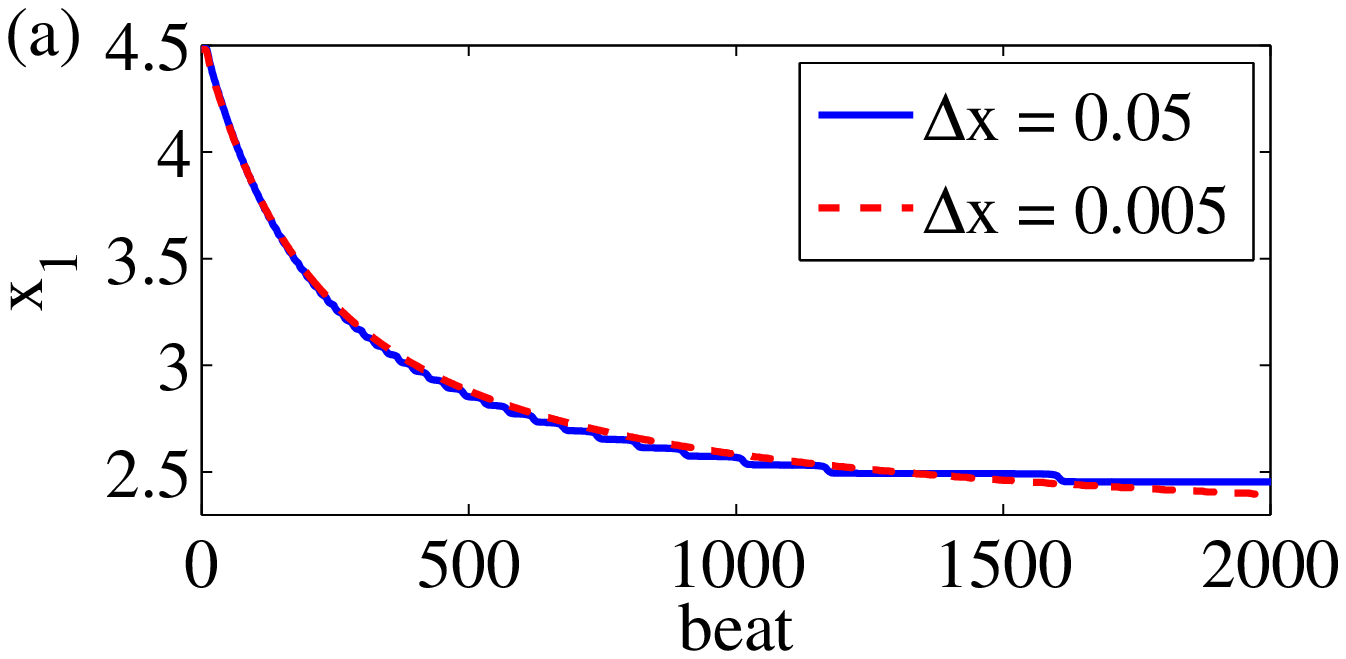, clip =,width=1.0\linewidth } \\
\epsfig{file =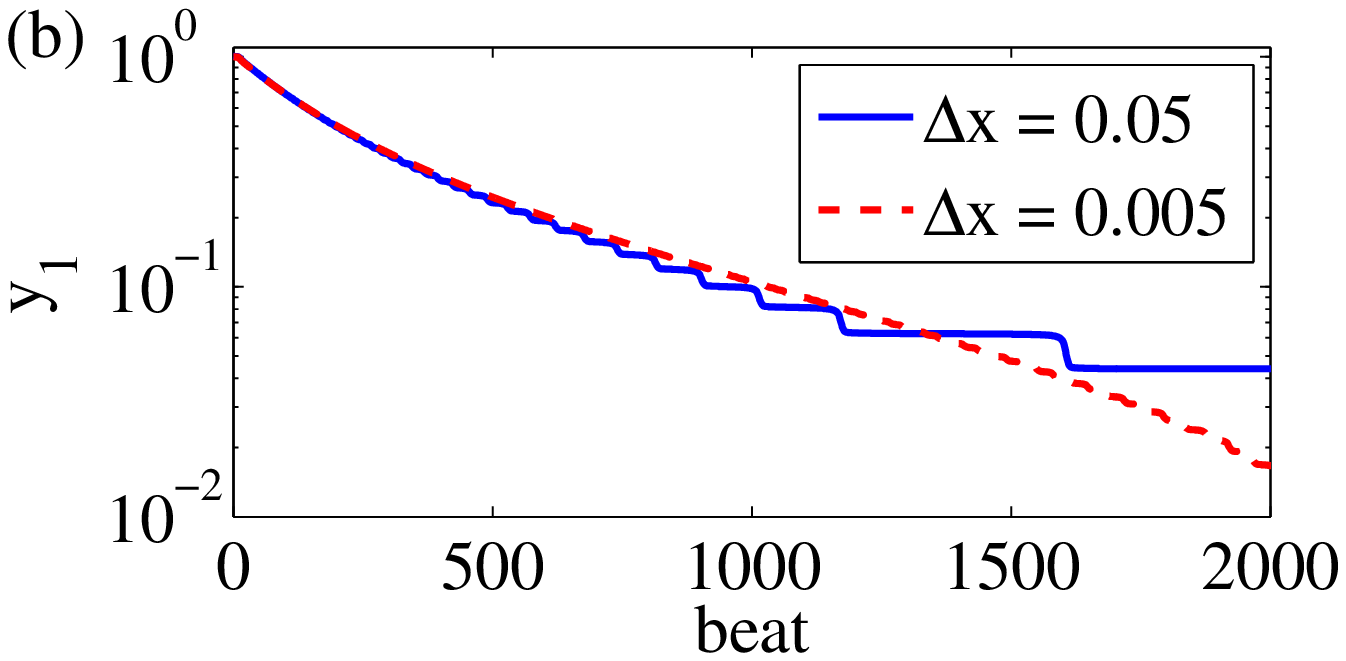, clip =,width=1.0\linewidth }
\caption{(Color online) (a) Transient dynamics of the first node $x_1$ after node movement is induced by decreasing $\Lambda=50$ to $\Lambda=20$ for coarse and refined spatial discretizations $\Delta x=0.05$ (solid blue) and $0.005$ (dashed red). (b) Decay of the transformed variable $y_1$ (see text for the definition of $y_1$). Other parameters are $r=1.15$, $\alpha,\gamma=\sqrt{0.3}$, $\beta=0$, and $\xi=1$ with $L=20$.}\label{fig:dx}
\end{figure}

We present here the results from an experiment where we track the first node location after node movement is induced towards the pacing site. Specifically, we consider a cable of length $L=20$ at steady-state with $r=1.15$ and $\Lambda=50$ and track the first node location $x_1$ after $\Lambda$ is decreased to $20$. Importantly, we run the experiment with two different spatial discretizations, $\Delta x=0.05$ and $0.005$, corresponding to the most coarse and most refined discretization used throughout this paper. Other parameters are $\alpha,\gamma=\sqrt{0.3}$, $\beta=0$, $\xi=1$, and $w=0$. In Fig.~\ref{fig:dx} (a) we plot the evolution of $x_1$ as a function of beat number after the change in $\Lambda$, plotting the results from $\Delta x=0.05$ and $0.005$ in solid blue and dashed red, respectively. We emphasize that the node location $x_1$ is computed using a linear interpolation between the adjacent point on either side of the node. Even with the interpolation, the node location $x_1$ decreases in a step-like fashion that is most evident for coarser $\Delta x$. We find that this nonlinear effect is a result of the following process. As described by the potential well framework in subsection~\ref{sec5subC}, node movement is driven by points switching from $c_-$ to $c_+$. This switching process occurs quickly, resulting in a sharp decrease in $x_1$, and is then followed by a slow nonlocal recovery of the surrounding profile, until the next point switches and the process repeats. 

Importantly, we also observe from Fig.~\ref{fig:dx} (a) that although the spatial discretization $\Delta x$ affects the transient behavior of $x_1$, it does not cause a significant change in the overall convergence rate. In Fig.~\ref{fig:dx} (b) we investigate the convergence rate further. Denoting the steady-state first node locations for $\Lambda=50$ and $\Lambda=20$ by $x_1'$ and $x_1''$, we define the new variable
\begin{align}
y_1=\frac{x_1-x_1''}{x_1'-x_1''},\label{eq:affine}
\end{align}
i.e., image of the affine transformation that maps $x_1=x_1'$ and $x_1=x_1''$ to one and zero, respectively. We note that the values $x_1'$ and $x_1''$ depend on the discretization used and should be computed separately for different discretizations. For instance, here we have found that $x_1'=4.4953$ and $x_1''=2.4241$ for $\Delta x=0.05$ and $x_1'=4.4817$ and $x_1''=2.3907$ for $\Delta x=0.005$. We note that the difference between the two values of both $x_1'$ and $x_1''$ is less than $0.05$, the largest of the two spatial discretizations used. The transformation $x_1\mapsto y_1$ allows us to investigate the convergence of $x_1$ to $x_1''$ by studying the decay of $y_1$. In Fig.~\ref{fig:dx} (b) we plot $y_1$ versus the beat number for $\Delta x=0.05$ (solid blue) and $\Delta x=0.005$ (dashed red), noting the logarithmic vertical axis, which reveals approximately exponential convergence. Panel (b) also highlights the effect of using a larger discretization as the step-like behavior becomes evident when $y_1$ decreases to be of the order of $\Delta x$. We find here that for $\Delta x=0.05$ and $0.005$ this occurs approximately at beat $700$ and $1800$, respectively.

\section{Analysis of the flat CV limit}\label{appC}

\begin{figure}[t]
\centering
\epsfig{file =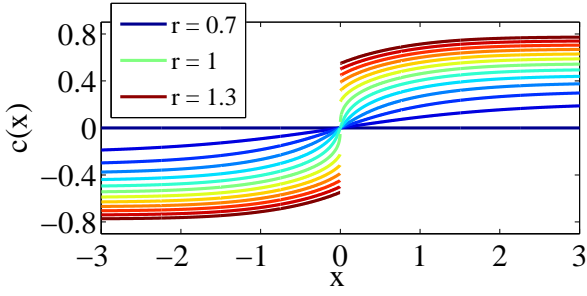, clip =,width=1.0\linewidth } 
\caption{(Color online) Example solutions for the limit of flat CV restitution ($\Lambda\to\infty$) on a bi-infinite cable with a node at $x=0$ for $r$ ranging from $0.7$ (blue curve with the smallest amplitude) to $1.3$ (red curve with the largest amplitude). Other parameters are $\alpha,\gamma=\sqrt{0.3}$, $\beta=0$, $\xi=1$, and $w=0$.}\label{fig:FlatCV}
\end{figure}

In this appendix we present the analysis of the dynamics of Eqs.~(\ref{eq:mapc}) and (\ref{eq:mapa}) in the flat CV limit where $\Lambda\to\infty$. In particular, we are interested in the steady-state behavior of solutions near a node. For simplicity we consider the case of a purely symmetric Green's function, i.e., $w=0$. In this case, Eqs.~(\ref{eq:mapc}) and (\ref{eq:mapa}) simplify to 
\begin{align}
c_{n+1}(x) &= -rc_n(x) + c_n^3(x) - \alpha a_n(x),\label{eq:mapcFlatCV}\\
a_{n+1}(x) &= \int G(x,x')\left[-\beta a_n(x')+\gamma c_{n+1}(x')\right]dx'.\label{eq:mapaFlatCV}
\end{align}
Do to the absence of the CV terms, node formation relies entirely on initial conditions. Analytical results can then be obtained by studying the case of a bi-infinite cable with a single node located at $x=0$. Thus, from this point we will use for the Green's function $G(x,x')=G(x'-x)$, assuming that the integration of Eq.~(\ref{eq:mapaFlatCV}) is over the whole real line. In Fig.~\ref{fig:FlatCV} we plot several example solutions obtained from simulations over a range from $0.7$ (blue curve with the smallest amplitude) to $1.3$ (red curve with the largest amplitude) using $\alpha,\gamma=\sqrt{0.3}$, $\beta=0$, and $\xi=1$. Note that after the onset of alternans $c(x)$ remains smooth through the node, but then develops a discontinuity at sufficiently large $r$, much like what we have seen for finite CV restitution. Furthermore, we observe that solutions are antisymmetric about $x=0$, i.e., $c(-x)=-c(x)$.

To find a steady-state description of solutions we begin by repeatedly inserting Eq.~(\ref{eq:mapaFlatCV}) into Eq.~(\ref{eq:mapcFlatCV}). After rearranging the order of integration and using the fact that the convolution of two Gaussians with variances $\sigma_a^2$ and $\sigma_b^2$ is another Gaussian with variance $\sigma_a^2+\sigma_b^2$, we obtain
\begin{align}
c_{n+1}(x) = -rc_n(x) + c_n^3(x) -\eta\int\widehat{G}_{\xi^2}(x'-x)c_n(x')dx',\label{eq:appB01}
\end{align}
where
\begin{align}
\widehat{G}_{\xi^2}(x) = \sum_{m=0}^\infty\beta^mG_{(m+1)\xi^2}(x),\label{eq:appB02}
\end{align}
and $G_{\xi^2}(x)$ denotes a Gaussian with variance $\xi^2$. We note that the sum in Eq.~(\ref{eq:appB02}) converges uniformly because $0\le\beta<1$. For the sake of calculating the critical bifurcation values and the shape (i.e., length scale) of the phase reversal, it is convenient to define
\begin{align}
c_0=\lim_{x\to0^+|c(x)|}\hskip2ex\text{and}\hskip2ex c_\infty=\lim_{x\to\infty}|c(x)|\label{eq:appB03}
\end{align}
as the limiting values of $c(x)$ near and far from the node, respectively. We next assume antisymmetric periodic steady-state solutions $-c_{n+1}(x) = c_n(x)=c(x)$ and use the fact that $\widehat{G}$ is symmetric to find
\begin{align}
&\lim_{x\to0^+}\int\widehat{G}_{\xi^2}(x'-x)c(x')dx' = 0,\\
&\lim_{x\to\infty}\int\widehat{G}_{\xi^2}(x'-x)c(x')dx'= \lim_{x\to\infty}\frac{c(x)}{1-\beta}=\frac{\pm c_\infty}{1-\beta}.
\end{align}
Thus, the steady-state values of $c_\infty$ and $c_0$ are given by
\begin{align}
c_\infty &= \left\{\begin{array}{ll}0 &\text{if }r\le1-\frac{\eta}{1-\beta},\\ \sqrt{r-1+\frac{\eta}{1-\beta}} &\text{if }r>1-\frac{\eta}{1-\beta},\end{array}\right.\\
c_0 &= \left\{\begin{array}{ll}0 &\text{if }r\le1,\\ \sqrt{r-1} &\text{if }r>1.\end{array}\right.
\end{align}
Thus, the critical bifurcation corresponding to the onset of alternans and the formation of the discontinuity are given by $r_1^\infty = 1-\frac{\eta}{1-\beta}$ and $r_2^\infty=1$.

We now derive the length scale $l$ of phase-reversals for smooth solutions assuming $r_1^\infty<r< r_2^\infty$. Recall that $l$ [defined in Eq.~(\ref{eq:phasereversal})] depends on the derivative $c'(x)$ evaluated at the node. To calculate $c'(0)$ we consider periodic solutions of Eq.~(\ref{eq:appB01}) and take a derivative with respect to $x$, obtaining
\begin{align}
(r-1)c'(x) = 3c^2(x) - \eta\int\partial_x\widehat{G}_{\xi^2}(x'-x)c(x')dx'.\label{eq:appB04}
\end{align}
Next, we use the fact that $\partial_x\widehat{G}_{\xi^2}(x'-x)=-\partial_{x'}\widehat{G}_{\xi^2}(x'-x)$, integrate Eq.~(\ref{eq:appB04}) by parts, and evaluate at $x=0$ to obtain
\begin{align}
c'(0) =\frac{\pm\eta}{1-r}\int\widehat{G}_{\xi^2}(x')c'(x')dx'.\label{eq:appB05}
\end{align}
Finally, expanding the Green's function to $\widehat{G}_{\xi^2}(x')=1/\sqrt{2\pi(m+1)\xi^2}+\mathcal{O}(x'^2)$ and integrating Eq.~(\ref{eq:appB05}) yields
\begin{align}
c'(0)\approx\frac{\pm2\eta c_\infty\text{Li}_{1/2}(\beta)}{(1-r)\beta\sqrt{2\pi\xi^2}},\label{eq:appB06}
\end{align}
where $\text{Li}_{s}(z) = \sum_{m=1}^\infty z^m/m^s$ is the polylogarithm function. Thus, the length scale $l$ of phase reversals is given by
\begin{align}
l\approx\frac{(1-r)\sqrt{2\pi\xi^2}}{\eta\text{Li}_{1/2}(\beta)/\beta}.\label{eq:appB07}
\end{align}
We note that the approximation of $\widehat{G}_{\xi^2}(x')$ as a constant used in obtaining the approximations for $c'(0)$ and $l$ in Eqs.~(\ref{eq:appB06}) and (\ref{eq:appB07}) is valid provided that $l$ remains smaller than the width of $\widehat{G}_{\xi^2}(x')$. Thus, Eq.~(\ref{eq:appB07}) is most accurate when $r$ approaches one. 

\end{appendix}

\bibliographystyle{plain}

\end{document}